\documentclass[12pt]{article}
\usepackage{fullpage,epsfig,graphics,amsbsy,amssymb,cancel,bbm,amsthm,cite}
\usepackage{lmodern}
\usepackage{slantsc}
\usepackage{psfrag,hyperref,color,slashed}
\usepackage{graphicx}
\usepackage{amsfonts,mathdots,dsfont}
\usepackage{amssymb,amsmath,amscd,mathrsfs,wasysym}
\usepackage{enumitem}
\usepackage[T1]{fontenc}
\usepackage{tikz}
\usetikzlibrary{arrows,chains,matrix,positioning,scopes,snakes,fadings}

\makeatletter
\tikzset{join/.code=\tikzset{after node path={%
\ifx\tikzchainprevious\pgfutil@empty\else(\tikzchainprevious)%
edge[every join]#1(\tikzchaincurrent)\fi}}}
\makeatother

\tikzset{>=stealth',every on chain/.append style={join},
         every join/.style={->}}
\tikzset{
    >=stealth',
    punkt/.style={
           rectangle,
           rounded corners,
           draw=black, very thick,
           text width=6.5em,
           minimum height=2em,
           text centered},
    pil/.style={
           ->,
           thick,
           shorten <=2pt,
           shorten >=2pt,}
}

\newcommand{\BB}{\mathbb}

\newcommand{\FR}{\mathfrak}

\def\N{{\cal N}}

\newcommand{\bea}{\begin{eqnarray}}
\newcommand{\eea}{\end{eqnarray}}
\newcommand{\be}{\begin{equation}}
\newcommand{\ee}{\end{equation}}
\newcommand{\nn}{\nonumber}
\newcommand{\Tr}{\operatorname{Tr}}

\newcommand{\sbullet}{\vcenter{\hbox{\tiny$\bullet$}}}

\newcommand{\bra}{\langle}
\newcommand{\ket}{\rangle}
\newcommand{\im}{\operatorname{Im}}
\newcommand{\re}{\operatorname{Re}}
\newcommand{\pf}{\operatorname{pfaff}}
\newcommand{\coker}{\operatorname{coker}}
\newcommand{\img}{\operatorname{img}}
\newcommand{\To}{\Rightarrow}

\newcommand{\ind}{\operatorname{ind}}
\newcommand{\sgn}{\operatorname{sgn}}
\newcommand{\reeb}{\textsl{\textsc{r}}}
\newcommand{\xeeb}{\textsl{\textsc{x}}}
\newcommand{\veeb}{\textsl{\textsc{v}}}

\newcommand{\rk}{\operatorname{rk}}

\newcommand{\sdet}{\operatorname{sdet}}
\newcommand{\End}{\operatorname{End}}
\newcommand{\Lie}{\operatorname{Lie}}
\newcommand{\ext}{\operatorname{Ext}}
\newcommand{\opn}{\operatorname}

\def\ga{\alpha}
\def\gb{\beta}

\def\Gc{\Gamma}
\def\gd{\delta}
\def\gdh{\textrm{\dh}}
\def\Gd{\Delta}
\def\ep{\epsilon}
\def\gt{\theta}
\def\vgt{\vartheta}

\def\gs{\sigma}
\def\ep{\epsilon}
\def\vrho{\varrho}
\def\Gs{\Sigma}
\def\gk{\kappa}
\def\gl{\lambda}

\def\Go{\Omega}
\def\go{\omega}
\def\Gu{\Upsilon}

\DeclareMathAlphabet{\mathpzc}{OT1}{pzc}{m}{it}

\newtheorem{theorem}{Theorem}[section]
\newtheorem{lemma}[theorem]{Lemma}
\newtheorem{proposition}[theorem]{Proposition}

\theoremstyle{definition}
\newtheorem{example}[theorem]{Example}
\newtheorem{remark}[theorem]{Remark}
\newtheorem{corollary}[theorem]{Corollary}

\numberwithin{equation}{section}
\begin{document}
\begin{flushright} \small
UUITP-16/19
 \end{flushright}
\smallskip
\begin{center} \Large
{\bf Transversally Elliptic Complex and Cohomological Field Theory}
 \\[12mm] \normalsize
{\bf Guido Festuccia${}^a$, Jian Qiu${}^{a,b}$, Jacob Winding${}^c$, Maxim Zabzine${}^a$} \\[8mm]
 {\small\it
   ${}^a$Department of Physics and Astronomy,
     Uppsala University,\\
     Box 516,
     SE-75120 Uppsala,
     Sweden\\
   \vspace{.5cm}
      ${}^b$ Mathematics Institute,  Uppsala University, \\
   Box 480, SE-75106 Uppsala, Sweden\\
   \vspace{.5cm}
      ${}^c$ Korea Institute for Advanced Study, \\
      85 Hoegiro, Dongdaemun-gu, Seoul 02455, Republic of Korea   }
\end{center}
\vspace{4mm}
\begin{abstract}
This work is a continuation of our previous paper arXiv:1812.06473 where we have constructed ${\cal N}=2$ supersymmetric Yang-Mills theory on
4D manifolds with a Killing vector field with isolated fixed points. In this work we expand on the mathematical aspects of the theory, with a particular focus on its nature as a cohomological field theory.
The well-known Donaldson-Witten theory is a twisted version of ${\cal N}=2$ SYM and can also be constructed using the Atiyah-Jeffrey construction \cite{ATIYAH1990119}. This theory is concerned with the moduli space of anti-self-dual gauge connections, with a deformation theory controlled by an elliptic complex. More generally, supersymmetry requires considering configurations that look like either instantons or anti-instantons around fixed points, which we call flipping instantons.
The flipping instantons of our 4D ${\cal N}=2$ theory are derived from the 5D contact instantons.  The novelty is that
 their deformation theory is controlled by a transversally elliptic complex, which we demonstrate here. We repeat the Atiyah-Jeffrey construction in the equivariant
 setting and arrive at the Lagrangian (an equivariant Euler class in the relevant field space) that was also obtained from our previous work arXiv:1812.06473. We show that
the transversal ellipticity of the deformation complex is crucial for the non-degeneracy of the Lagrangian and the calculability of the theory.
 Our construction is valid on a large class of quasi toric 4~manifolds.
 \noindent

 \noindent
\end{abstract}

\eject
\normalsize

\tableofcontents%

\section{Introduction}\label{sec_intro}

This work is the sequel to our previous paper \cite{Festuccia:2018rew} where we have constructed a ${\cal N}=2$ supersymmetric gauge theory on any 4D manifold with a Killing vector field with isolated fixed points. In this work we expand on mathematical aspects of this theory and approach the framework in  \cite{Festuccia:2018rew} from its relation with 5D susy gauge theories. Indeed many aspects of the 4D construction can be encoded and understood very naturally through a 5D theory.

In  \cite{Festuccia:2018rew}  we have constructed a theory for the ${\cal N}=2$ vector multiplet, and reformulated it as a cohomological field theory.
This cohomological theory is a generalization and unification of the equivariant Donaldson-Witten theory and Pestun's construction \cite{Pestun:2007rz} on $S^4$. Let us sketch below some general ideas behind this construction.

\subsection{Donaldson-Witten theory}\label{sec_CcoimsaTS}

Donaldson-Witten (DW) theory \cite{Witten:1988ze} was constructed as a cohomological field theory and can be interpreted as a topological twist of ${\cal N}=2$ supersymmetric Yang-Mills theory. The theory localizes to the moduli space ${\cal M}$ of instantons which are anti-self-dual (ASD) connections of the gauge bundle $E\to X$. The deformations of instantons are controlled by an elliptic complex and thus ${\cal M}$ is finite dimensional.

The theory computes intersections of some cohomological classes on the moduli space ${\cal M}$ of anti-self-dual connections of some vector bundle $E\to X$.
The said cohomological classes are constructed out of the 1st Pontryagin class of the universal bundle over $X\times{\cal M}$. It is the trace of the square of the universal curvature. From the physics point of view, the universal curvature has the expression
\bea {\cal F}=F+\Psi+\phi~,\label{univ_curv}\eea
where $F$ is the usual curvature of $E$, $\Psi$ is an odd 1-form coming from twisting the fermions of ${\cal N}=2$ susy gauge theories, and finally $\phi$ is one of the scalars. There are subtleties in the scalar sector that we gloss over in this introduction.

One can think of the above expression of ${\cal F}$ as saying that it can have 0,1 or 2 'legs' along $X$ and the remaining legs on ${\cal M}$, corresponding to $\phi,\Psi,F$ respectively.
The role of the slant product is also important: from a homology class of $X$, one may construct from the Pontryagin class some descendant cohomology classes on ${\cal M}$ as was explained in the physics context already in sec.3 of \cite{witten1988} (recently these descendants have also been shown to lead to higher operations in susy theories \cite{BeemBenzviBullimoreDimofteNeitzke}).
One can think of the slant product as follows: pick $\Go^p$ a closed form Poincare dual to a cycle $\Gs_{4-p}\subset X$, take the wedge and integrate over $X$:
\bea
\int\limits_X\Tr[{\cal F}^2]\wedge\Go \in H^{\sbullet}({\cal M})~.\label{DW_class}\eea
The whole process is equivalent to slanting $\Tr[{\cal F}^2]$ with $\Gs_{4-p}$. For example, the traditional $\gt$-term corresponds to picking $\Go=\gt={\rm const}$, while picking the volume form as $\Go$ gives $\Tr[\phi^2]$, interpreted as the Pontryagin class on ${\cal M}$.

The instanton moduli space ${\cal M}$ is a submanifold of the gauge equivalence class of connections ${\cal A}/{\cal G}$, so integrating a cohomology class along ${\cal M}$ can be done by first wedging it with the Poincare dual of ${\cal M}$ and integrate over all of ${\cal A}/{\cal G}$. Note that, to convey the main idea here, we are being callous about possible reducible connections which are a source of singularities.
The Poincare dual mentioned here can be written using the Mathai-Quillen (MQ) formalism \cite{Mathai:435793}.
This formalism replaces cohomology of compact support with exponentially decaying support, e.g. the MQ representative of the Poincare dual of ${\cal M}$ inside of ${\cal A}$ has leading term $\exp(-||F^+||^2)$, which peaks precisely at $F^+=0$, i.e. at ${\cal M}$. The Yang-Mills kinetic term can be rewritten as
\bea \Tr\big[\int\limits_X\,-F\wedge F+2|F^+|^2\big]=\Tr\big[\int\limits_X\,|F|^2\big]\label{4D_bound_std}\eea
 linking the calculation of characteristic classes to the more standard gauge theory partition function.
Indeed Atiyah and Jeffrey \cite{ATIYAH1990119} showed a detailed match of this construction with the Lagrangian of Witten's topologically twisted $\N=2$ SYM \cite{Witten:1988ze}.

\subsection{Adding Equivariance and the Omega Background}

If $X$ has some isometry, then ${\cal M}$ may also inherit these isometries. It makes sense then to study the equivariant cohomology ring of ${\cal M}$,
constructed as
\bea
 \int\limits_X\Tr[{\cal F}_{eq}^2]\wedge\Go_{eq} \in H_{eq}^{\sbullet}({\cal M})~,\label{DW_class_eq}
 \eea
where ${\cal F}_{eq}$ still has the decomposition \eqref{univ_curv} but, due to the change of susy algebra, it has now the meaning of an equivariant class.
This is how the equivariant DW theory was first presented  \cite{Losev:1997tp, Moore:1997dj, Lossev:1997bz, Moore:1998et}, i.e. at the level of the instanton moduli space.

One can also try to incorporate directly the equivariance at the level of ${\cal N}=2$ SYM, this goes under the name of the omega-background \cite{Nekrasov:2002qd, Nekrasov:2003rj} (also see earlier works \cite{Losev:1997tp, Moore:1997dj, Lossev:1997bz, Moore:1998et}) and for more general toric manifolds see \cite{MR2227881,Gottsche:2006tn,Gottsche:2006bm,Gasparim:2008ri,Bershtein:2015xfa,Bershtein:2016mxz,Bawane:2014uka,Sinamuli:2014lma,Rodriguez-Gomez:2014eza} (for related work in 5D see \cite{Hosseini:2018uzp,Crichigno:2018adf}).

Equivariant DW theory was defined as a cohomological field theory \cite{Nekrasov:2002qd, Nekrasov:2003rj}, here we treat it within the equivariant Atiyah-Jeffrey (AJ) framework. We also take a detour via 5D, which allows us to package many formulae in a  cleaner way, especially in the case of compact manifolds for which complications that are not present in \cite{Nekrasov:2003rj} arise. We find that the instanton equations are deformed by certain scalars, in agreement with \cite{Nekrasov:2003rj} (see the equations (2.11) there and we shall comment more on this in sec.\ref{sec_RtEDWt}). They read
\bea &&0=\frac{1}{2}(1+*)(F+\Go),~~~0=D^{\dag}(\iota_{\veeb}F-D\varphi),\nn\eea
where $\Go$ is a 2-form built from $\varphi$ and $\veeb$, as dictated by susy.
The deformation problem of this system of pde's is still controlled by en elliptic complex.

\subsection{More general ${\cal N}=2$ SYM on curved manifolds}\label{sec_Fgoei}

    Our previous construction of ${\cal N}=2$ SYM theory on curved manifolds was inspired from the reduction \cite{Festuccia:2016gul} from 5D. The corresponding rigid supergravity background is presented in \cite{Festuccia:2018rew}. The construction requires a Killing vector field with isolated fixed points, and unifies equivariant DW theory and Pestun's construction on $S^4$.
    The main feature is a generalization of the notion of (anti)-self-duality on $X$. Using the Killing vector field $\veeb$ we can build a projector $P$ acting on the space of 2-forms such that $P$ approaches $1/2(1\pm *)$ at the fixed points of $\veeb$.
    Then $P\Go^2\subset \Go^2$ is a sub-bundle \footnote{By the lower-semi-continuity of the rank of the image of a linear map, both $\rk\opn{img}(P)$ and $\rk\opn{img}(1-P)$ can only suddenly jump up. As the sum $P+(1-P)=id$, we conclude that both ranks are locally constant, constant if the space is connected.} of rank 3.

    Instantons for which the gauge field solves $PF=0$ are in fact well-motivated from 5D YM (this is the subject of sec.4, in particular sec.4.4), where smooth solutions do exist \cite{Baraglia:2014gma}.
    Another piece of evidence that the space of solutions to $PF=0$ is non-empty is the following: since $P$ approaches the ASD or SD at fixed points, one can imagine placing usual (anti-)instantons of very small size centred round the fixed points thus producing approximate solutions. This is the usual philosophy of "gluing" small instantons and small anti-instantons distributed over fixed points.

 The subbundle $P \Go^2$ is of rank 3 and $P F=0$ involves three constraints, which, plus the gauge fixing, is the right number of constraints for a gauge field. However this does not lead to an elliptic system of PDEs on the gauge field, which is needed to make sure that the instanton moduli space is finite dimensional. Thus in contrast to DW theory, the space of solutions $PF=0$ is infinite dimensional and intractable.

We can improve upon the moduli problem by adding equivariance. We will show that imposing equivariance, and then adding scalar fields and appropriately deforming the instanton equation to $P(F+\Go)=0$ gives a system of PDEs which  are transversally elliptic (i.e., elliptic transverse to the action of the isometry). One can then compute the equivariant index, which gives us the tangent space of the moduli space. The equivariant index has the feature that, at each given weight (grading), its dimension is finite, thanks to transversal ellipticity.

In view of this, let us boldly assume that the moduli space ${\cal M}_{\rm new}$ of equivariant solutions to $P(F+\Go)=0$ exists in some shape or form (most likely infinite dimensional), and point out certain features that make it more approachable.

 The toric symmetry gives the tangent space of ${\cal M}_{\rm new}$ a graded vector space structure. From the index calculation $\dim T{\cal M}_{new}<\infty$ at each grading, making its infinite dimensionality more amenable.
    Hence we expect that the entire moduli space ${\cal M}_{\rm new}$ is a graded manifold away from reducible connections. The zero grading part $({\cal M}_{\rm new})_0$ corresponds to the zero modes in our cohomological theory, studied in the text, while the nonzero grading parts are fibred over $({\cal M}_{\rm new})_0$.

 \subsection{Proposal of a new moduli problem}

 It is important to remark that differently from DW theory, where one can zoom onto the moduli space of $(1+*)F=0$ first and add equivariance later, in our construction, equivariance is an integral part of the moduli problem: it cannot be turned off. Aside from the main results of the localisation calculation presented in this paper, we propose the new moduli problem as a mathematical theory to be studied.

 Apart from the possible graded manifold structure already pointed out earlier, it has another nice feature.
 The equality \eqref{4D_bound_std} says that the YM term is bounded by the second Chern number, which is a topological invariant. This bound is saturated when $F$ is ASD and thanks to topological invariance, the instanton moduli space ${\cal M}$ is a robust object to study. We will have a similar bound here.
As equivariance is indispensable for our moduli problem, let us restrict ourselves to the domain of equivariant bundles (w.r.t the isometry group of $X$). We let $\veeb$ be a fixed Killing vector and $F_{eq}=F+\varphi$ be closed under equivariant differential $D_{eq}=D-\iota_{\veeb}$, then we can write
 \bea
  &&\int\limits_Xh(|F+\varphi d\gb|^2+h^{-2}|D\varphi|^2)=\int\limits_X -\go_{eq}F_{eq}\wedge F_{eq}+h(1+f^2)|B|^2~,\label{4D_bound_eq}\\
&&\hspace{2cm}B=P\big(F+\varphi d\gb+h^{-1}*\big((g\veeb)\wedge D\varphi\big)\big)~,\nn\eea
we will explain the various terms above in the main text. For now we need only know that $h>0$, $f\in[-1,1]$ and $P$ is our projector.
The lhs action is bounded by $-\int\go_{eq}\Tr[F_{eq}^2]$, and this bound is saturated when $P(F+\Go)=0$ with $\Go=\varphi d\gb+h^{-1}*((g\veeb)\wedge D\varphi)$.

Expanding $\go_{eq}F_{eq}\wedge F_{eq}$ one gets a term
\bea -\Tr\int\limits_Xf [F\wedge F],\nn\eea
which resembles the usual $\gt$ term with a sort of varying $\theta$-angle $f$.
Indeed, our new instanton swings from being ASD where $f=-1$ to being SD where $f=+1$.
Our new theory is expected to compute the classes
\bea
 \int\limits_X\Tr[{\cal F}_{eq}^2]\wedge\go_{eq}\in H_{eq}^{\sbullet}({\cal M}_{\rm new})\label{observable}\eea
with ${\cal M}_{\rm new}$ being the new equivariant instanton moduli space.

A potential use of this new theory is the construction of equivariant smooth invariants of quasi-toric 4-manifolds. It is known that the equivariant cohomology ring considered as an algebra over $H^{\sbullet}_{eq}(pt)$ actually classifies quasi toric 4-manifolds up to homeomorphism.
As our exotic instanton moduli space does not have a non-equivariant counterpart, it is interesting to ask if such moduli space brings us anything new about the toric geometry of these manifolds? 
Already at the zero-instanton level, as we shall see in the main text, the partition function is a special function called $\Upsilon$ (very similar to the triple sine, but in one dimension lower). This special function has complicated analytic and 'wall-crossing' behaviour.
This is already richer than the usual equivariant DW theory, where the zero instanton part gives no more than the Euler number and signature of $X$.

In analogy with the standard instanton treatment we certainly expect the usual non-compactness issue of the moduli space coming from instantons of zero size. Within the scope of this paper we do not attempt
 to investigate the Uhlenbeck compactification systematically. 
But we believe that in the equivariant setting these zero size instantons will only appear at the torus fixed points. Thus for example, the calculations of the cohomology of the instanton deformation complex
 involving these objects can be carried out using algebraic means as we do in  sec.\ref{sec_DitIS}. 
It is generally  expected that the equivariance provides a much cleaner treatment of compactification. We believe the same will be true for our exotic instanton theory.

\subsection{Summary of the results}
Let us give a schematic summary of our results. The localization computation requires a non-trivial mix of supersymmetry and BRST-transformations.
Hence we include the latter in our discussion here. Taking into account the gauge fixing sector, all transformations can be organised as follows:
\bea
\begin{tikzpicture}
  \node at (-3,0) {$\gd\;\uparrow$};
  \node at (0,1.3) {$\gdh\;\rightarrow$};
  \matrix (m) [matrix of math nodes, row sep=1.6em, column sep=3em]
    {  & {c} & a & \vphantom{F} & { \chi\oplus \bar c}  \\
      &  \phi & \Psi & \vphantom{F} & H\oplus b \\ };
  \path[->]
  (m-2-2) edge (m-1-2)
  (m-2-3) edge (m-1-3)
  (m-2-5) edge (m-1-5);

  \path[->]
  (m-1-2) edge node[above] {} (m-1-3)
  (m-2-2) edge node[above] {} (m-2-3)
  (m-1-3) edge node[above] {} (m-1-5)
  (m-2-3) edge node[above] {}(m-2-5);
\end{tikzpicture}.
\label{full-susy-double}\eea
where we presented the fields in 5D terms. The detailed definition of each field will be given in sec.\ref{sec_5DN=1SYM}.
The 4D version follows from dimensional reduction: the 1-forms $a,\Psi$ reduce straightforwardly $a \rightarrow (a, \varphi)$ and $\Psi \rightarrow (\Psi, \eta)$), while that of $H,\chi$ will be presented in sec.\ref{sec_RteDt}.
Here  $\gd$ and $\gdh$ are two odd anti-commuting symmetries which satisfy $\gdh^2=0$ while $\gd$ squares to a $\veeb$ action. The full supersymmetry with BRST-symmetry\footnote{To the best of our knowledge the similar algebraic structures were first observed in the context of the Donaldson-Witten theory in 1989 in \cite{Ouvry:1988mm}.} is given by the sum $\delta + \gdh$.

Once the cohomological theory is constructed, one can proceed with localisation. The equivariant AJ construction then gives that
the action is minimised at the configuration where the gauge bundle is equivariant with curvature satisfying $P (F+\Go) = 0$.  The standard procedure tells us that we need only compute a Gaussian integral over the fluctuations round such a background.

For the Gaussian integral, we truncate all fields to linear order around a chosen background and the susy algebra is likewise truncated to be linear in the fluctuating fields. We denote the linearized transformations by the same letters: $\gd$ and $\gdh$ satisfying the same algebra.

All fields organise into a chain complex with differential $\gdh$ causing massive cancellations in the Gaussian integral: pairs of fermion and boson modes related by $\gdh$ cancel. What is left after the cancellation are of course modes in the cohomology of $\gdh$.
In fact the chain complex with $\gdh$ is the one controlling the deformation of the exotic instantons, and the cohomology of $\gdh$ gives the super dimension of the tangent space $T{\cal M}_{\rm new}$. We mentioned that $\gdh$ is not elliptic, but only so transverse to the isometries of $X$. This has the fortuitous consequence that the cohomology, albeit infinite dimensional, is of finite dimension at each grading (we restrict to $X$ with $U(1)^2$ isometry and so we have $\BB{Z}^2$ grading coming from the $U(1)$ weights).
This procedure resembles virtual localisation\footnote{This technique is used in computing Gromov-Witten invariants and Donaldson-Thomas invariants on toric varieties \cite{GraberPandharipande}.} in the sense that we do not actually have a description of ${\cal M}_{\rm new}$ apart from its tangent complex. In this paper we will proceed along this same idea without delving into the algebraic details of the technique.

The analysis of $\gdh$-cohomology at an instanton background is in fact not new. In the study of 4D monopoles (our 5D instantons are more appropriately called solitons), one studies the semi-classical quantisation round a monopole or dyon background and in particular the super-multiplet formed by the zero modes. For example, in \cite{Osborn:1979tq} it was shown that in the 4D $N=4$ theory the zero modes form a short massive multiplet with 1 spin 1, 4 spin 1/2 and 5 spin zero states, as was expected from the Montonen\&Olive duality \cite{MONTONEN1977117}. What we are doing here is a twisted version of the same analysis.

The supersymmetry action truncated to the quadratic order around a chosen background has the schematic expression
\be
S_2=\textrm{field}(-L_{\veeb}^2+\{\gdh^{\dag},\gdh\})\textrm{field}~.\nn
\ee
The Gaussian integral then gives $\sdet^{1/2}(-L_{\veeb}^2+\gdh^{\dag}\gdh)$.
We see again that pairs of fields related by $\gdh$ give no contribution and so the super-determinant collapses to $\sdet^{1/2}_{H_{\gdh}}(-L_{\veeb}^2)$.
This is the central theme we would like to promote in our paper:
\begin{center}
  \it A choice of supersymmetry dictates that one can decompose the kinetic energy operator as $\Gd=-L_{\veeb}^2+\{\gdh,\gdh^{\dag}\}$ with $\gdh^2=0$.
  One can then compute the Gaussian integral as $\sdet\Gd$, which in turn is the equivariant Euler character of a (possibly transversally) elliptic operator $\gdh$. Moreover, supersymmetric localisation says that this computation is exact.
\end{center}
 Here the ellipticity of the second order operator $\Gd$ is ultimately related to transverse ellipticity of $\gdh$.

Next we discuss the backgrounds around which one computes the above superdeterminant. These correspond to the torus fixed points of ${\cal M}_{\rm new}$.
Here in order for the torus fixed point set of ${\cal M}_{\rm new}$ be nonempty, one may be forced to consider toric equivariant sheaves in addition to the toric equivariant vector bundles. On top of that, if one performs the computation on the Coulomb branch, i.e. one has a scalar field $\gs$ with $D\gs=0$, one is necessarily dealing with vector bundles with reducible connections (or invertible sheaves).
We shall prove that such configurations are completely determined given the value of the 0-form in the equivariant first Chern-Class at the fixed points of $X$. We denote these values as $\hat\varphi_i$, one $i$ for each fixed point of $X$. These values themselves are determined by the flux along toric invariant 2-cycles. Note that our focus here is quite different from the idea of point like instantons that has been shopped about in the literature.
We do not actually prove that the gauge or scalar field is zero away from the fixed points of $X$, even though this may well be true.  Rather we prove a weaker but more manageable statement that the value $\varphi_i$ at the fixed points completely determines everything. As our way of fixing $\hat\varphi_i$ uses the $c_1$ which is also defined for sheaves, we expect that our result remains valid even in that case, but a proof is beyond the scope of this paper.

Thus schematically the 4D partition function reads
\be
Z=\sum\limits_{\hat\varphi_i}\int d\sigma ~Z^{clas} Z^{pert}\prod_iZ^{Nek}_{\BB{C}^2}(s_i \hat\varphi_i+\gs,s_i\ep_i,s_i\ep_i')~,\nn
\ee
where $s_i=+$ (respectively $-$) at a fixed point $x_i$ means that the projector $P$ approaches the ASD (resp. SD) projector at $x_i$, $\ep_i,\ep_i'$ are equivariant parameters to be read off from the toric data and finally $Z^{pert}$ is a special function generalising the $\Upsilon$ function in Pestun's $S^4$ computation.
Hence, the final answer is not surprising, but the structural result concerning the double complex, (i.e. the ideas espoused in italic above) are more central and, we believe, underlie all localisation computations in supersymmetric gauge theories (for a review see \cite{Pestun:2016zxk}).

\bigskip
{\bf Acknowledgements:} We thank Nikita Nekrasov and Vasily Pestun for discussions over many issues. A special thanks goes to Maksim Maydanskiy for providing the proof in sec.\ref{sec_TGCS}.
The work of GF is supported by the ERC STG grant 639220. The work of MZ  is supported in part by Vetenskapsr\r{a}det under grant \#2014-5517, by the STINT grant, and by the grant  "Geometry and Physics"  from the Knut and Alice Wallenberg foundation.

\newpage

\section{Double Chain Complex and Supersymmetry}\label{sec_Sla}

As we mentioned in the introduction, we would like to design a cohomological theory with a Lagrangian that is Poincare dual to the moduli space of certain flipping instantons. This theory should have an equivariant differential $\gd$ and another differential $\gdh$ that controls the deformation problem of our instantons. Around such an instanton background, we can truncate all fields up to first order deviations therefrom, then both $\gd,\gdh$ become linear in the fields giving a double chain complex with $\gd$ and $\gdh$. Note that $\gdh$ squares to zero but $\gd$ squares to a Lie derivative, but we still use the word chain complex. We hope this does not cause confusion.

The remaining computation is then a Gaussian integral around the instanton background, and we shall show that such computation gives rise to a super-determinant that is often insensitive to the details of the theory.
 In this section we introduce the linear algebra setup and consider some toy models relevant for our main discussion.

\subsection{Basic Model}
We start with the problem of computing a Gaussian integral on a symplectic vector space $T^*\BB{R}^{n|n}$. We denote the coordinates of $\BB{R}^{n|n}$ as $(a,\chi)$ and $(\tilde \psi,\tilde H)$ that of the fibre (the Greek ones are fermionic). One has the following supersymmetry algebra
\bea
\begin{array}{l|l}
  \gd a=\tilde \psi~, & \gd \tilde\psi=V a~, \\
  \gd \chi=\tilde H ~, & \gd \tilde H=V \chi~, \\
\end{array}\label{basic_cplx}\eea
where $V$ is an endomorphism acting on $\BB{R}^n$.
The Gaussian integral in question has quadratic terms
\bea S=\frac12\left[\begin{array}{cc}
                  a & \tilde H \\ \end{array}\right]\underbrace{\left[
         \begin{array}{cc}
           S_1 & P \\
           P^T & S_2 \\ \end{array}\right]}_B\left[
                \begin{array}{c}
                  a \\
                  \tilde H \\ \end{array}\right]+\frac12\left[\begin{array}{cc}
                  \tilde \psi  & \chi \\ \end{array}\right]\underbrace{\left[
         \begin{array}{cc}
           A_1 & Q \\
           -Q^T & A_2 \\ \end{array}\right]}_F\left[
                \begin{array}{c}
                  \tilde\psi \\ \chi \\ \end{array}\right]~,\label{Quad_term}\eea
where all $P,Q,S_i,A_i$ are $n\times n$, the $S_i$ are symmetric while the $A_i$ are anti-symmetric. We assume that $\gd S=0$, which implies
\bea
 S_1+V^TA_1=0~,~~PV+V^TQ=0~,~~P^T-Q^T=0~,~~S_2V+A_2=0~.\label{relations_II_diag}
 \eea
 The Gaussian integral gives a determinant from the $a,\tilde H$ part and a Pfaffain from the $\tilde\psi,\chi$ part. First if $V$ is zero identically then
\bea
V=0~:~~~~~ \frac{\pf F}{(\det B)^{1/2}}=\frac{\det Q}{|\det P|}=\frac{\det Q}{|\det Q|}=\sgn\det Q~.\nn\eea
Secondly if $\det V|_a\neq0$ we get
\bea
\pf F\cdotp\det V|_a=\pf(-BV)~,\nn
\eea
and the ratio gives
\bea
 \frac{\pf F}{(\det B)^{1/2}}=\frac{\pf(-BV)}{\det V|_a\cdotp {\det}^{1/2}(B)}~.\nn
 \eea
Then up to a sign
\bea
\frac{\pf F}{(\det B)^{1/2}}=\pm\frac{{\det}^{1/2} V|_{\chi}}{{\det}^{1/2} V|_a}=\pm \operatorname{sdet}^{1/2}V~.\label{used_VII}
\eea
In particular if $\det V|_{\chi}=0$ then the result is zero, or in field theory parlance, the $\chi$ zero modes are not soaked up.
In the infinite dimensional setting we will have to take care of zero modes.
\begin{remark}
This simplest model shows that the Gaussian integral is largely independent of the Gaussian term and gives the super-determinant of $V=\gd^2$.
Furthermore, there can be cancellations in the super-determinant, i.e. the determinant of $V$ over $a$ may cancel that over $\chi$.
This is a satisfactory story in finite dimension,  where one can just cancel the determinants 'by hand'.
But if the dimension is infinite, to keep track of the cancellation, one must have a map between $a$ and $\chi$ in order to set up correspondences between pairs of subspaces of $a$ and $\chi$ whose contribution to ${\rm sdet}$ cancels.
To this end we extend the previous story to a double chain complex furnishing us with the required map in between.
\end{remark}

\subsection{Multi-Level Model}\label{sec_Mlm}

Consider the generalized picture, we introduce  the   complex
\bea
\begin{tikzpicture}
  \node at (-3.5,0) {$\gd\;\uparrow$};
  \node at (0,1.3) {$\gdh\;\rightarrow$};
  \matrix (m) [matrix of math nodes, row sep=1.2em, column sep=1.2em]
    { 0 & {\color{blue}E_1} & E_0 & {\color{blue}E_{-1}} & E_{-2} &\vphantom{E_{-3}} \\
      0&  F_2 & {\color{blue}F_1} & F_0 & {\color{blue}F_{-1}} & \vphantom{F_{-2}}\\
              };
  \path[->]
  (m-2-2) edge (m-1-2)
  (m-2-3) edge (m-1-3)
  (m-2-4) edge (m-1-4)
  (m-2-5) edge (m-1-5);

  \path[->]
  (m-1-1) edge (m-1-2)
  (m-2-1) edge (m-2-2)
  (m-1-2) edge node[above] {\scriptsize{$R_1$}} (m-1-3)
  (m-2-2) edge node[above] {\scriptsize{$-R_1$}} (m-2-3)
  (m-1-3) edge node[above] {\scriptsize{$-R_0$}} (m-1-4)
  (m-2-3) edge node[above] {\scriptsize{$R_0$}}(m-2-4)
  (m-1-4) edge node[above] {\scriptsize{$R_{-1}$}} (m-1-5)
  (m-2-4) edge node[above] {\scriptsize{$-R_{-1}$}} (m-2-5)
  (m-1-5) edge (m-1-6)
  (m-2-5) edge (m-2-6);
\end{tikzpicture}\label{toy_cplx}\eea
where $\gdh$ is a differential that acts horizontally and $\gd$ vertically. We denote by $s_p$ the generator of $E_p$, $t_p$ that of $F_p$, and the fermionic components are in blue. We use the passive notation to write $\gd$, \dh,
\bea
&\gd s_p=t_{p+1}~,~~~\gd t_p=V_ps_{p-1}~;\nn\\
     &\gdh s_p=(-1)^pR_{p+1}s_{p+1}~,~~~\gdh t_p=(-1)^pR_pt_{p+1}~,\nn
\eea
where $V_p\,,R_p$ are just some matrices of appropriate dimension, and, when confusion is unlikely, we write simply $V$ and $R$ to avoid clutter. We demand that
\be
 \{\gd,\gdh\}=0=\gdh^2~, \label{double_diff}
\ee
leading to
\be
 V_pR_p=R_pV_{p+1},~~~R_pR_{p+1}=0, \label{V_cancel}
\ee
for all $p$. The differentials can be gathered in two big matrices
{\scriptsize\bea  (\gd+\gdh)\left[
                \begin{array}{c}
                  t_2 \\
                  s_0 \\
                  t_0 \\
                  s_{-2} \\
                  \colon \\
                \end{array}
              \right]=\left[
         \begin{array}{ccccc}
           V & 0 & 0 & 0 &\\
           R & 1 & 0 & 0 & \colon\\
           0 & R & V & 0 & \colon\\
           0 & 0 & R & 1 & \colon\\
           & .. & .. & .. & \\
         \end{array}
       \right]\left[
                \begin{array}{c}
                  s_1 \\
                  t_1 \\
                  s_{-1} \\
                  t_{-1} \\
                  \colon \\ \end{array}\right],~~~
 (\gd+\gdh)\left[
                \begin{array}{c}
                  s_1 \\
                  t_1 \\
                  s_{-1} \\
                  t_{-1} \\
                  \colon \\
                \end{array}
              \right]=\left[
         \begin{array}{ccccc}
           1 & 0 & 0 & 0 &\\
           -R & V & 0 & 0 & \colon\\
           0 & -R & 1 & 0 & \colon\\
           0 & 0 & -R & V & \colon\\
           & .. & .. & .. & \\
         \end{array}
       \right]\left[
                \begin{array}{c}
                  t_2 \\
                  s_0 \\
                  t_0 \\
                  s_{-2} \\
                  \colon \\ \end{array}\right]\label{big_matrix}, \eea}
where the diagonal terms come from $\gd$ and off diagonal ones from $\gdh$.

We now want to perform the Gaussian integral of $e^{-S}$ with $S$ quadratic in all variables. First for clarity, we shift
\bea \tilde t_p=t_p+(-1)^{p+1}R_ps_p\nn\eea
so that the two matrices become diagonal: $(\gd+\gdh)s_p=\tilde t_{p+1}$, $(\gd+\gdh)\tilde t_p=V_ps_{p-1}$.
One then can gather together the bosonic elements $s_0,s_{-2},\cdots$ and $\tilde t_2,\tilde t_0,\cdots$ and call them $a$ and $\tilde H$. Similarly one gathers the fermionic elements $\tilde t_1,\tilde t_{-1},\cdots$ and $s_1,s_{-1},\cdots$ and call then $\tilde\psi$ and $\chi$. The complex now collapses into a 2-level one, exactly like \eqref{basic_cplx},
\be
 (\gd+\gdh)\left[
                \begin{array}{c}
                  a \\
                  \tilde H \\ \end{array}\right]=\left[
         \begin{array}{cc}
           1 & 0 \\
           0 & V \\ \end{array}\right]\left[
                \begin{array}{c}
                  \tilde\psi \\
                  \chi \\ \end{array}\right], \qquad (\gd+\gdh)\left[
                \begin{array}{c}
                  \tilde\psi \\
                  \chi \\ \end{array}\right]=\left[
         \begin{array}{cc}
           V & 0 \\
           0 & 1 \\ \end{array}\right]\left[
                \begin{array}{c}
                  a \\
                  \tilde H \\ \end{array}\right].\label{second_type_diag}
\ee
We assume that the quadratic form $S$ is the same as in \eqref{Quad_term} and $(\gd+\gdh)S=0$. Then we can proceed just like
in the previous section. We get the Gaussian integral
\bea
 V=0:&&~~\frac{\pf F}{(\det B)^{1/2}}=\sgn\det Q,\label{V=0}\\
\det V|_a\neq0:&&~~\frac{\pf F}{(\det B)^{1/2}}=\pm {\sdet}^{1/2}V\big|_{E_{\sbullet}}.\label{V_nondeg}
\eea
Now for the second case, the super-determinant of \eqref{used_VII} turns into one taken over the chain complex $E_{\sbullet}$.

As this complex has chain maps $\gdh$ that commute with $V$ from equations \eqref{double_diff},\eqref{V_cancel}, the contribution to the super-determinant cancels between fields
in $E_{\sbullet}$ and $E_{\sbullet+1}$ linked together by $\gdh$. In the end only those in the cohomology of $\gdh$ actually make a contribution
\be
  {\sdet}V\big|_{E_{\sbullet}}={\sdet}V\big|_{H_{\sbullet}}~,\label{conclusion_earlier}
 \ee
where $H_{\sbullet}$ denotes the $\gdh$-cohomology of $E_{\sbullet}$.

\begin{remark}
  As we already said in the preamble to this section, we intend to use this double complex to model the fields in a supersymmetric gauge theory, truncated to linear order around a super-symmetric background (strictly speaking, it is supersymmetry plus BRST combined that will give the two differentials $\gd,\gdh$).

  This is also a reason why ${\cal N}=2$ supersymmetric theories are amenable to localisation calculations: we need the equivariant localisation (with $\gd$) to get us down to the moduli space of certain supersymmetric configurations. Then we need $\gdh$ to control the deformation around such configurations and to keep track of cancellations in the super-determinant. Thus the double chain complex is a universal model for a tractable localisation calculation.

\end{remark}

\subsection{Constructing the Action}\label{sec_FMn}

So far in the two-level or multilevel case we have assumed a quadratic term \eqref{Quad_term}, which is arbitrary provided that it is non-degenerate. But using the maps denoted $R_{\sbullet}$ in \eqref{toy_cplx}, one has already a good candidate for $S$. We will spell out some details to illustrate a philosophical point of the current construction.

We take only the second and third column of the complex \eqref{toy_cplx} for simplicity. The coordinates are called $a,\psi,\chi,H$ for $E_0,F_1,E_{-1},F_0$ respectively. In writing the action we make a Wick rotation $H\to iH$
\be
 S_0=\frac12(\gd+\gdh)\big(\bra \psi, V a \ket-\bra \chi,iH+R_0a\ket\big)~,\nn
\ee
which is killed by $\gd+\gdh$ when $\bra,\ket$ is $V$ invariant. We have
\be
S_0=\frac12\big(\bra Va,Va\ket-\bra\psi,V\psi\ket-\bra iH-R_0a,iH+R_0a\ket+\bra \chi, V\chi+2R_0\psi\ket\big)~.\nn
\ee

For the $V=0$ case, one needs to identify the matrix denoted $Q$ in \eqref{V=0}, which comes from $\bra \chi, R_0\psi\ket$ only. Thus the $\sgn\det Q=\sgn\det R_0$, and we see that the Gaussian integral is controlled solely by the supersymmetry algebra.

For the $V$ non-degenerate case, one just uses \eqref{V_nondeg}. But to see more clearly what has happened, we integrate out $H$ from $S_0$
\bea
 S_0\to \frac12\big(\bra Va,Va\ket-\bra\psi,V\psi\ket+\bra R_0a,R_0a\ket+\bra \chi, V\chi+2R_0\psi\ket\big)~.\nn
\eea
We see that the quadratic term for $a$ and for $(\chi,\psi)$ reads
\bea
&S_{bos}=\frac12\big(\bra Va,Va\ket+\bra R_0a,R_0a\ket\big) = \frac 1 2 \bra a , (-V^2 +R_0^T R_0 ) a\ket  \equiv \frac 1 2 \bra a , \Delta a \ket ~,\nn \\
&S_{ferm}=
\left[\begin{array}{cc}
                \psi & \chi \\
              \end{array}\right]\left[
                                  \begin{array}{cc}
                                    -V & -R^T_0 \\
                                    R_0 & V \\
                                  \end{array}\right]\left[
                                                      \begin{array}{c}
                                                        \psi \\
                                                        \chi \\
                                                      \end{array}\right]~.
\label{fermi_kinetic_ferm}\eea
Here we use the fact that $V$ is antisymmetric.
If one squares the fermionic matrix, one gets
\bea \left[
                                  \begin{array}{cc}
                                    -V & -R^T_0 \\
                                    R_0 & V \\
                                  \end{array}\right]^2=\left[
       \begin{array}{cc}
         V^2-R^T_0R_0 & 0 \\
         0 & V^2-R_0R_0^T \\
       \end{array}\right]~.\nn\eea

Integrating over $a$ gives the determinant ${\det}^{-1/2}(-V^2+R^T_0R_0) \equiv \det^{-1/2} \Delta $, while the fermions give a Pfaffian as usual.
 Altogether $a,\psi,\chi$ give
\bea
\sdet=\pm\frac{{\det}^{1/4}(-V^2+R^T_0R_0){\det}^{1/4}(-V^2+R_0R^T_0)}{{\det}^{1/2}(-V^2+R^T_0R_0)}
=\pm\frac{{\det}^{1/4}(-V^2+R_0R_0^T)}{{\det}^{1/4}(-V^2+R^T_0R_0)}~,\label{above_det}
\eea
where the upper determinant is taken over $E_{-1}\simeq F_0$ while the lower one over $E_0\simeq F_1$.

The above discussion encapsulates the gist of topological twist: we named the bosonic quadratic term as $\Gd$ to suggest that it is the Laplacian.
The fermionic quadratic term \eqref{fermi_kinetic_ferm} squares to $\Gd$, meaning that one can think of \eqref{fermi_kinetic_ferm} as the Dirac operator.
We see that the decomposition $\Gd=-V^2+R^T_0R_0$ is determined from $\gd$ and $\gdh$, i.e. from the supersymmetry algebra.
One has in fact great freedom in how to do this decomposition by picking different supersymmetry algebras.
In this work, our supersymmetry algebra is such that we have a novel decomposition $\Gd=-V^2+R^T_0R_0$, with $R_0$ being transversaly elliptic.

To see the significance of transversal ellipticity, let us inspect again the structure of the super-determinant \eqref{above_det}. First both terms in $\Gd$ are positive definite: this is obvious for $R^T_0R_0$, while $V$ is antisymmetric hence $-V^2$ has non-negative spectrum as well.
$R_0$ establishes an isomorphism between $(\ker R_0)^{\perp}$ and $\img R_0$, pairing up vectors in the two spaces with the same $-V^2$ eigenvalue. Therefore the contribution of $(\ker R_0)^{\perp}$ and $\img R_0$ to \eqref{above_det} cancels.
The remaining two subspaces $\ker R_0$ and $\coker R_0$ are not linked and make a contribution.
Note finally that $R_0^TR_0$ acts as zero on $\ker R_0$ while $R_0R_0^T$ acts as zero on $\coker R_0$. So \eqref{above_det} again collapses to the
\be
\pm\frac{{\det}^{1/4}\big|_{\coker R_0}(-V^2)}{{\det}^{1/4}\big|_{\ker R_0}(-V^2)}
=\pm{\sdet}^{1/4}\big|_{H_{R_0}}(-V^2)~,\nn
\ee
where $H_{R_0}$ is the cohomology of $R_0$ in our two-level example. This is of course just the conclusion we saw earlier in \eqref{conclusion_earlier}.

Getting back to the multi-level model with $\gdh$ given by $R_{\sbullet}$, we also assume that $V$ comes from the fundamental vector field of some group action $G$, and $G$ commutes with $\gdh$.
Then one can compute the equivariant index of $(E_{\sbullet},\gdh)$
\bea
 \ind_g(E_{\sbullet})=\sum_i(-1)^i\Tr[gE_i]=\sum_i(-1)^i\Tr[gH_i],~~~g\in G~.\nn
 \eea
and the super-determinant is the equivariant Euler-character evaluated at $V$.

Since the operator $\Gd$ is elliptic, a decomposition $\Gd=-V^2+\gdh^T\gdh$ guarantees
that $\gdh$ is elliptic transverse to the $G$-action.
If $\gdh$ is actually elliptic then its cohomology is of finite dimension. One can then compute the equivariant index using methods of \cite{atiyah1966}. More generally for the transversal elliptic case, one can compute the equivariant index using the method of \cite{Ellip_Ope_Cpct_Grp}.
The point is that even though $\gdh$ has infinite dimensional cohomology, one can decompose the cohomology into representations of $G$, with finite multiplicity. Since for our application $G$ is products of $U(1)$,  computing the index equivariantly amounts to introducing a grading on the cohomology, with finite dimensionality at each given grading.

Having figured out the structure of the algebra close to a supersymmetric configuration and the resulting Gaussian integral, we next turn to the full fledged supersymmetry algebra and consider how to write an action that counts the desired supersymmetric configurations.

\section{Review of Mathai-Quillen formalism}\label{sec_RoMQroTE}

It is shown in \cite{ATIYAH1990119} that Witten's Lagrangian for the topologically twisted ${\cal N}=2$ gauge theory is the Euler class of an infinite dimensional vector bundle presented in the Mathai-Quillen (MQ) formalism. We shall quickly review this point and show that even the Lagrangian we propose is of this type.

\subsection{Basic construction}
We use the formulation of \cite{kalkman1993}.
Start from a finite rank vector bundle (VB) over a manifold
\bea
\begin{tikzpicture}
   \matrix (m) [matrix of math nodes, row sep=1.2em, column sep=1.2em]
    { V^{2n} & E \\
       &  M \\ };
  \path[->]
  (m-1-1) edge (m-1-2)
  (m-1-2) edge (m-2-2);
\end{tikzpicture}\nn\eea
Denote by $x^i$ the coordinate of $M$ and $\psi^i$ be its fermionic companion. Let also $e_a$ be the local basis of $V$, $z^a$ be the local fibre coordinate and $\zeta^a$ its fermionic companion. Finally let $\chi^a,~H^a$ be a fermionic/bosonic pair that will play the role of the anti-ghost and Lagrange multiplier later. We denote with $A$ and $F$ the connection and curvature of $E$, then one can write down the (de Rham) complex
\bea
\begin{array}{l|l}
 \gd x=\psi, & \gd\psi=0, \\
 \nabla_{\gd}\chi=iH, & i\nabla_{\gd}H=F\chi, \\
 \nabla_{\gd} z=\zeta, & \nabla_{\gd}\zeta=Fz, \\
\end{array}.\label{dR_basic}\eea
where we use the notation
\bea &\nabla_{\gd}\chi^a=H^a+A_{ib}^a\psi^i\chi^b,~~\nabla_{\gd}z^a=\zeta^a+A_{ib}^a\psi^iz^b,\nn\\
&F\chi^a=(F_{ij})^a_{~b}\psi^i\psi^j\chi^b,~~Fz^a=(F_{ij})^a_{~b}\psi^i\psi^jz^b\nn\eea
It is always helpful to think of $\psi,\zeta,H$ as $dx,dz,d\chi$ and so $\gd$ is just like the de Rham differential.
The main difference here is that \eqref{dR_basic} is a covariantised version of the more basic one \eqref{basic_cplx} (with $V=0$) to take into account the non-triviality of $E$. Nonetheless the variation satisfies $\gd^2=0$ or equivalently $\nabla_{\gd}^2=F$.

We recall that the Thom class is a closed form of degree $2n={\rm rk}\,E$, with compact support along the fibre and fibrewise integral equalling 1. The Mathai-Quillen form is an integral representative of the Thom class, except that the compact support along the fibre is replaced with an exponential decaying support, which is just as good.

To construct the MQ form, denote with $\bra\,,\,\ket$ a pairing of $V$, and consider
\bea
&&S=-\frac{1}{2}\gd\bra \chi,iH+2z\ket=\frac12\bra H,H-2iz\ket+\frac12\bra \chi,F\chi+2\nabla_{\gd} z\ket~.\label{quad_thom}\eea
Its integral
\bea
 	u=(2\pi)^{-2n}\int d\chi dH e^{-\frac{1}{\hbar}S}~,\nn
\eea
gives a representative of the Thom class of $E$. To see this we integrate out $H$
\bea
 u=(\frac{\hbar}{2\pi})^n\int d\chi e^{-\frac{1}{\hbar}(\frac12|z|^2+\bra\chi,\nabla z\ket+\frac12\bra\chi,F\chi\ket)}\label{u_imme}
\eea
one sees that $u$ is of Gaussian support $e^{-|z|^2/2\hbar}$ centred along the zero section. Performing the $\chi$ integral will bring down a series of terms from the exponent
\bea u=\frac{1}{(2\pi\hbar)^n}e^{-\frac{1}{2\hbar}|z|^2}(\wedge^{2n}\nabla_{\gd}z+\cdots+\pf(\hbar F)).\nn\eea
The leading term $\wedge^{2n}\nabla_{\gd}z$ gives the volume form along the fibre with fibrewise integral 1. The sub-leading
terms are not written out, but they are of type $F^k\wedge^{2n-2k}\nabla_{\gd}z$ and provide the necessary correction to make $u$ closed.

One can pull back the Thom class with any section $\gs\in\Gc(E)$ and get a representative of the Euler class. To do so one simply replaces $z$ with $\gs$ in \eqref{u_imme}. In particular setting $z=\gs_0=0$, the last term in $u$ alone survives
\bea \gs_0^*u=\pf(\frac{1}{2\pi}F)\nn\eea
which is the Euler class.

Of course the real advantage is to choose a section that vanishes transversally so we can localise the integral.
As preparation for later, we do so in a more systematic way. We use a section $\gs$ to define a second differential $\gdh$ as
\bea \gdh\chi=-\gs,~~~i\gdh H=\nabla_{\gd}\gs\nn\eea
so that $\{\gd,\gdh\}=0$ and $\gdh^2=0$ trivially.

The Gaussian term $S$ in \eqref{quad_thom} (now with $z$ replaced with $\gs$) can be written as something $(\gd+\gdh)$-exact
\bea S&=&-\frac{1}{2}(\gd+\gdh)\bra \chi,iH+\gs\ket\label{quad_thom_I}\\
&=&-\frac{1}{2}\bra iH-\gs,iH+\gs\ket+\frac{1}{2}\bra \chi,F\chi+2\nabla_{\gd} \gs\ket
=\frac{1}{2}H^2+\frac{1}{2}\gs^2+\frac{1}{2}\bra \chi,F\chi\ket+\bra\chi,\nabla \gs\ket.\nn\eea
Integrating over $H$ we get the Euler class
\bea e=\gs^*u=(\frac{\hbar}{2\pi})^n\int d\chi e^{-\frac{1}{2\hbar}(|\gs|^2+2\bra\chi,\nabla \gs\ket+\bra\chi,F\chi\ket)}.\nn\eea

When $\dim M=\rk E$, we can choose $\gs$ with isolated zeroes. As the action \eqref{quad_thom_I} is exact, tuning $\hbar$ will not change the cohomology class of $e=\gs^*u$. By sending $\hbar\to0$, the integral is dominated by points where $\gs=0$. Letting $x_0$ be one such zero of $\gs$ and around this point we truncate the susy algebra to linear order by writing $x=x_0+a$ up to the first order deviation from $x_0$
\bea
\begin{array}{l|l}
 \gd a=\psi, & \gdh a=0, \\
 \gd\psi=0, & \gdh\psi=0, \\
 \gd\chi=iH, & \gdh \chi=-\gs'a, \\
 \gd H=0 & i\gdh H=\gs'\psi \\ \end{array}.\label{dR_basic_sec}\eea
At linear order the differentials fit into a complex
\bea
\begin{tikzpicture}
  \node at (-2.5,0) {$\gd\;\uparrow$};
  \node at (0,1.3) {$\gdh\;\rightarrow$};
  \matrix (m) [matrix of math nodes, row sep=1.2em, column sep=1.2em]
    { 0 &  \left[a\right]_0 & {\color{blue}[\chi]_{-1}} & 0 \\
      0 &  {\color{blue}[\psi]_1} & \left[iH\right]_0 & 0 \\ };
  \path[->]
  (m-2-2) edge (m-1-2)
  (m-2-3) edge (m-1-3);

  \path[->]
  (m-1-1) edge (m-1-2)
  (m-2-1) edge (m-2-2)
  (m-1-2) edge node[above] {\scriptsize{$-\gs'$}} (m-1-3)
  (m-2-2) edge node[above] {\scriptsize{$\gs'$}} (m-2-3)
  (m-1-3) edge (m-1-4)
  (m-2-3) edge (m-2-4);
\end{tikzpicture}\label{toy_cplx_Thom}\eea
conforming to \eqref{toy_cplx} with two levels and $V=0$. As we have promised the double complex in sec.\ref{sec_Mlm} models the susy algebra round $x=x_0$.

We truncate now $S$ to quadratic order
\bea S|_2=\frac{1}{2}H^2+\frac{1}{2}\bra\gs'a,\gs'a\ket+\bra\chi,\gs'\psi\ket.\nn\eea
That $x_0$ is an isolated zero of $\gs$ means that $\gs'$ is a non-degenerate matrix. Therefore
we can read off from \eqref{V=0} (now with $Q=\gs'$) the result of performing the integrals in the Euler class above
\bea \int_Me=\sum_{x_p}\sgn\det (\gs'(x_p)).\label{int_e}\eea
This is the well-known statement that the Euler number gives the number of zeros of any section counted with sign.
\begin{example}
  As a pedestrian example, we take $\BB{C}P^1$ and a line bundle ${\cal O}(n)$ and write the Thom/Euler class.
  Parameterise $\BB{CP}^1$ using $t\in[0,1]$ and $\gt\in[0,2\pi]$. Let $A=td\gt$ (for $t<1$) so that $\nabla=d-inA$ is the covariant derivative of ${\cal O}(n)$. Using \eqref{u_imme}, but making allowance for the necessary changes in going from a real bundle to a complex one, we get
  \bea u=\frac{1}{2i\pi\hbar}e^{-\frac{|\gs|^2}{\hbar} }\big(-\nabla\gs \wedge \nabla\bar \gs-\hbar F\big),~~~F=-indA.\nn\eea
  Here $\gs$ is the section and $u$ is written valid away from the south pole $t=1$. But under the transition from north to south hemisphere $\gs\to \gs (t/(1-t))^{-n/2}e^{-in\gt}$ and the expression of $u$ is actually globally valid.

  If we pick a bad section, say, $\gs=(t/(1-t))^{n/2}e^{in\gt}$ with non-simple zeros, the $\int \gs^*u$ gets its main contribution close to $t=0$. An integration by part argument gives $n$ exactly. On the other extreme, set $\gs=0$ one gets $\int \gs^*u=i/(2\pi)\int F=n/(2\pi)\int dA=n$ again. Finally if the section has simple zeros the formula \eqref{int_e} also gives $n$.
\end{example}

\subsection{Adding Equivariance}\label{sec_Ae}
Let $K$ be a compact Lie group and assume that $E$ is a $K$-equivariant vector bundle. If we choose a section $\gs$ preserved by the $K$-action, then the zeros of $\gs$ are no longer isolated. But then in computing the Euler class we can first localise to the zero of $\gs$ and then localise using the $K$-action. This is a trade off that can be crucial in infinite dimension.

We denote with $\FR{k}$ the Lie algebra of $K$, then there is an action of $\FR{k}$ on the total space of $E$. We use the same letter $\xi\in\FR{k}$ to denote a Lie-algebra element and the fundamental vector field of the action on $E$ generated by $\xi$.
By definition this action is a fibrewise isomorphism and covers an action of $v_{\xi}$ on the base $M$.

Furthermore the $K$-equivariance implies that there is a function $\mu\in \FR{k}^*\otimes {\rm End}\,E$, i.e. $\mu(\xi)\in{\rm End}\,E$ for $\xi\in\FR{k}$. It satisfies
  \bea &\nabla\mu(\xi)=-\iota_{v_{\xi}}F\label{moment_prin},\\
  &F(v_{\eta},v_{\xi})=\mu([\eta,\xi])-[\mu(\eta),\mu(\xi)].\label{moment_fail}\eea

Assuming that $K$ acts as an isometry, we fix a choice of $\xi\in\FR{k}$ and write simply $v$ for $v_{\xi}$.
We can then replace the de Rham algebra \eqref{dR_basic_sec} with an equivariant version
\bea
\begin{array}{l|l}
  \gd x=\psi & \gdh x=0 \\
  \gd\psi=v & \gdh \psi=0 \\
  \nabla_{\gd}\chi=iH & \gdh\chi=-\gs \\
  i\nabla_{\gd}H=F\chi+\mu(\xi)\chi & i\gdh H=\nabla_{\gd} \gs\\
\end{array}.\label{dR_euiv}\nn\eea
The section $\gs$ has to be invariant under $v$
\bea (\nabla_v-\mu(\xi))\gs=0.\label{inv_section}\eea
One can check that $\gd$ squares to the $v$-action if one uses \eqref{moment_prin} and \eqref{moment_fail}.

The Gaussian term \eqref{quad_thom_I} will be modified to accommodate the equivariance
\bea S&=&\frac{1}{2}(\gd+\gdh)\big(-\bra \chi,iH+\gs\ket+\bra \psi,v\ket)\nn\\
&=&\frac{1}{2}\big(-\bra iH-\gs,iH+\gs\ket+\bra \chi,(F+\mu(\xi))\chi+2\nabla\gs\ket+\bra v,v\ket-\bra \psi,L_v\psi\ket\big),\nn\eea
where in the last term we used the fact that $v$ is Killing.
The integral is now concentrated at $\gs=0$ \emph{and} $v=0$. Provided that the locus $\{\gs=0,v=0\}$ is isolated,
we can then ensure that the Gaussian is non-degenerate and can therefore use the universal formula \eqref{conclusion_earlier}.
\begin{example}
  Again we take ${\cal O}(n)$ bundle over $\BB{C}P^1$ as example. This bundle is $U(1)$ equivariant: we realise ${\cal O}(n)$ as the quotient of $[z_1,z_2;u]$ under the $\BB{C}^*$ action $[z_1,z_2;u]\to [\gl z_1,\gl z_2;\gl ^nu]$. Choose a $U(1)$ that rotates $z_2$ alone. Then in the patch $z_2\neq 0$, this $U(1)$ acts on the base coordinate $z_1/z_2$ with weight $-1$, and on fibre $uz_2^{-n}$ with weight $-n$. But on the patch $z_1\neq 0$, it acts on the base with weight 1 but 0 on the fibre.

  To illustrate the advantage of having equivariance, we can again pick the zero section and yet according to \eqref{conclusion_earlier} we still only get ${\det}^{1/2}V|_{\chi}\cdotp{\det}^{-1/2}V|_a$ from the north and south pole where the $U(1)$ degenerates. Note that since $\gs=0$, its derivative $\gs'$ which serves as $\gdh$, vanishes and so the $\gdh$-cohomology is everything. But at the south pole $z_2=0$, the action on the fibre is trivial and so that ${\det}^{1/2}V|_{\chi}$ vanishes. The only contribution is from $z_1=0$ and the super-determinant gives $n$ (again, one needs to switch from complex bundle to the underlying real bundle which offsets the square root).
\end{example}

\subsection{Lifting to the Principal Bundle}\label{sec_Lttpb}
We have constructed the integral representative of the Thom/Euler class, with integral over both fibre and base of a vector bundle $E\to M$. But it is sometimes more desirable not to work directly on the base manifold $M$. This is in view of our eventual application where $M$ should be the gauge equivalence class of connections of a given VB, which is singular due to the existence of reducible connections. However the space of all connections for a given VB is an affine space and much easier to work with (even though of infinite dimension). So we want to work over $P$ instead of $M$ with some necessary changes: we let $E=P\times_G V$, i.e. expressing $E$ as the associated VB of a principal bundle $P$ and express quantities such as the connection, curvature and sections from the perspective of $P$ instead of $M$.

\noindent {\bf Notations} We collect some notations used in this section:
\begin{enumerate}
\item $G$ a Lie group and $\FR{g}$ its Lie algebra, with basis $\{\xi_a\}$.
\item For $g\in G$ its action on a manifold is denoted by $R_g$, in the same way for $\xi\in\FR{g}$ its infinitesimal action is denoted $R_{\xi}$.
But use the same notation for $\FR{g}$ and the fundamental vector field generated by $\FR{g}$.
\item Given a vector space $V$ with a $G$ action, we use $\cdotp$ for the action, be it finite $v\to g\cdotp v$ or infinitesimal $v\to\xi\cdotp v$ for a $v\in V$.
But if $V$ transforms in the adjoint then the action is written as $gvg^{-1}$ or $[\xi,v]$.
\item $L_v$ and $\iota_v$ are the Lie derivative along a vector field $v$ and the contraction with $v$ respectively.
\end{enumerate}
Over $P$ the connection $\vgt$ is a $\FR{g}$-valued equivariant 1-form satisfying
\bea \iota_{R_\xi}\vgt=\xi,\label{eq_I}\eea
where we use the right $G$ action generated by $\xi\in\FR{g}$ since our convention is that the structure group $G$ acts from the left, and hence there is a natural right $G$ action on $P$. The equivariance here means simply
\bea L_{R_\xi}\vgt=-[\xi,\vgt],\label{eq_II}\eea
Note that it is only after pulling back $\vgt$ to $M$ with a local section that $s^*\vgt$ acquires the familiar inhomogeneous transformation property of connections.
The curvature $\varrho=d\vgt+\vgt\vgt$ is a $\FR{g}$-valued horizontal 2-form satisfying
\bea \iota_{R_{\xi}}\vrho=0,~~~L_{R_{\xi}}\vrho=-[\xi,\vrho].\label{eq_III}\eea

Next we need to write the de Rham complex \eqref{dR_basic} on $P\times_GV$, but we first work over the direct product $P\times V$ and in the end pass to the quotient $(P\times V)/G=P\times_GV$.
Denote with $a$ the even coordinate of $P$ and $\psi$ its odd companion, $\chi$ the odd coordinate of $V$ and $H$ its even companion.
We have
\bea \begin{array}{l|l}
       \gd a=\psi & \gd\psi=0, \\
       \gd\chi=iH & \gd H=0.\end{array}
\nn\eea

Compared to \eqref{dR_basic} there is no longer any connection/curvature term in the differential, this is because we are on the product space $P\times V$.

Now we need to take the quotient $P\times V\to P\times_GV$, for this we need some knowledge of the Weil/Kalkman algebra. For readers familiar with the BRST procedure, he will immediately recognise the resemblance, otherwise we refer to appendix \ref{sec_WKMfEC} for a quick review.

We introduce $\FR{g}$-valued coordinates $c,\phi$ with degree 1 and 2 and the differential
\bea
\begin{array}{ll|ll}
  \gd a=\psi, & \gdh a=-L_{R_c}a & \gd\vgt=\vrho-\vgt\vgt, & \gdh\vgt=\{c,\vgt\}-\phi,\\
  \gd\psi=0, & \gdh\psi=-L_{R_c}\psi+L_{R_{\phi}}a, & \gd\vrho=[\vrho,\vgt], & \gdh\vrho=[c,\vrho]\\
  \gd\chi=iH, & \gdh\chi=\chi\cdotp c {\color{blue}-s}, & \gd c=\phi, & \gdh c=c^2 \\
  i\gd H=0, & i\gdh H=-iH\cdotp c+\chi\cdotp\phi{\color{blue}+ds}, & \gd\phi=0, & \gdh \phi=[c,\phi]\\
\end{array}\label{dR_Weil}\eea
For now the reader may set to zero the blue terms, which will be explained later. We have written the action of $c,\phi$ on $\chi,H$ from the right since $\chi,H$ are coordinates of $V$.
One can observe that $\gd+\gdh$ corresponds exactly to \eqref{Kalkman_diff} i.e. de Rham coupled to the Weil/Kalkman algebra.

To write down eventually the Euler class one needs to pick a section $\gs$, which now is formulated as an equivariant map $s:\,P\to V$, i.e. $s$ satisfies
\bea s(R_ga)=g^{-1}\cdotp s,\label{equiv_sec}\eea
see the list of notations earlier. The relation \eqref{equiv_sec} makes sure that $\gs:~M\to P\times_GV$ defined as
\bea \gs(x)=[a,s(a)],~~{\rm where~}x=\pi(a),\nn\eea
is independent of the choice of the lift $a$ over $x$ and so is a well-defined section.

Using $s$ we can modify $\gdh$ as in the last line of \eqref{dR_basic_sec} and arrive back at \eqref{dR_Weil} with the blue terms incorporated.
We remark that the equivariance of $s$ \eqref{equiv_sec} is crucial to make $\gdh^2=0$.

\smallskip

We are now ready to write the Euler class. Assume that $V$ has a $G$-invariant pairing  $\bra\,\ket$, we write $S$ very much like \eqref{quad_thom_I}
\bea S_0=-\frac{1}{2}(\gd+\gdh)\bra \chi,iH+s\ket=-\frac{1}{2}\bra iH-s,iH+s\ket+\frac{1}{2}\bra \chi,\chi\cdotp\phi+2ds\ket.\label{thom_0}\eea
The absence of $c$ here is due to the $G$-invariance of $\bra,\ket$. Hence $e^{-S}$ produces an element of $\Go^{\sbullet}(P\times V)\otimes S^{\sbullet}\FR{g}$ with the second factor coming from $\phi$.
We will add a few more coordinates to perform three tasks
\begin{enumerate}
\item We need to plug the actual curvature $\varrho$ into $\phi$ to get a form on $P$. This can be achieved by adding to $S_0$ the term
$i\bra \varphi,\vrho-\phi\ket$ with $\varphi$ being $\FR{g}$-valued and even. Then integrating over $\varphi$ sets $\phi=\vrho$.
\item Enforcing $G$-invariance involves integrating along $G$, so we add to $S$ the term $\bra \eta,\vgt\ket$ with $\eta$ being a $\FR{g}$-valued odd variable. Integrating over $\eta$ produces a top power of $\vgt$ serving as the volume form along the $G$-orbit of $P$.
\item To factor out the volume of $G$, we need to 'fix the gauge' by adding $i\bra b,F(a)\ket$ stipulating that the $F=0$ slice be transverse to the $G$-orbit. Integrating $b$ gives us a delta function enforcing $F(a)=0$, but the incurred determinant factor will be cancelled by the Fadeev-Popov trick.
\end{enumerate}

Altogether we need to add four coordinates $\varphi,\eta,\bar c,b$ with differentials
\bea
\begin{array}{l|l}
  \gd \varphi=\eta, & \gdh \varphi=[c,\varphi], \\
  \gd \eta=0, & \gdh \eta=\{c,\eta\}-[\phi,\varphi], \\
  \gd \bar c=b, & \gdh \bar c=\{c,\bar c\}+\varphi, \\
  \gd b=0, & \gdh b=[c,b]-[\phi,\bar c]-\eta, \\
\end{array}.\nn\eea
After steps 1 and 2, the action is modified as
\bea S&=&S_0+(\gd+\gdh)\bra\vgt,\varphi\ket=S_0+\bra \vrho-\phi-\vgt\vgt+\{c,\vgt\},\varphi\ket-\bra\vgt,\eta+[c,\varphi]\ket\nn\\
&=&S_0+\bra \vrho-\phi-\vgt\vgt,\varphi\ket-\bra\vgt,\eta\ket.\label{thom_P}\eea
Integrating $\eta$ in the last term provides the top power of $\vgt$.
This also implies that the $\vgt\vgt$ term will never contribute since we have already saturated the vertical direction. Finally integrating over $\varphi$ sets $\phi=\vrho$ as desired.
The treatment so far follows closely \cite{ATIYAH1990119}, where it was shown further that the action \eqref{thom_P} leads to the twisted $N=2$ susy gauge theory upon identifying each term in the gauge theory setting. We shall review this in sec.\ref{sec_DWt}.

We also need to fix the gauge i.e. finish step 3. We add a gauge fixing term, where we write $\tilde b=b+\varphi$
\bea S_{gf}&=&-(\gd+\gdh)(\bra \bar c,\tilde b+F\ket)=S-\bra \tilde b+\{c,\bar c\},\tilde b+F\ket+\bra\bar c,[c,\tilde b]-[\phi,\bar c]+F'(\psi-L_{R_c}a)\ket\nn\\
&=&-\bra \tilde b,\tilde b+F\ket-\bra \{c,\bar c\},F\ket
+\bra\bar c,-[\phi,\bar c]+F'(\psi-L_{R_c}a)\ket.\label{S_gf}\eea
Integrating over $\tilde b$ gives a $|F|^2$ fixing the gauge, but this leaves us with the Fadeev-Popov determinant, which is offset by integrating over $c,\bar c$. Thus the Fadeev-Popov trick allows us to factor out ${\rm Vol}\,G$.

Putting together $S$ and $S_{gf}$. The integral $e^{-S-S_{gf}}$ is still concentrated at $s=0$, which is by construction not isolated: it has a $G$ action. We proceed by assuming that $\{s=0\}/G$ is isolated, implying that the gauged fixed locus $\{s=0,F=0\}$ is isolated.
Now we work up to first order deviation from the locus $\{s=0,F=0\}$, then the $\gd$ and $\gdh$ algebra collapses to a chain complex
\bea
\begin{tikzpicture}
  \node at (-2.6,0) {$\gd\;\uparrow$};
  \node at (0,1.3) {$\gdh\;\rightarrow$};
  \matrix (m) [matrix of math nodes, row sep=1.2em, column sep=2em]
    {  & {\color{blue}[c]_1} & \left[a\right]_0 & {\color{blue}\left[\chi\right]_{-1}}  \\
      &  \left[\phi\right]_2 & {\color{blue}[\psi]_1} & \left[H\right]_0 \\
       };
  \path[->]
  (m-2-2) edge (m-1-2)
  (m-2-3) edge (m-1-3)
  (m-2-4) edge (m-1-4);

  \path[->]

  (m-1-2) edge node[above] {\scriptsize{$-R_c$}} (m-1-3)
  (m-2-2) edge node[above] {\scriptsize{$R_{\phi}$}} (m-2-3)
  (m-1-3) edge node[above] {\scriptsize{$-s'$}} (m-1-4)
  (m-2-3) edge node[above] {\scriptsize{$s'$}}(m-2-4);
 \end{tikzpicture}~~
\begin{tikzpicture}
  \node at (-2,0) {$\oplus$};

  \matrix (m) [matrix of math nodes, row sep=1.2em, column sep=2em]
    {  \left[\varphi\right]_0 & {\color{blue}\left[\bar c\right]_{-1}}  \\
       {\color{blue}\left[\eta\right]_1} & \left[b\right]_0 \\           };
  \path[->]
  (m-2-2) edge (m-1-2)
  (m-2-1) edge (m-1-1);

  \path[->]
  (m-1-1) edge node[above] {\scriptsize{$id$}} (m-1-2)
  (m-2-1) edge node[above] {\scriptsize{$-id$}}(m-2-2);
\end{tikzpicture}\label{toy_cplx_Thom_G}\eea
where we have also labelled the ghost number for each variable.
The horizontal complex is exact: the left is so due to our assumption that $\{s=0\}$ is isolated mod the $G$ action, while the right one is trivially so.

The Gaussian integral itself is read off from our earlier labour in \eqref{V_nondeg}, where the matrix $Q$ comes from only two terms (those quadratic in fermions, one from first row and one from second row in \eqref{toy_cplx_Thom_G})
\bea \bra\chi,s'\psi\ket+\bra \bar c,F'\psi\ket.\label{used_I}\eea
Due to our assumption that $\{s=0,F=0\}$ is isolated, the matrix $F'$ can be regarded as the adjoint of $R$, the $\FR{g}$-action. Furthermore the complex
\bea a\stackrel{(s',F')}{\longrightarrow} \chi\oplus c \nn\eea
is the folding of the 3-level one (the top left in \eqref{toy_cplx_Thom_G})
\bea c\stackrel{R}{\to}a\stackrel{s'}{\to}\chi.\label{used_II}\eea
So the $\sgn\det Q$ that we seek is expressed entirely from the data of \eqref{toy_cplx_Thom_G}, i.e.
\begin{center}
 \it the sign of the torsion of the complex \eqref{used_II}.
\end{center}
In infinite dimensional setting the above punch line will be taken as the definition of the path integral.

Before leaving this section let us do a sanity check. Computing the torsion requires (see e.g. \cite{RAY1971145}) that one decompose the middle term of \eqref{used_II} into the base direction and fibre direction so that $R$ and $s'$ becomes square matrices. The torsion is the just the ratio of the two determinants.
Assuming that the orientation of base and fibre is chosen correctly, the sign of the torsion in fact equals $\sgn\det(s')$ \emph{with det taken transverse to the $G$-orbit}, as expected.

\begin{remark}
A natural question is: seeing that we obtained a result that is well-known, what has our labour achieved. The answer is threefold, 1. in the infinite dimensional case, one cannot separate the base and fibre of $P$ easily and so one must work over $P$. 2. replacing $\phi$ with $\vrho$ and integrating over $G$-orbit will bring about some determinant factors. These can cause serious problem in the infinite dimensional setting unless they are designed to cancel amongst each other, for this one needs various odd symmetry ($\gd$ or $\gdh$) linking bosons with fermions into pairs. 3. most importantly, the section will certainly not vanish transversally in infinite dimension setting, but the integral representative of the Euler class takes care of this. A famous example is the computation of genus 0 Gromov-Witten invariant when the target is an isolated rational curve. The non-transversality appears in the multiple covering case, but is treated successfully in \cite{Aspinwall1993} using the MQ formalism.
\end{remark}

\subsection{Non-free Action and Ghost Zero Modes}\label{sec_NfAaGZM}
As we intend to use $P$ to model the space of connections of a VB, the action of the gauge group is non-free at a reducible connection. So we need to modify the treatment above to take this into account.

This treatment again follows \cite{ATIYAH1990119}.
In the discussion above, we needed to obtain the connection and curvature. By definition, when evaluated on a vector field on $P$, a connection is supposed to return to  its vertical component. If we have a $G$-invariant metric on $P$, then we can achieve the vertical projection by using orthogonal projection. We again denote with the same $\xi$ the Lie algebra of $G$ and the fundamental vector field of the right-$G$-action on $P$, and with $g(-,-)$ the metric on $P$. We can write
\bea \iota_v\vgt:=\sum_a(C^{-1})^{ab}\bra v,\xi_b\ket \xi_a,~~~v\in{\rm vect}\,(P),~~~C_{ab}=g(\xi_a,\xi_b)\nn\eea
viz. we define a vector field to be horizontal if it is perpendicular to the action of all $\xi_a\in\FR{g}$.
When the $G$-action is free, one could have identified $C_{ab}$ with the Killing form on $\FR{g}$, but as we anticipate eventual non-free action, we denote the killing form as $k_{ab}$. We will freely use this last Killing form to raise or lower Lie algebra indices next.

Continuing, we have $\vgt^a=(C^{-1})^{ab}(g\xi_b)$ \footnote{here $g\xi_b$ is a 1-form on $P$, we hope it will not be confused with the left invariant vector field on $G$}. The curvature is obtained in the usual way $\varrho=d\vgt+\vgt^2$.
However this way of writing $\vgt$ involves the unsavory inversion of a matrix $C_{ab}$ and in particular, we cannot extend this to the case of non-free action.
But we note that $\vgt$ and $\varrho$ appears in the action \eqref{thom_P} as
\bea \bra \vrho-\phi-\vgt\vgt,\varphi\ket-\bra\vgt,\eta\ket.\label{used_V}\eea
We recall that in the first term anything that is vertical (e.g. $\bra\vgt\vgt,\varphi\ket$) can be ignored due to the saturation of vertical forms coming from the second term.
We can do a change of variables
\bea \tilde\varphi^a=(C^{-1})^{ab}k_{bc}\varphi^c,~~~~\tilde\eta^a=(C^{-1})^{ab}k_{bc}\eta^c\label{chg_var}\eea
This incurs no determinant factor since $\varphi,\eta$ are a pair of even/odd variables. Then \eqref{used_V} above can be written as
\bea (d(g\xi_a)-C_{ab}\phi^b)\tilde\varphi^a-(g\xi_a)\tilde\eta^a\label{used_VI}\eea
We will omit the tilde on $\varphi,\eta$ next.
In sec.\ref{sec_DWt}, we will see an explicit construction of this where horizontal projection is done through Hodge decomposition and the $C$-matrix will correspond to the Laplacian $\Gd$.

The equation \ref{used_VI} is clearly valid even for non-free actions. One sees further that if a subgroup $H\subset G$ acts trivially at some locus $a=a_0$, then the corresponding $\phi^a$ will not appear in \eqref{used_VI} and so is not replaced with $d(g\xi_a)$ by the path integral. It will instead remain as a parameter in the path integral. In the gauge theory context, this parameter is the Coulomb branch parameter. Similarly the ghosts corresponding to the subgroup that acts trivially will not appear in the gauge fixing action \eqref{S_gf} and hence will not appear anywhere in the action. These ghost components will not be integrated over.

Let $\FR{h}\subset\FR{g}$ be the Lie algebra of $H$, we denote the $\FR{h}$ components of $c$, $\phi$ as the \emph{zero modes}. One important change that ensues is that when we truncate the action and the susy algebra at a given susy configuration, we should expand $\phi$ from its 'expectation value'
\bea c=c_0+c'=c',~~\phi=\phi_0+\phi',\nn\eea
with $c',\phi'$ should be regarded as valued in the quotient $\FR{g}/\FR{h}$ or in $\FR{h}^{\perp}$ using the Killing form. There is a natural adjoint action of $\FR{h}$ on $c',\phi'$. This change of expansion also changes the look of the resulting double complex \eqref{toy_cplx_Thom_G}. It is easy to see that $\gd+\gdh$ no longer squares zero but to an action by $\phi_0$.
Indeed up to linear order
\bea
\begin{tikzpicture}
  \node at (-2.6,0) {$\gd\;\uparrow$};
  \node at (0,1.3) {$\gdh\;\rightarrow$};
  \matrix (m) [matrix of math nodes, row sep=1.2em, column sep=2em]
    {  & {\color{blue}[c']} & \left[a'\right] & {\color{blue}\left[\chi\right]}  \\
      &  \left[\phi'\right] & {\color{blue}[\psi]} & \left[H\right] \\ };
  \path[->]
  (m-2-2) edge (m-1-2)
  (m-2-3) edge (m-1-3)
  (m-2-4) edge (m-1-4);

  \path[->]
  (m-1-2) edge node[above] {\scriptsize{$-R$}} (m-1-3)
  (m-2-2) edge node[above] {\scriptsize{$R$}} (m-2-3)
  (m-1-3) edge node[above] {\scriptsize{$-s'$}} (m-1-4)
  (m-2-3) edge node[above] {\scriptsize{$s'$}}(m-2-4);
\end{tikzpicture}~~
\begin{tikzpicture}
  \node at (-2,0) {$\oplus$};
  \matrix (m) [matrix of math nodes, row sep=1.2em, column sep=2em]
    {  \left[\varphi\right] & {\color{blue}\left[\bar c\right]}  \\
       {\color{blue}\left[\eta\right]} & \left[b\right] \\           };
  \path[->]
  (m-2-2) edge (m-1-2)
  (m-2-1) edge (m-1-1);

  \path[->]
  (m-1-1) edge node[above] {\scriptsize{$id$}} (m-1-2)
  (m-2-1) edge node[above] {\scriptsize{$-id$}}(m-2-2);
\end{tikzpicture}\label{toy_cplx_Thom_G_zero}\eea
with susy algebra
\bea
\begin{array}{cc|cc}
 \gd a'=\psi, & \gdh a'=-L_{R_{c'}}a_0, & \gd \varphi=\eta,  & \gdh \varphi=0,\\
 \gd\psi=L_{R_{\phi_0}}a', & \gdh\psi=L_{R_{\phi'}}a_0, & \gd \eta=-[\phi_0,\varphi], & \gdh \eta=0,\\
 \gd\chi=iH, & \gdh\chi=-a's'(a_0), &\gd \bar c=b , & \gdh \bar c=\varphi,\\
 i\gd H=\chi\cdotp\phi_0, & i\gdh H=\psi s'(a_0), &\gd b=-[\phi_0,\bar c], & \gdh b=-\eta, \\
\end{array}\nn\label{diff_P}\eea
where $\gdh$ remains nilpotent (to check this keep in mind $s(a_0)=0$) while
\bea \gd^2=L_{R_{\phi_0}}\nn\eea
that is, the Coulomb branch parameter $\phi_0$ modifies $\gd$ from being de Rham to an equivariant differential.

In comparison to the treatment of ghost zero modes in \cite{Pestun:2007rz}, ours is more rough. For example, Pestun added extra fields to saturate the  integral over the zero modes i.e. the $\FR{h}$ component of $c,\phi,\bar c,b,\eta,\varphi$. But here we simply note that the zero modes of $\eta,\varphi$ cancel by pair so are that of $\bar c,b$ and $c,\phi$. This is viable so long as the zero modes are of finite dimension.

\subsection{Adding Equivariance bis}\label{sec_Ae_II}
We assume that the bundle $E=P\times_GV$ is equivariant w.r.t another action by $\Gc$. The point of adding this new ingredient is that in case $\{s=0\}/G$ is not isolated, one needs to integrate over it and adding $\Gc$-equivariance facilitates this integral. Indeed the $\Gc$-action on $E$ descends to $\{s=0\}/G$, then one can use equivariant localisation on the latter loci. Of course the non-triviality is that one does not want to first localise to $\{s=0\}$, mod by $G$ and then localise with $\Gc$, rather one has to do the all the steps in one go. We have seen such an example already at the end of sec.\ref{sec_Ae}.

The technical issue here is that the $\Gc$-action on $E=P\times_GV$ may not lift to $P\times V$ i.e. there is no $\Gc$-action on $P$.
On the other hand, we hope to have convinced the reader that it is paramount that we work on $P$. To solve this conflict, we note that for writing down the equivariant complex, one needs only the infinitesimal $\Gc$-action. So if $Y_{\xi}$ is the fundamental vector field corresponding to the $\xi\in\Lie \Gc$-action on $M$, one can always lift it to $P$ (say, by using a connection on $P$).
We still call the lift $Y_{\xi}$, and we write likewise an infinitesimal action $T_{\xi}$ on $V$. However let us stress again that
since $Y_{\xi},\,T_{\xi}$ do not stem from a group action, there are no straightforward commutation relations e.g. $[Y_{\xi},Y_{\eta}]\neq Y_{[\xi,\eta]}$. Instead $Y_{\xi},T_{\xi}$ will satisfy some coherence relations that come from the fact that $\Gc$ does act on $E=P\times_GV$. We will relegate the discussion of such coherence relations in the appendix. Besides, in concrete susy theory models, the complex looks much more natural than the abstract version we present here.
\bea
\begin{array}{ll|ll}
  \color{black}\gd a=\psi, & \gdh a=-L_{R_c}a & \gd\vgt=\vrho-\vgt\vgt+\iota_{Y_{\xi}}\vgt, & \gdh\vgt=\{c,\vgt\}-\phi,\\
  {\color{black}\gd\psi=Y_{\xi}}, & \gdh\psi=-L_{R_c}\psi+L_{R_{\phi}}a, & \gd\vrho=[\vrho,\vgt]+\iota_{Y_{\xi}}\vrho, & \gdh\vrho=[c,\vrho]\\
  {\color{black}\gd\chi=iH}, & \gdh\chi=\chi\cdotp c{\color{blue}-s}, & {\color{black}\gd c=\phi}, & {\color{black}\gdh c=c^2} \\
  i\gd H=-T_{\xi}\chi, & i\gdh H=-iH\cdotp c+\chi\cdotp\phi{\color{blue}+ds}, & \gd\phi=C_{\xi,c}, & \gdh \phi=[c,\phi]\\
\end{array}\label{dR_Weil_equiv}\eea
where $R_{\sbullet}$ denotes the right $G$-action on $P$ while the $G$ action on $V$ is shortened to a dot. More importantly $C_{\sbullet,\sbullet}:\,\textrm{Lie}\,\Gc\times \textrm{Lie}\,G\to \textrm{Lie}\,G$ measures the non-commutativity between $\Gc$-'action' and the right $G$-action on $P$.

One can also add the blue terms involving a section $s$, which we recall is a $G$-equivariant map $P\to V$, whose infinitesimal version reads $L_{R_c}s=s\cdotp c$. Now that $P\times_GV$ is also $\Gc$ equivariant, so we choose the $s$ to be $\Gc$-invariant
\bea Y_{\xi}\circ s=-T_{\xi}s\label{inv_section_II}.\eea
This relation replaces \eqref{inv_section} when we work over the principal bundle. The coherence relations replacing \eqref{moment_prin} and \eqref{moment_fail} are given in sec.\ref{sec_Efotpb}. With these conditions, one can check that $\gdh^2=\{\gd,\gdh\}=0$
while $\gd$ squares to the $\xi$-action.

One can write the action
\bea S_0&=&\frac{1}{2}(\gd+\gdh)\big(-\bra \chi,iH+s\ket+\bra \psi,Y_{\xi}+R_{\phi}\ket)\nn\\
&=&\frac{1}{2}\big(\bra iH-s,iH+s\ket+\bra \chi,-T_{\xi}\chi+\chi\cdotp\phi+2ds\ket+\bra Y_{\xi}+R_{\phi},Y_{\xi}+R_{\phi}\ket-\bra \psi,L_{Y_{\xi}+R_{\phi}}\psi\ket\big).\nn\eea

Next one adds the gauge fixing sector $\eta,\varphi,b,\bar c$ as in sec.\ref{sec_Lttpb}, however this step involves a slight snag.
The reason is that we know we can write \eqref{dR_Weil_equiv}, since the fields $a,\psi,\chi,H,c,\phi$ make up the bundle
$P\times_GV$ which \emph{is} $\Gc$-equivariant. But the gauge fixing sector $\eta,\varphi,b,\bar c$ is extraneous to the bundle and we cannot write an equivariant complex without stipulating further properties of say $C_{\xi,-}$. Indeed, by averaging over $G$-orbit, we can make $C_{\xi,-}$ invariant along the $G$-direction. Now we can write
\bea
\begin{array}{ll}
  \gd \varphi=\eta, & \gdh \varphi=[c,\varphi], \\
  \gd \eta=C_{\xi,\varphi}, & \gdh \eta=\{c,\eta\}-[\phi,\varphi], \\
  \gd \bar c=b, & \gdh \bar c=\{c,\bar c\}+\varphi, \\
  \gd b=C_{\xi,\bar c}, & \gdh b=[c,b]-[\phi,\bar c]-\eta, \\
\end{array}\nn\eea
the properties $\gdh^2=\{\gd,\gdh\}=0$ relies crucially on our assumption that $C_{\xi,-}$ be invariant.

Finally one can add to $S_0$ as in \eqref{thom_P} the terms $(\gd+\gdh)\bra\vgt,\varphi\ket=\bra \vrho+\phi+\iota_{Y_{\xi}}\vgt-\vgt\vgt,\varphi\ket-\bra\vgt,\eta\ket$. Also one adds the gauge fixing sector
\bea (\gd+\gdh)(\bra \bar c,\tilde b+F\ket)=-\bra \tilde b,\tilde b+F\ket-\bra \{c,\bar c\},F\ket
+\bra\bar c,-[\phi,\bar c]+F'(\psi-L_{R_c}a)\ket+\bra\bar c,C_{\xi,\bar c}\ket.\nn\eea
Now the bosonic term would contain $||Y_{\xi}||^2$ which forces us onto the fixed point set of $Y_{\xi}$ which is $G$-invariant. We assume next that the locus $\{s=0,F=0\}\cap\{Y_{\xi}=0\}$ is a set of isolated points.
The rest of the discussion is similar to the non-equivariant case: at each such point we can get a complex of type \eqref{toy_cplx_Thom_G},
and \eqref{conclusion_earlier} says that the result is a superdeterminant of $Y_{\xi}$ taken over the $\gdh$ cohomology.
Therefore, the right complex of \eqref{conclusion_earlier} contributes nothing, as $\gdh$ is exact in that part. This is as it should be since that part of the complex is extraneous to the problem and is only used to fix gauge.

\section{Application to Gauge Theory}\label{sec_AtGt}
The general setting is the following, let $X$ be a 4-manifold and $E\to X$ be a $G$-vector bundle. Take ${\cal A}$ the space of connections on $E$, and ${\cal G}$ the group of gauge transformation on $E$, which then acts on ${\cal A}$. The gauge equivalence class of connections ${\cal M}={\cal A}/{\cal G}$ is not a manifold due to the reducible connections. Therefore we will work over ${\cal A}$ and make heavy use the discussion of sec.\ref{sec_Lttpb}. To construct a bundle associated to ${\cal A}\to{\cal M}$, we pick a vector space ${\cal V}$ to be the space of sections of a rank 3 sub-bundle of adjoint 2-forms on $X$. Different choice of the sub-bundle leads to different theories.

\subsection{Donaldson-Witten Theory}\label{sec_DWt}
Consider the bundle $\Go^{2+}(\End E)$, consisting of $\End(E)$-valued self-dual 2-forms, and a vector space ${\cal V}=\Gc(\Go^{2+}(\End E),M)$ consisting of sections of such, then we can form the associated bundle
\bea {\cal E}={\cal A}\times_{\cal G}{\cal V}.\label{asso_bdl}\eea
Recall from \eqref{equiv_sec} that a section of ${\cal E}$ is a ${\cal G}$-equivariant map ${\cal A}\to {\cal V}$.
So for a connection $A$ of $E$ regarded as a point in ${\cal A}$, we assign $F_A^{2+}$, the self-dual component of the curvature of $A$. One checks the equivariance easily.
The zeroes of this section are exactly those anti-self-dual connections. We will write down the Euler class which is Poincare dual to the zeros locus, i.e. the moduli space of anti-self-dual connections.

Before writing down the complex we summarise the notations.
\begin{enumerate}[label=\textbf{n.\arabic*}]
  \item $A$: connection of $E$; $A_0$: connection of $E$ chosen as a reference, all other connections are written as $A=A_0+a$ with $a\in\Go^1(X,ad)$.
  \label{notation 1}
      \item
      $d_A=D=d-A$ (resp. $d_0=d_{A_0}=d-A_0$) the covariant derivative using $A$ (resp. $A_0$), with $F_A$ (resp. $F_{A_0}=F_0$)  the corresponding curvature. \label{notation 2}
      \item even variables: $\phi,\varphi\in\Go^0(M,ad)$, $H\in\Go^{2+}(X,ad)$. \label{notation 3}
      \item odd variables: $c,\eta\in\Go^0(M,ad)$, $\Psi\in\Go^1(X,ad)$, $\chi\in\Go^{2+}(X,ad)$. \label{notation 4}
\end{enumerate}
Next we write down the susy algebra as a result of twisting the $N=2$ susy gauge theory see \cite{witten1988}.
First, for each topological type of $E$, we pick a connection $A_0$ as a reference and write any other connection as $A=A_0+a$. This way $a$ is an adjoint 1-form.
\bea \begin{array}{l|l}
       \gd a=\Psi, & \gdh a=d_0c-[a,c], \\
       \gd\Psi=0, & \gdh\Psi=d_0\phi+\{c,\Psi\}-[a,\phi],  \\
       \gd c=-\phi, & \gdh c=c^2, \\
       \gd\phi=0, & \gdh\phi=[c,\phi], \\
       \gd\chi=H, & \gdh\chi=\{c,\chi\}+F_{a+A_0}^+, \\
       \gd H=0, & \gdh H=[\phi,\chi]+[c,H]-(d_{a+A_0}\Psi)^+, \\ \end{array}~~
       \begin{array}{l|l}
       \gd \bar c=b, & \gdh \bar c=\{c,\bar c\}+\varphi{\color{blue}\,\mapsto\Gd_A\varphi}, \\
       \gd b=0, & \gdh b=[c,b]+[\phi,\bar c]-\eta{\color{blue}\,\mapsto\gd\Gd_A\varphi}, \\
       \gd \varphi=\eta, & \gdh \varphi=-[\varphi,c], \\
       \gd \eta=0, & \gdh\eta=\{c,\eta\}-[\varphi,\phi], \\        \end{array}\label{tbl_DW}\,.\eea
One should ignore the blue term for now, but later we shall explain the replacement $\varphi\mapsto\Gd_A\varphi$, $\eta\mapsto\gd\Gd_A\varphi$.

The division of the complex into two parts is meant to parallel that of \eqref{dR_Weil}, with the section $s$ given by $a\to F_{a+A_0}^+$.
One may check that \eqref{tbl_DW} satisfies
\bea \gd^2=\gdh^2=\{\gd,\gdh\}=0.\nn\eea

We first write down an action that would parallel that of $S_0$ in \eqref{thom_0} with Wick rotation $H\to iH$
\bea S_0=-\frac{1}{2}(\gd+\gdh)\Tr\int_X\bra \chi,iH-F^+\ket=\frac{1}{2}\Tr\int_X-\bra iH+F^+,iH-F^+\ket+\bra \chi,[\phi,\chi]-2(D\Psi)^+\ket.\nn\eea
Integrating over $H$ gives us term $|F^+|^2$ and so the Euler class corresponds to the solution to the anti-self-dual connections, i.e. those $A$ with $F^+=0$.

Continuing with the parallel
 with sec.\ref{sec_Lttpb}, we need to understand how to write the connection of the bundle ${\cal A}\to{\cal M}$. We follow the construction of \cite{ATIYAH1990119}, as the tangent space of ${\cal A}$ are sections of adjoint 1-forms $\Go^1(M,ad)$, there is a ${\cal G}$ invariant metric
\bea \bra a_1,a_2\ket:=\int_X \Tr a_1*a_2,~~~a_{1,2}\in\Go^1(X,ad).\nn\eea
With a metric one can define the horizontal subspace of $T{\cal A}$ by orthogonal projection: we declare $\gd A\in T_A{\cal A}$ horizontal if $\gd A$ is orthogonal to $d_A\ep$ for any $\ep\in \Go^0(X,ad)$, i.e. orthogonal to any gauge transformation. Concretely we let the connection $\vgt$ be
\bea \vgt(\gd A):=\Gd_A^{-1}d^{\dag}_A\gd A.\nn\eea
Note that $d_A^2\neq 0$ nevertheless
\bea (\gd A-d_A\Gd_A^{-1}d^{\dag}_A\gd A)\perp d_A\ep,~~~~\forall\ep\in\Go^0(X,ad)\nn\eea
for $\gd A\in\Go^1(X,ad)$ due to degree reasons. Written in supermanifold language, i.e. writing $\Psi$ for $\gd A$ we have
\bea \vgt=\Gd_A^{-1}d^{\dag}_A\Psi.\nn\eea

Parallel to \eqref{thom_P}, we need to insert the term $(\gd+\gdh)\bra\vgt,\varphi\ket$ into the action, in the gauge theory setting, this would read
\bea (\gd+\gdh)\Tr\int_X \bra\Gd_A^{-1}d_A^{\dag}\Psi,\varphi\ket.\label{used_III}\eea
But certainly a term with $\Gd_A^{-1}$ cannot appear in a local action. So as explained in sec.\ref{sec_NfAaGZM} we make a change of variable parallel to \eqref{chg_var}
\bea \varphi\to \Gd_A^{-1}\varphi,~~~\eta\to \gd\Gd_A^{-1}\varphi=\Gd_A^{-1}\eta+\cdots\nn\eea
where $\cdots$ denote terms coming from varying $\Gd_A^{-1}$. The determinant in such change of variable cancels between $\eta$ and $\varphi$.
This change of variable also explains the blue terms in the table \ref{tbl_DW}.

The net effect is that we get the action
\bea S_0&=&\frac{1}{2}\Tr\int_X |H|^2+|F^+|^2+\bra \chi,[\phi,\chi]-2(D\Psi)^+\ket
+\bra\{\Psi,*\Psi\}-D^{\dag}D\phi,\varphi\ket-\bra  D^{\dag}\Psi,\eta\ket\nn\\
&=&\frac{1}{2}\Tr\int_X |H|^2+|F^+|^2+\chi\wedge[\phi,\chi]-2\chi\wedge D\Psi
+[\varphi,\Psi]*\Psi-(D\phi)*(D\varphi)-(D^{\dag}\Psi)*\eta.\nn\eea
One observes that even though the complex \eqref{tbl_DW} contains the ghost, the action above does not, and it is the twisted $N=2$ Lagrangian of Witten \cite{witten1988}, and also following the argument there, one needs to declare $\varphi$ and $\phi$ a complex conjugate pair so as to get a positive definite  action (see the discussion at the end of sec.2. of \cite{witten1988}).

The action is dominated at $F^+=0$, where upon truncating to linear order the susy algebra \eqref{tbl_DW} collapses into a chain complex
\bea
\begin{tikzpicture}
  \node at (-2,0) {$\gd\;\uparrow$};
  \node at (0,1.3) {$\gdh\;\rightarrow$};
   \matrix (m) [matrix of math nodes, row sep=1.6em, column sep=2.4em]
    {  & {\color{blue}\left[c\right]_1} & \left[a\right]_0 & {\color{blue}\left[\chi\right]_{-1}}  \\
      &  \left[\phi\right]_2 & {\color{blue}\left[\Psi\right]_1} & \left[H\right]_0 \\ };
  \path[->]
  (m-2-2) edge (m-1-2)
  (m-2-3) edge (m-1-3)
  (m-2-4) edge (m-1-4);

  \path[->]
  (m-1-2) edge node[above] {\scriptsize{$d_0$}} (m-1-3)
  (m-2-2) edge node[above] {\scriptsize{$-d_0$}} (m-2-3)
  (m-1-3) edge node[above] {\scriptsize{$(d_0{\textrm-})^+$}} (m-1-4)
  (m-2-3) edge node[above] {\scriptsize{$-(d_0{\textrm -})^+$}}(m-2-4);
\end{tikzpicture}~~
\begin{tikzpicture}
  \node at (-2,0) {$\oplus$};
  \matrix (m) [matrix of math nodes, row sep=1.2em, column sep=2em]
    {  \left[\varphi\right]_0 & {\color{blue}\left[\bar c\right]_{-1}}  \\
       {\color{blue}\left[\eta\right]_1} & \left[b\right]_0 \\           };
  \path[->]
  (m-2-2) edge (m-1-2)
  (m-2-1) edge (m-1-1);

  \path[->]
  (m-1-1) edge node[above] {\scriptsize{$\Gd_0$}} (m-1-2)
  (m-2-1) edge node[above] {\scriptsize{$-\Gd_0$}}(m-2-2);
\end{tikzpicture}\label{instanton_cplx}\eea
We would like to draw the readers attention to the horizontal part of the complex, assuming that $A_0$ is anti-self-dual, then the complex is just the well-known instanton deformation complex
\bea 0\to\Go^0(ad)\stackrel{d_0}{\longrightarrow}\Go^1(ad)\stackrel{(d_0{\textrm{-}})^+}{\longrightarrow}\Go^{2+}(ad)\to0.\label{AHS_cplx}\eea
The cohomology in the middle step are the small deformations $a$ from $A_0$ such that the anti-self-duality is maintained (the kernel of the second map) while modulo the gauge transformation (the image of the first map).

In \eqref{toy_cplx_Thom_G} we assumed that the horizontal complex is exact, so that the zeros of a section is isolated. Now in this infinite dimensional setting, exactness is replaced with \emph{ellipticity}. We spend some time review briefly this point.

A differential operator $D:\, E\to F$ induces a bundle map $\pi^*E\to \pi^*F$ (where $\pi: T^*X\to X$ is the projection) as follows. The symbol of $D$, which is obtained by retaining in $D$ only the highest order derivatives and replacing each $\partial_i$ with $\xi_i$, gives a bundle map over $T^*X$ from $\pi^*E\to \pi^*F$ (from now on we shall omit $\pi^*$). The operator $D$ is elliptic iff the bundle map thus constructed is an isomorphism away from the zero section. \footnote{the reason for retaining only the highest order derivative is that far out along the fibre of $T^*M$, only the terms with the highest order of $\xi$ dominates. Thus one has that the pair $E\stackrel{\gs(D)}{\to} F$, as an element of the K-group, has compact support along the fibre. Once this is ascertained, the lower order terms in $\gs(D)$ can be homotoped away using a linear homotopy, without changing the class of $E\stackrel{\gs(D)}{\to} F$ in the $K$-group.}

The symbol for the complex \eqref{AHS_cplx} is
\bea 0\to\Go^0(ad)\stackrel{\xi\wedge\textrm{-}}{\longrightarrow}\Go^1(ad)\stackrel{(\xi\wedge{\textrm{-}})^+}{\longrightarrow}\Go^{2+}(ad)\to0.\label{AHS_cplx_symbol}\eea
We need to show that this complex is exact. That the first map is injective and the first two maps compose to zero is obvious. In the middle step.
Suppose $(\xi\wedge a)^+=0$, then $\xi\wedge\ga=0$ since decomposable 2-forms cannot be ASD or SD. Indeed $0=(\xi\wedge\ga)\wedge(\xi\wedge\ga)=-(\xi\wedge\ga)*(\xi\wedge\ga)=-|\xi\wedge\ga|^2$.
But now $\xi\wedge \ga=0$ means $\ga=\xi\wedge \phi$ for some $\phi$ by elementary linear algebra. Thus the kernel of the second map is contained in the image of the first map. Finally for any SD $\go$, we consider (we used the same symbol for $\xi$ and $g\xi$)
\bea\frac12(1+*)\xi\wedge\iota_{\xi}\go=\frac12(\xi\wedge\iota_{\xi}\go+\iota_{\xi}\xi\wedge*\go)
=\frac12(\xi\wedge\iota_{\xi}\go+\iota_{\xi}\xi\wedge\go)=\frac12|\xi|^2\go.\nn\eea
So if $\xi\neq 0$, the last map is onto and the whole complex is exact.

The ellipticity of \eqref{AHS_cplx} implies that the cohomology at each step is of finite dimension. In particular, tangent space at a given ASD connection is of finite dimension. It was also shown in \cite{AtiyahHitchinSinger} that the deformation is unobstructed, and so away from the reducible connections, the moduli space of ASD connection is a smooth manifold.

Back to the computation of the path integral, the general formula \eqref{conclusion_earlier} is readily applicable. The cohomology of \eqref{AHS_cplx} (or rather its Euler character) can be computed using the index theorem, while the cohomology of the right half of \eqref{instanton_cplx} involve the Laplacian and so the cohomology is empty if the connection is irreducible. In summary the super-determinant \eqref{conclusion_earlier} can be obtained easily.

In the next section, the deformation complex will no longer be elliptic but transversally elliptic.

\subsection{5D $N=1$ SYM}\label{sec_5DN=1SYM}
We add equivariance to the discussion of last section. However as we will also soon discuss a 5D $N=1$ theory, we will write the complex in 5D terms to avoid repetition. Note that upon reduction, we will get the equivariant DW theory or the exotic 4D $N=2$ SYM, all depending on the relative alignment of certain Killing vector fields.

Let now $M$ be a 5-manifold with a nowhere vanishing Killing vector field $\reeb$ (typically a Reeb vector field for contact $M$). Now we are interested in looking for connections which are anti-self-dual transverse to $\reeb$. More concretely, let $F$ be the curvature and $F_H$ be its component perpendicular to $\reeb$, which can be written as
\bea F_H=F-\gk\wedge\iota_{\reeb}F,\nn\eea
where $\gk$ is the 1-form obtained by applying the metric to $\reeb$. Since $\reeb$ is Killing, we can normalise its length to be 1, and hence $\iota_{\reeb}\gk=|\reeb|^2=1$. We are interested in those $F$ that has
\bea F_H^+=P_+F=\frac12(1+\iota_{\reeb}*_5)F_H=\frac12(F-\gk\iota_{\reeb}F+\iota_{\reeb}*F)=0,\label{trans_asd}\eea
where we have introduced the projector $P_+$ that projects a 2-form to its horizontal and self-dual component.

As in sec.\ref{sec_DWt}, we still let ${\cal A}$ be the space of connections and
\bea {\cal V}=\Gc(\Go_H^{2+}(\End(E)),M).\nn\eea
The association $A\to F_H^+$ gives a section of the associated bundle, whose zeros are exactly solutions to \eqref{trans_asd}.

The following complex was motivated from twisting the 5D $N=1$ supersymmetry, which contains an adjoint scalar field $\gs$. The combination $\gs-A_{\reeb}$ is special in that its susy transformation is zero, except it is not an adjoint scalar since $A_{\reeb}$ is not. We need to split $A=A_0+a$ and use instead the combination
\bea \phi=\gs+a_{\reeb}\nn\eea
as an adjoint scalar. In \eqref{tbl_5D} $\phi$ becomes the partner of the ghost and is the degree 2 generator of the Weil complex (with $\gdh$).

Item \ref{notation 1}, \ref{notation 2} in the list of notations in sec.\ref{sec_DWt} remain unchanged. However $\chi,H$ are now sections of $ \Go^{2+}_H(M,ad)$, which is certainly not surprising.
\bea \begin{array}{l|l}
       \gd a=\Psi & \gdh a=d_Ac  \\
       \gd\Psi=-\iota_{\reeb}F_0-L_{\reeb}^0a & \gdh\Psi=d_A\phi+\{c,\Psi\}  \\
       \gd c=-\phi & \gdh c=c^2 \\
       \gd\phi=L_{\reeb}^0c & \gdh\phi=[c,\phi] \\
       \gd\chi=H & \gdh\chi=\{c,\chi\}+(F_A)_H^+ \\
       \gd H=-L_{\reeb}^0\chi & \gdh H=[\phi,\chi]+[c,H]-(d_A\Psi)_H^+ \\ \end{array}~~
       \begin{array}{l|l}
       \gd\bar c=b & \gdh \bar c=\{c,\bar c\}+{\color{blue}d_A^{\dag}\iota_{\reeb}F} \\
       \gd b=-L^0_{\reeb}\bar c & \gdh b=[c,b]+[\phi,\bar c]-{\color{blue}\gd d_A^{\dag}\iota_{\reeb}F} \\ \end{array}
       \label{tbl_5D}\eea
Note that $L^0_{\reeb}$ means Lie derivative with connection $A_0$ and $\iota_{\reeb}F_0+L_{\reeb}^0a=\iota_{\reeb}F_A+d_Aa_{\reeb}$.
The section $(F_A)_H^+$ satisfies the equivariance analogous to \eqref{inv_section_II}
\bea (F_A)_H^+\xrightarrow{A\to A+\iota_{\reeb}F_0+L_{\reeb}^0a}(F_A)_H^++L_{\reeb}^0(F_A)_H^+.\nn\eea
The complex satisfies $\{\gd,\gdh\}=0$, $\gdh^2=0$ and $\gd$ squares to the $\reeb$ action.
Comparing with the table \ref{tbl_DW}, the term $\Gd_A\varphi$ should naively be replaced with $\Gd_Aa_{\reeb}$, but this does not result in a nilpotent $\gdh$. So the correct 5D covariantisation is $\Gd_A\varphi\mapsto d_A^{\dag}\iota_{\reeb}F$, as in the blue terms.

It might be surprising that we introduced equivariance $L_{\reeb}^0$ into $\gd$ based on a choice of reference connection. But in fact this is the manifestation of the complication of complex \eqref{dR_Weil_equiv}, where we had no $\Gc$-equivariance on the principal bundle. The term $\gd\phi=C_{\xi,c}$ there corresponds here to $\gd\phi=L_{\reeb}^0c$. The $G$-invariance of $C_{\xi,c}$ corresponds here to the fact that $A_0$ is a fixed reference connection.

For all purposes, only the combination $\gd+\gdh$ will appear and so the choice of split $A=A_0+a$ will not matter: e.g.
\bea &(\gd+\gdh)\Psi=
-\iota_{\reeb}F_0-L_{\reeb}^0a+d_A\phi+\{c,\Psi\}=-\iota_{\reeb}F_A+d_A(-a_{\reeb}+\phi)+\{c,\Psi\},\nn\\
&(\gd+\gdh) H=-L_{\reeb}^0\chi+[\phi,\chi]+[c,H]-(d_A\Psi)^+=-L^A_{\reeb}\chi+[-a_{\reeb}+\phi,\chi]+[c,H]-(d_A\Psi)^+\nn\eea
On the rhs only $A$ appears, while the combination $-a_{\reeb}+\phi=\gs$
is actually an adjoint scalar in the untwisted 5D supersymmetric theory. We see that the choice of split drops out.

\begin{remark}\label{rmk_equivariance}
We make a lengthy remark about equivariance.

Confusion might arise due to the plethora of the word 'equivariant bundle'. Here we do \emph{not} assume that the gauge bundle is equivariant. Nor is there any natural way to make ${\cal A}$ equivariant, but the associated bundle ${\cal A}\times_{\cal G}{\cal V}$ is \emph{always} equivariant if $M$ has some isometry, as we explain next.

If $\xi\in{\rm Lie}\,\Gc$ acts on $M$, we lift the action on $M$ to ${\cal A}$ and ${\cal V}$ arbitrarily using $A_0$:
\bea &\xi\circ a=\iota_{\xi}F_0+L_{\xi}^0a,~~~a\in\Go^1(ad),\nn\\
&\xi\circ s=L_{\xi}^0s,~~~s\in\Go^{2+}_H(ad)\nn\eea
This descends to an action on ${\cal A}\times_{\cal G}{\cal V}$. Indeed one can rewrite $\xi\circ a=\iota_{\xi}F+d_Aa_{\xi}$, and $\xi\circ s=L^A_{\xi}s+[a_{\xi},s]$. The terms $d_Aa_{\xi}$ and $[a_{\xi},s]$ are gauge transformation on $A$ and on $s$ with parameter $a_{\xi}$, and so can be cancelled across ${\cal A}\times_{\cal G}{\cal V}$, by definition of the associated bundle. The commutator of the actions of $\xi_1,\xi_2$ on $a$ can be computed too:
\bea \xi_{[1}\circ\xi_{2]}\circ a&=&\xi_{1}\circ(\iota_{\xi_{2}}F_0+L^0_{\xi_2}a)-[1\leftrightarrow2]=L^0_{\xi_2}(\iota_{\xi_1}F_0+L_{\xi_1}^0a)-[1\leftrightarrow2]\nn\\
&=&\iota_{[\xi_{2},\xi_1]}F_0+d_0F(\xi_1,\xi_2)-[F_0(\xi_2,\xi_1),a]+L_{[\xi_2,\xi_1]}^0a=[\xi_2,\xi_1]\circ a-d_A(F_0(\xi_2,\xi_1)),\nn\\
\xi_{[1}\circ\xi_{2]}\circ s&=&\xi_1\circ L_{\xi_2}^0s-[1\leftrightarrow2]=[\xi_2,\xi_1]\circ s-[F_0(\xi_2,\xi_1),s]\nn.\eea
Again we see that for both $a$ and $s$, the commutation relation fails up to a common gauge transformation with parameter $F_0(\xi_2,\xi_1)$, and hence the commutation relation can be restored.

To summarise the section $(F_A)_H^+$ is not an invariant section of $\Go^{2+}_H(M,ad)$, but it is an invariant section of ${\cal A}\times_{\cal G}{\cal V}$.
\end{remark}

Next we will reduce the 5D theory to 4D theories, assume that $M$ admits a free $U(1)$ and $X=M/U(1)$ is a smooth 4-manifold. We have more notations
\begin{enumerate}
  \item $\xeeb$ is the fundamental vector field of $U(1)$ and it has $[\xeeb,\reeb]=0$,
  \item $\hat\xeeb=\xeeb/|\xeeb|$ and we also use $\hat \xeeb$ to denote $g\hat\xeeb$,
  \item $\gb$ is the 1-form $g\xeeb/|\xeeb|^2$, note that $\iota_Xd\gb=0$,
  \item $\rho=\bra \xeeb,\reeb\ket$, $f=\bra\hat\xeeb,\reeb\ket$, $h=|\xeeb|$, $\rho={f/h}$,
  \item $\veeb$ is the vector field on $X$ that $\reeb$ descends to.
\end{enumerate}\label{tab_notations}

\subsection{Reduction to Equivariant DW Theory}\label{sec_RtEDWt}
Comparing \eqref{tbl_DW} \eqref{tbl_5D}, we see that the scalars $\varphi$ and $\eta$ disappear, their roles are taken over by the component of the gauge field $A$ and $\Psi$ along the $\reeb$ direction. In fact it is well known that the 4D $N=2$ susy actually is the reduction of 5D $N=1$ susy, and one component of the gauge field turns into a scalar.

We reduce along a free $U(1)$ generated by $\xeeb$, which is assumed to be nowhere anti-parallel with $\reeb$ (that is $\bra \hat\xeeb,\reeb\ket>-1$).
If $X$ equals $\reeb$ everywhere, then we shall get the DW theory without equivariance since $\reeb$ simply descends to zero.
In general $\reeb$ will descend to a Killing vector $\veeb$ on $X$ providing the equivariance in DW theory.

One can demand that no gauge transformation have $\xeeb$ dependence, and the
5D gauge bundle is pulled back from the 4D base. We simply take the connection to be $A_5=\pi^*A_4+\varphi \gb$, where $\pi:\,M\to X$ is the projection. This way $\iota_XA$ becomes an adjoint scalar $\varphi$ and $\iota_X\Psi$ becomes $\eta$. One can now get the susy complex of the equivariant DW theory, e.g. $F_5=F_4+d_{A_4}\varphi\wedge\gb+\varphi d\gb$ and
\bea &&(\gd+\gdh) a_5=\Psi_5+d_{A_5}c~\To\nn\\
&&\hspace{2cm} (\gd+\gdh)a_4=\Psi_4+d_{A_4}c,~~~(\gd+\gdh)\varphi=\eta+[c,\varphi],\nn\\
&&(\gd+\gdh) \Psi_5=-\iota_{\reeb}F_5+d_{A_5}\gs+\{c,\Psi_5\}~\To\nn\\
&&\hspace{2cm}(\gd+\gdh) \Psi_4=-\iota_{\veeb}F_4+d_{A_4}(\gs+\varphi\bra\reeb,\gb\ket)+\{c,\Psi_4\},~~~\gd\eta=-L_{\veeb}^{A_4}\varphi+\{c,\eta\}-[\varphi,\gs].\nn\eea

For the reduction of $H,\chi$ sector, one notes that $\reeb$ and $\xeeb$ are assumed never anti-parallel $\bra \hat \xeeb,\reeb\ket\gneq -1$, the space of self-dual 2-forms transverse to $\reeb$ and to $\xeeb$ are isomorphic.
Indeed let $B\in\Go_H^{2+}(M)$ we map it to
\bea B\mapsto \frac12\big(B-\hat \xeeb\iota_{\hat \xeeb}B+\iota_{\hat \xeeb}(\gk\wedge B)\big),\label{iso}\eea
one may check that the rhs is perpendicular to $\xeeb$ and self-dual transverse to $\xeeb$. Conversely let a 2-form $C$ be perpendicular to $\xeeb$ self-dual transverse to $\xeeb$, we can map it to
\bea C\mapsto\frac{2}{(1+\bra \hat\xeeb,\reeb\ket)^2}\big(C-\gk\iota_{\reeb}C+\iota_{\reeb}(\hat\xeeb\wedge C)\big)\label{iso_inv}\eea
where the rhs is in $\Go_H^{2+}(M)$ and the two maps are inverses of each other.
Here we see that $\bra \hat\xeeb,\reeb\ket\gneq-1$ is crucial for such an isomorphism to exist. In the next section, this condition would fail and we get exotic DW theory.

We apply the map \eqref{iso} to $\chi$ and $H$ and get their susy transformation

\bea &&(\gd+\gdh)\chi_4=H_4+\{c,\chi_4\}+(\cdots),\nn\\
&&(\gd+\gdh)H_4=-L_{\veeb}^{A_4}\chi_4+[c,H_4]-\gd(\cdots),\nn\\
&&(\cdots)=\frac{1}{4}(1+*_4)\big((1+f)(F_4+\varphi d\gb)+h^{-1}*_4(g_4\veeb\wedge D\varphi)-g_4\veeb\wedge(\iota_{\veeb}F_4-d_{A_4}(\varphi\rho/h^2))\big).\nn\eea
The complexity of the transformation should convince one that it is far better to package the susy rules of equivariant DW theory in 5D terms, then the large bracket above is just one term.

\subsection{Reduction to Exotic DW Theory}\label{sec_RteDt}
Now we consider the scenario where $\xeeb,\reeb$ can be anti-parallel and that $\reeb$ has irregular orbits.
Then the maps \eqref{iso} \eqref{iso_inv} are not isomorphisms. Nonetheless if $B\in\Go^{2+}_H$ then
\bea B^{\perp}=B-\gb\iota_{\xeeb}B\label{iso_P}\eea
is a section of another rank 3 subbundle of the rank 6 bundle $\Go^2_H$ defined by a new projector.

To explain this, we need to relate some 5D quantities such as the metric to 4D, we refer the reader to the beginning of sec.\ref{sec_Mgs}. A simple calculation shows that $B^{\perp}$ satisfies $PB^{\perp}=B^{\perp}$ with $P$ being
\bea P=\frac{1}{(1+f^2)}(1+f*_4-(g_4\veeb)\wedge\iota_{\veeb}),~~~~f=\bra \reeb,\hat \xeeb\ket_5.\nn\eea
This is a 4D projector $P^2=P$. As both $\reeb,\,\hat\xeeb$ have unit length, one sees $f\in[-1,1]$, and it takes $\pm1$ when $\xeeb,\reeb$ are (anti)-parallel (hence $\veeb=0$). Thus the projector is an extrapolation of self-dual and anti-self-dual projectors.

Conversely if $C$ satisfies $PC=C$, then
\bea C\to C-|\xeeb|\gb\wedge\iota_{\veeb}*_4C\in\Go^{2+}_H\label{iso_P_inv}\eea
is the inverse map.

In fact, a greater generality is possible \cite{Festuccia:2018rew}
\bea P_{\go}=\frac{1}{(1+\cos^2\go)}(1+\cos\go*_4-\frac{\sin^2\go}{|{\veeb}|_4^2}(g_4\veeb)\wedge\iota_{\veeb}),~~~P_{\go}^2=P_{\go}\nn\eea
where $\go\in [0,\pi]$ and goes to $0,\pi$ sufficiently fast when $\veeb=0$.

Applying the isomorphisms \eqref{iso_P} \eqref{iso_P_inv} to \eqref{tbl_5D} we can get the complex for the exotic DW theory.
\bea \begin{array}{l|l}
       \gd a_4=\Psi_4 & \gdh a=d_{A_4}c  \\
       \gd \varphi=\eta & \gdh \varphi=[c,\varphi]  \\
       \gd\Psi_4=-\iota_{\veeb}F_0-L_{\veeb}^0a_4 & \gdh\Psi_4=d_{A_4}\phi+\{c,\Psi_4\}  \\
       \gd\eta=-L_{\veeb}^0\varphi & \gdh\eta=[\phi,\varphi]+\{c,\eta\}  \\
       \gd c=-\phi & \gdh c=c^2 \\
       \gd\phi=L_{\veeb}^0c & \gdh\phi=[c,\phi] \\
       \gd\chi=H & \gdh\chi=\{c,\chi\}+(F_H^+)^{\perp} \\
       \gd H=-L_{\veeb}^0\chi & \gdh H=[\phi,\chi]+[c,H]-\gd(F_H^+)^{\perp} \\ \end{array}~~~~
              \label{tbl_4D}.\eea
Here $\chi,H\in\Go^2(X)$ satisfy $P\chi=\chi,\,PH=H$. We record the term $(F_H^+)^{\perp}$ in the transformation $\gdh\chi$
\bea (F_H^+)^{\perp}=\frac12(1+f^2)P\Big(F_4+\varphi d\gb+h^{-1}*_4\big((g_4\veeb)\wedge d_{A_4}\varphi\big)\Big)\nn.\eea
The 2-form $d\gb$ is the curvature of the $U(1)$-bundle $M$ over $X$, but otherwise has no intrinsic meaning in 4D.
So one can think of the zero of the section $(F_H^+)^{\perp}$ as deformations of the 4D exotic instanton
\bea P(F_4+\Go)=0.\nn\eea
One deforms this equation (such as using $P_{\go}$ instead of $P$ or add terms involving scalar fields as above) in the hope of getting a more manageable moduli space of exotic instantons. However, if one wants to study this problem purely in 4D, one loses any insight for such deformation, thus whenever possible we will take the 5D point of view by using \eqref{tbl_5D}.

{The reduction procedure can be applied to a range of examples and arrive at 4D theories on manifolds such as: $\#_k(S^2\times S^2)$, $\BB{C}P^2\#\overline{\BB{C}P^2}$ and $(S^2\times S^2)\#\overline{\BB{C}P^2}$, see sec.\ref{sec_Eotc5mwfU}. In fact we believe we can reach all of the quasi-toric 4-folds
\bea p(S^2\times S^2)\#q\BB{C}P^2\#r\overline{\BB{C}P^2},~~p,q,r\in\BB{Z}_{\geq0}.\nn\eea}

\subsection{Action and localisation of 5D $N=1$ SYM}
We hope to have demonstrated in the previous two sections that it is much neater to encode both the equivariant DW theory and the exotic DW theory in terms of 5D $N=1$ theory. In this section, we will continue along this line and discuss the action and localisation locus in 5D language.

To write the action we need to Wick rotate $\gs\to i\gs$, $H\to iH$.
\bea S&=&\frac{1}{2}(\gd+\gdh)\Tr\int_M-\bra \chi,iH-F_H^+\ket-\bra\Psi,\iota_{\reeb}F+iD\gs\ket\nn\\
&=&\frac{1}{2}\Tr\int_M |H|^2+|F_H^+|^2+\bra \chi,[\phi,\chi]-2(D\Psi)_H^+\ket+|\iota_{\reeb}F|^2+|D\gs|^2+\bra\Psi,i[\gs,\Psi]+[\iota_{\reeb},D]\Psi\ket
\nn\\
\label{5D_action_I}\eea
Here one still sees the vestige of $\bra \vgt,\varphi\ket$ term in the toy model \eqref{thom_P} of sec.\ref{sec_Lttpb}.
For example we have explained the term \eqref{used_III}, then under the current replacement $\varphi\to\Gd_A\varphi\to d_A^{\dag}\iota_{\reeb}F$, it  transmogrified into
$(\gd+\gdh)\bra \Psi,\iota_{\reeb}F\ket$.

{The action above is written as the equivariant Thom/Euler class of sec.\ref{sec_RoMQroTE} in particular sec.\ref{sec_Ae_II}, the discussion there shows that the path integral localises on
\bea F_H^+=0,~~~\iota_{\reeb}F-D\gs=0.\label{loc_loc}\eea
The first equation is the 5D instanton, the second condition is the analogue of $Y_{\xi}=0$ in sec.\ref{sec_Ae_II}. In fact  $-\iota_{\reeb}F+D\gs$ is a vector field on ${\cal A}$ deforming the gauge connection, its fixed point precisely says that the gauge bundle is equivariant.

So to summarise, the localisation loci of the 5D theory are the equivariant contact instantons. Finally, due to the Wick rotation $\gs\to i\gs$, we get an even more restrictive fixed point $\iota_{\reeb}F=D\gs=0$, i.e. not only is the instanton equivariant, it is reducible.}

Having identified the localisation locus, the gauge fixing is done by adding to $S$ the terms
\bea S_{gf}=(\gd+\gdh)\int_M\bra\bar c,\tilde b+d_A^{\dag}a\ket.\nn\eea
Again we know the Gaussian integral around a configuration $A_0$ satisfying \eqref{loc_loc} will be insensitive to the details of $S_{gf}$ and is given by \eqref{conclusion_earlier}. Sections \ref{sec_Sla} and  \ref{sec_RoMQroTE} say that everything hinges now on the analysis of the complex linearised at $A_0$
\bea
\begin{tikzpicture}
  \node at (-3,0) {$\gd\;\uparrow$};
  \node at (0,1.3) {$\gdh\;\rightarrow$};
   \matrix (m) [matrix of math nodes, row sep=1.6em, column sep=3em]
    {  & {\color{blue}c} & a & \vphantom{F} & {\color{blue}\chi\oplus \bar c}  \\
      &  \phi & \Psi & \vphantom{F} & H\oplus b \\ };
  \path[->]
  (m-2-2) edge (m-1-2)
  (m-2-3) edge (m-1-3)
  (m-2-5) edge (m-1-5);

  \path[->]
   (m-1-2) edge node[above] {\scriptsize{$d_0$}} (m-1-3)
  (m-2-2) edge node[above] {\scriptsize{$-d_0$}} (m-2-3)
  (m-1-3) edge node[above] {\scriptsize{$(d_0{\textrm-})_H^+\oplus d_A^{\dag}\iota_{\reeb}d_A{\textrm{-}}$}} (m-1-5)
  (m-2-3) edge node[above] {\scriptsize{$-(d_0{\textrm -})_H^+\oplus d_A^{\dag}\iota_{\reeb}d_A{\textrm{-}}$}}(m-2-5);
\end{tikzpicture}.
\label{instanton_cplx_5}\eea
In particular the top row gives us the deformation complex
\bea 0\to \Go^0(ad)\stackrel{d_0}{\longrightarrow}\Go^1(ad)\xrightarrow{(d_0{\textrm-})_H^+\oplus d_0^{\dag}\iota_{\reeb}d_0{\textrm{-}}}\Go_H^{2+}\oplus\Go^0(ad)\to 0\label{key_cplx}.\eea
{Note that the kernel of the second map says that the deformation of $A$ must maintain $F_H^+=0$, but not necessarily equivariance. This will be important for the transversal ellipticity later and be responsible for the graded manifold structure of the instanton moduli space.}

\subsection{Transversally Elliptic Complex}
In order for the super-determinant \eqref{conclusion_earlier} to be computable, one needs that
the deformation complex \eqref{key_cplx} be elliptic. Unfortunately it is not.

To see this, it is more convenient to use the folding trick to fold the complex into two level one by placing the first $\Go^0(ad)$ at the last place and replace $d_0$ with its adjoint
\bea 0\to \Go^1(ad)\xrightarrow{(d_0{\textrm-})_H^+\oplus d_0^{\dag}\iota_{\reeb}d_0{\textrm{-}}\oplus d_0^{\dag}{\textrm{-}}}\Go_H^{2+}\oplus\Go^0(ad)\oplus\Go^0(ad)\to 0\label{key_cplx_fold}\eea
The symbol of the map is then (where $\xi_{\reeb}:=\bra \reeb,\xi\ket$)
\bea \gs(\xi):\,a\mapsto (\xi\wedge a)_H^+\oplus \big(-\bra \xi,a\ket \xi_{\reeb}+|\xi|^2\iota_{\reeb}a\big)\oplus(-\bra\xi,a\ket), \label{5D symbol}\eea
where we have abused notation in using $\xi$ to denote both a vector and covector by using the metric tacitly.
Here we see the problem, letting $0\neq \xi\|\reeb$, the symbol map collapses
\bea  a\mapsto 0\oplus 0 \oplus(-\bra\xi,a\ket),\nn\eea
which certainly cannot be an isomorphism.

\smallskip

The solution will be provided by using the transversally elliptic complex.
We have assumed that $\reeb$ is killing (but not necessarily a Reeb vector field), so there are two cases:
\begin{enumerate}
  \item the orbit of $\reeb$ is closed then $\reeb$ is the fundamental vector field of a $U(1)$ isometry
  \item the orbit of $\reeb$ is irregular then the isometry of $M$ is at least $U(1)^2$ and $\reeb$ is the fundamental vector field of a certain linear combination of the two $U(1)$'s.
\end{enumerate}
In either case we denote the isometry group as $K$ and let $T^*_KM$ be the subspace of $T^*M$ transverse to the $K$-action. More concretely $T^*_KM$ consists of pairs
\bea T^*_KM=\{(x,\xi)|x\in M,\,\xi\in T_x^*M,\, \bra\xi,V\ket=0 \textrm{ if $V$ is the fundamental vector field of $K$-action}\}.\nn\eea
As in the earlier discussion the symbol of a differential operator $D:\, E\to F$ induces a bundle map $\gs(D):\,E\to F$ over $T^*M$ (as usual $\pi^*$ is omitted). But as we have a $K$-action, we can restrict the bundle map to $T^*_KM$. The operator $D$ is transversally elliptic iff the bundle map restricted to $T^*_KM$ is an isomorphism away from the zero section.

Back to our symbol map \eqref{5D symbol}. In either of the two cases above, we have the component $\xi_{\reeb}=\bra\xi,\reeb\ket=0$. The symbol becomes
\bea \gs(\xi)\big|_{T^*_KM}:\,a\mapsto (\xi\wedge a)_H^+\oplus (|\xi|^2\iota_{\reeb}a)\oplus(-\bra\xi,a\ket).\nn\eea
Suppose that $a$ is sent to zero under this symbol map, then $(\xi\wedge a_H)^+=0$, and using the same 4D reasoning (but now trasnsverse to the $\reeb$ direction), we have $a_H=\xi\wedge\phi$ for some zero form $\phi$. Combining with $\bra\xi,a\ket=0$, we get $\phi=0$, i.e. $a_H=0$.
While $|\xi|^2\iota_{\reeb}a=0$ says $\iota_{\reeb}a=0$, and so all components of $a$ are zero. So $\gs(D)$ is injective. For the surjectivity, given $\go\oplus\phi_1\oplus \phi_2$, we just choose $a_{\reeb}=\phi_1/|\xi|^2$, as for $a_H$, choose its component along $\xi$ to be $\phi_2/|\xi|^2$ and its component perpendicular to $\xi$ as $\iota_{\xi}\go/|\xi|^2$.

Summarising the discussion above, we have proved
\begin{proposition}
  The following 5D system of equations has a transversally elliptic deformation complex
  \bea & F_H^{2+}=0,\nn\\
  &d_A^{\dag}\iota_{\reeb}F_A=0\label{5D_eqn}\eea
where $H$ denotes transverse w.r.t. $\reeb$.
\end{proposition}
{We see that the second equation does not enforce $\iota_{\reeb}F=0$ which would be the equivariance condition, but something weaker. The equivariant index of this system gives the tangent space to the moduli space of \eqref{5D_eqn} and has the following structure. It is typically of infinite dimension, but at each fixed $K$-representation, the multiplicity is finite. Since our $K$ is a product of $U(1)$, representations are just $\BB{Z}$-gradings. This gives the moduli space a graded manifold structure. In particular, the zero grading part has full $K$-invariance and so corresponds to the moduli space of equivariant 5D instantons. The nonzero grading parts are fibred over the degree zero part, at least round the smooth locus.}

\section{Modified Localisation Locus}\label{sec_Tzmamll}
We explained in sec.\ref{sec_RteDt} that the localisation locus is $P(F_4+\Go)=0$ with $\Go$ there resulting from the 5D equation $F_H^+=0$. The choice of $\Go$ will not affect the symbol and so it has no bearing on the ellipticity. In the DW theory, such an $\Go$ is chosen so that $P(F_4+\Go)$ should vanish 'cleanly' (in the finite dimensional setting this amounts to perturbing a section so it vanishes transversally). In the physics parlance, this means we choose $\Go$ to soak up possible fermionic zero modes.

In this section we discuss the necessity of deforming $F_H^+=0\mapsto F_H^++\gs d\gk=0$ due to fermion zero modes, which will result in even more deformations to $\Go$. 
In the following we assume that the 5D manifold $M$ is Sasaki with scalar curvature $s>-4$ everywhere.
In particular when $M$ is Sasaki-Einstein then $s=20$.

\subsection{The Fermionic Zero Modes}

We first explain what do we mean by zero modes. Recall from sec.\ref{sec_FMn} that
for a pair of fields $q,p$ with $\gd p=Vq$ and $p=\gd q$,
the kinetic term is schematically
\bea \gdh^{\dag}\gdh-V^2\nn\eea
Thus the zero modes of $q$ or $p$ are those in the $\gdh$-cohomology that satisfy $Vq=0$ or $Vp=0$.
Put in another way, the zero modes are precisely those that would contribute $0$ or $\infty$ in the super-determinant \eqref{conclusion_earlier}.

Let us see the concrete example of $\chi$ zero modes.
In the action \eqref{5D_action_I} we chose the $(\gd+\gdh)$-exact term (where we put off Wick rotation for a bit)
\bea (\gd+\gdh)\bra \chi,-H+F_H^+\ket=\bra H+F_H^+,-H+F_H^+\ket-\bra \chi,L^A_{\reeb}\chi-[\gs,\chi]+2(D\Psi)^+_H\ket.\nn\eea
In the fermion kinetic term, a particular mode of $\chi$ is missing: it is the one perpendicular to $(d_A\Psi)^+_H$ and has zero eigenvalue under $L^A_{\reeb}-Ad_\gs$. We shall prove in sec.\ref{sec_Vt} that if such modes existed, they are proportional to $d\gk$.

A possible way out is to modify the $(\gd+\gdh)$-exact term with some non-derivative terms, which will not affect the symbol of the differential operators but will soak up some zero modes. In view of the combination $L_{\reeb}^A-Ad_{\gs}$, we define a new connection $\bar A=A+\gk\gs$. We also make a redefinition of fields
\bea
\bar H=H-\gs d\gk\nn\eea
in \eqref{tbl_5D}. Then we rewrite the $(\gd+\gdh)$-exact term using $\bar H$
\bea (\gd+\gdh)\bra \chi,-\bar H+\bar F_H^+\ket=-\bar H^2+(\bar F_H^+)^2-\bra\chi,-L_{\reeb}^{\bar A}\chi+2(\bar D\Psi)^+_H-2\Psi_{\reeb}d\gk\ket.\nn\eea
This change modifies the localisation locus to
\bea \iota_{\reeb}\bar F=0,~~\bar F_H^+=F_H^++\gs d\gk=0,\label{mod_inst}\eea
The particular mode of $\chi$ of concern earlier now is no longer missing from the action: it gets paired with $\Psi_{\reeb}d\gk$.

In practice, we also performed a Wick rotation in \eqref{5D_action_I} causing the path integral to localise on reducible connections and $\gs$ becomes a covariant constant. Deformations of the type \eqref{mod_inst} often appear in the treatment of Donaldson theory in 4D.

One has to do similar analysis to $\bar c$ and $c$, we do so systematically by recognising that the $\chi,\bar c,c$ zero modes are the cohomology at the last step of the folded complex \eqref{key_cplx_fold}, and invariant under $\reeb$.
The background gauge field is $\bar A$ satisfying \eqref{mod_inst}, and we omit the bar next.

We first obtain the adjoint to the map in the complex \eqref{key_cplx_fold}: from the inner product
\bea \bra P_+Da,\chi\ket+\bra D^{\dag}\iota_{\reeb}Da,\bar c\ket+\bra D^{\dag}a,c\ket,\nn\eea
we get the adjoint as
\bea (\chi,\bar c,c)\mapsto D^{\dag}\chi-D^{\dag}(\gk D\bar c)+Dc.\nn\eea
The zero modes of $(\chi,\bar c,c)$ are in the kernel of this map and they must be invariant under $\reeb$.
Thanks to \eqref{mod_inst}, we have $D^{\dag}(\chi-\gk D\bar c)\perp Dc$ and so both terms must be zero. The equation $Dc=0$ has no solution unless at a reducible connection, we focus on
\bea 0=D^{\dag}(\chi-\gk D\bar c)=D^{\dag}\chi+\gk \Gd \bar c-JD\bar c\label{zero_modes}\eea
where one needs to use both \eqref{mod_inst} and $L_{\reeb}^A$-invariance. We analyse this equation next.

\subsection{Vanishing theorems}\label{sec_Vt}
Keeping to the setting of a Sasaki-manifold we review some vanishing theorems. In particular we are interested in a $B\in \Go^{2+}_H(M,ad)$ that has $D^{\dag}B=-\gk \Gd h+JDh$ for some adjoint scalar. Further both $B,h$ are $L_{\reeb}^A$-invariant at a background connection $A$ satisfying \eqref{mod_inst}.

To derive an effective vanishing theorem one needs to separate the components of $\Go^{2+}_H$ that are along $d\gk$ and those $\Go_H^{0,2}\oplus \Go_H^{2,0}$. We denote the former as $B_3$ and latter as $\tilde B$.
Under the current geometric setting $D^{\dag}\tilde B$ is horizontal and orthogonal to $JDh$ for any $h$, and so $D^{\dag}\tilde B=0$ by itself. But $D^{\dag}B_3$ does not enjoy such properties. So it is natural to discuss $B_3,\tilde B$ separately, nonetheless we have a slightly broader result
\begin{proposition}\label{prop_broader}
Assume the connection has $\iota_{\reeb}F=P_+F=0$ and $L_{\reeb}^AB=0$
\bea ||(D^{\dag}B)^H||^2=||(D^{\dag}B_3)^H||^2+\frac14\Tr\int ((\nabla\tilde B)^H\cdotp(\nabla \tilde B)^H+(s/2+2)\tilde B\cdotp\tilde B)~\textrm{vol}_g.\nn\eea
\end{proposition}
Apply this to the case $B=\tilde B$, then since the rhs contains positive definitive terms if $s+4>0$, we get a
\begin{corollary}\label{thm_vanish}
  Assuming that the Sasaki manifold has $s+4>0$ everywhere, then $B$ has no zero modes in $\Go_H^{2,0}\oplus\Go_H^{0,2}$.
\end{corollary}
The proof of prop.\ref{prop_broader} needs an important
\begin{lemma}(Weitzenb\"ock formula)
\bea &&\bra D^{\dag}C,D^{\dag}B\ket=-\frac12\bra L^A_{\reeb}C,L^A_{\reeb}B\ket+\frac12\bra J\cdotp C,J\cdotp B\ket+\bra F,C\times B\ket\nn\\
&&\hspace{2.15cm}+\Tr\int (\frac{1}{4}D C\cdotp D B-\frac14CXB){\rm vol}_g,~~~~{\rm where}~B,C\in\Go^{2+}_H(M,ad).\label{Weizenbock}\eea
\end{lemma}
The notations used here are
\begin{enumerate}
\item $J=-1/2 d\gk$ serving as the K\"ahler form transverse to $\reeb$,
\item $J\cdotp B=J^{pq}B_{pq}$, $D C\cdotp D B=(D_i C_{jk})(D^iB^{jk})$ with $D$ containing both the Levi-Civita and gauge connections, \item $X$ is a rank 4 tensor
\bea
X_{pqrs}=R_{pqrs}-\frac12(Ric\bar\wedge g)_{pqrs}=W_{pqrs}-\frac1{6}(Ric\bar\wedge g)_{pqrs}-\frac{s}{24}(g\bar\wedge g)_{pqrs}~,\label{B_bilinear}\eea
$R$: Riemann tensor, $Ric$: Ricci tensor, $s$: Ricci scalar, $W$: Weyl-tensor.
\item $\bar\wedge$ and $\times$ are the operations
 \bea & (B\bar\wedge C)_{ijkl}=B_{ik}C_{jl}-B_{jk}C_{il}-B_{il}C_{jk}+B_{jl}C_{ik},\nn\\
 &(B\times C)_{ij}=B_{ik}C_j^{~k}-B_{jk}C_i^{~k}~~~B,C\in\Go^{2+}(M,ad).\nn\eea
\end{enumerate}
We use exactly the same notation as in \cite{Qiu:2014cha}, we also collect some facts proved there
\begin{enumerate}
  \item $(D B_3)\cdotp D\tilde B=(D B_3)_{ijk}(D\tilde B)^{ijk}=0$
  \item $B_3^{ij}X_{ijkl}\tilde B^{kl}=0$
  \item $B_3^{ij}X_{ijkl}B_3^{kl}=-6(B_3)_{ij}(B_3)^{ij}$
  \item $(\tilde B)^{ij}X_{ijkl}(\tilde B)^{kl}=(4-s/2)(\tilde B)_{ij}(\tilde B)^{ij}$
\end{enumerate}
Next we take the reader through the proof of proposition \ref{prop_broader} quickly, missing details are in \cite{Qiu:2014cha}.
We separate $\tilde B$ from $B_3$ in $||(D^{\dag}B)^H||^2$,
\bea ||(D^{\dag}B)^H||^2=||(D^{\dag}B_3)^H||^2+||D^{\dag}\tilde B||^2+2\bra D^{\dag}\tilde B,(D^{\dag}B_3)^H\ket,\nn\eea
where we note that $D^{\dag}\tilde B$ is automatically horizontal. Apply the Weitzenb\"ock formula to the last two terms
\bea ||D^{\dag}\tilde B||^2+2\bra D^{\dag}\tilde B,D^{\dag}B_3\ket=\frac14\Tr\int (\nabla\tilde B\cdotp\nabla \tilde B-\tilde BX\tilde B)\textrm{vol}_g+(F,\tilde B\times \tilde B)+2(F,\tilde B\times B_3)-\frac12||L^A_{\reeb}\tilde B||^2.\nn\eea
As $B_3\times B_3=0$ we can write
\bea ||D^{\dag}\tilde B||^2+2(D^{\dag}\tilde B,D^{\dag}B_3)=\frac14\Tr\int (\nabla\tilde B\cdotp\nabla \tilde B-\tilde BX\tilde B)~\textrm{vol}_g+(F,B\times B)-\frac12||L^A_{\reeb}\tilde B||^2.\nn\eea
We further decompose $\nabla\tilde B\cdotp\nabla \tilde B$
\bea (\nabla B)_{ijk}=(\nabla B)^H_{ijk}+\reeb_i(\nabla_{\reeb}B)_{jk}+\reeb_{[j}\reeb^l\nabla_iB_{|l|k]}
=(\nabla B)^H_{ijk}+\reeb_i(\nabla_{\reeb}B)_{jk}-\reeb_{[j}J_{~i}^lB_{|l|k]}\nn\eea
Thus
\bea (\nabla B)\cdotp(\nabla B)=(\nabla B)^H\cdotp(\nabla B)^H+(\nabla_R B)\cdotp(\nabla_R B)+2B_{lk}B^{lk}.\nn\eea
On the other hand for any $B$
\bea L^A_{\reeb}B_{jk}=\nabla_{\reeb}B_{jk}+B_{l[k}\nabla_{j]}\reeb^l=\nabla_{\reeb}B_{jk}+B_{l[k}J^l_{~j]}\nn\eea
and so
\bea (\nabla B)\cdotp(\nabla B)=(\nabla B)^H\cdotp(\nabla B)^H+(L^A_{\reeb} B-J\times B)\cdotp (L^A_{\reeb} B-J\times B)+2B_{lk}B^{lk}.\nn\eea
We get finally
\bea &&||D^{\dag}\tilde B||^2+2(D^{\dag}\tilde B,D^{\dag}B_3)\nn\\
&&\hspace{1cm}=\frac14\Tr\int ((\nabla\tilde B)^H\cdotp(\nabla \tilde B)^H-\tilde BX\tilde B+(L^A_{\reeb}\tilde B-J\times \tilde B)\cdotp (L^A_{\reeb} \tilde B-J\times \tilde B)+2\tilde B\cdotp\tilde B)~\textrm{vol}_g\nn\\
&&\hspace{1cm}+\bra F,B\times B\ket-\frac12||L^A_{\reeb}\tilde B||^2.\nn\eea
Now we set $F$ such that $P_+F=0$, and $L_{\reeb}^AB=0$
\bea ||(D^{\dag}B)^H||^2=||(D^{\dag}B_3)^H||^2+\frac14\Tr\int ((\nabla\tilde B)^H\cdotp(\nabla \tilde B)^H+(s/2-4)\tilde B\cdotp\tilde B+4\tilde B\cdotp\tilde B+2\tilde B\cdotp\tilde B)~\textrm{vol}_g\nn\eea
we finish the proof.

\smallskip

Now consider $B_3$, so we write $B_3=bJ$ for some adjoint scalar $b$. It satisfies $D^{\dag}(bJ)=-\gk \Gd h+JDh$, or
\bea -4b\gk+JDb=-\gk \Gd h+JDh.\nn\eea
As both $b,h$ are $L_{\reeb}^A$-invariant one has $b=h$ and $\Gd h=4h$.
 It is not clear that $\Gd h=4h$ will never have a solution. But in the case the gauge bundle is trivial and $M=S^5$ it is well-known that the eigenvalue for the Laplacian on $S^n$ is $\ell(\ell+n-1)$ and so for $n=5$ the available eigen-values are $0,5,12$ etc, so 4 is not viable. Hence we conjecture that if $M$ is Sasaki-Einstein, then with the simple shift in \eqref{mod_inst}, all $\chi,\bar c$ zero modes are removed. As for $c$, it has a constant zero mode when the connection is reducible and zero otherwise.

\section{Reducible connection and the flux}
\subsection{A moduli problem}
After Wick rotation our localisation locus includes the equation
\bea
 \iota_{\reeb}F=0=D\gs~.\label{reducible}\eea
This puts us on the reducible connection i.e. Coulomb branch.
In the discussion next, we assume that we are at a generic point of the Coulomb branch, that is $\gs$ is a generic element in the Cartan and so the structure group is reduced into the maximal torus. Thus it makes sense to compute the first Chern-class (we assume that the gauge group is simple, so $c_1=0$ unless the structure group is reduced). More invariantly, we can construct from the covariantly constant scalar $\gs$ the combination
\bea
A^{\gs}=\frac{1}{2|\gs|}\bra A,\gs\ket,~~~F^{\gs}=\frac{1}{2|\gs|}\bra F,\gs\ket~.\label{sigma_version}\eea
where $\bra,\ket=-2\Tr$ is the Killing form of the Lie algebra.
These are regarded as $U(1)$ connection and curvatures, and the factor $2$ accounts for the difference in the normalisation of generators of $SU(2)$ and $U(1)$,

Since we merely use the 5D Sasaki geometry to encode the 4D (exotic-)DW, we are only interested in those bundles that can be pushed down along the free $U(1)$ to 4D. Therefore the connection is of type
\bea A_5=\pi^*A_4+\gb\varphi\label{special_type}\eea
see the list of notation on page \pageref{tab_notations}.
As a consequence one has
\bea \iota_{\xeeb}F=-d_{A_4}\varphi~.\label{equiv_54}\eea
As a slight digression if one is given a bundle over $M$ with a connection s.t.
\bea & \iota_{\xeeb}F=-d_A\varphi\nn\\
& P\exp\int_{\rm fibre}(A+\gb\varphi)\label{holonomy_condition}=id\eea
then the bundle can be pushed down to $M/U(1)=X$, see \cite{Festuccia:2016gul}.

Next we shall consider the infinitesimal deformation problem of \eqref{reducible}, \eqref{mod_inst}, \eqref{equiv_54} i.e. deformation of
\bea \iota_{\reeb}F=0=D\gs,~~F_H^++\gs d\gk=0,~~\iota_{\xeeb}F=-d_{A_4}\varphi.\label{abe_eqns}\eea
We assume that the background fields $A_0,\varphi_0,\gs$ are in the Cartan and they solve \eqref{abe_eqns}, we write the deformation as (dispensing with the subscript 4 on $A_4$)
\bea
A=A_0+a+\hat a~,~~\varphi=\varphi_0+f+\hat f~,~~\gs=\gs_0+s~,\nn\eea
where unhatted $a,f,s$ are in the Cartan and the hatted one are orthogonal to the Cartan.
We did not write $\hat s$ because the so called Cartan is defined by $\gs$ (assuming that it is generic). Besides we may also assume that $s$ is not a constant.

Expanding \eqref{abe_eqns} to first order in deformations
\bea &\iota_{\reeb}(da+d_0\hat a)=0=ds-[\hat a,\gs_0]~,\nn\\
&(da+d_0\hat a)_H^++sd\gk=0~,\\
&\iota_{\xeeb}(da+d_0\hat a)=-df-d_0\hat f+[\hat a,\varphi_0]~,\nn\eea
where $d$ is the ordinary derivative and $d_0$ is the covariant derivative with connection $A_0$.
For the hatted fields, if $\gs_0$ is generic
\bea
\hat a=0~,~~d_0\hat f=0~.\nn\eea
For those fields in the Cartan
\bea
 \iota_{\reeb}da=0~,~~ds=0~,~~(da)_H^++sd\gk=0~,~~\iota_{\xeeb}da=-df~.\label{used_IV}\eea
First $ds=0$ says $s$ is constant, which is just an overall shift of the Coulomb branch parameter $\gs_0$. So we can assume $s=0$.
Then one gets that $da$ is horizontal anti-self-dual, and it admits equivariant extension $da\to da-f$ so that
\bea (d-\iota_{\xeeb})(da-f)=0.\nn\eea
The next proposition shows that the deformation is completely fixed by the value of $f$ at the isolated closed Reeb orbits, which we denote as $S_i$.
\begin{proposition}\label{prop_local}
  In our current geometric setting and assuming \eqref{abe_eqns}, the value of $\varphi$ at the loci $S_i$ (the isolated closed Reeb orbits) determine the $A,\varphi,\gs$ completely.
\end{proposition}
Note \eqref{equiv_54} and \eqref{reducible} says $L_{\reeb}^A\varphi=0$, so it makes sense to talk about the value of $\varphi$ along a closed Reeb orbit.
\begin{proof}
  First for $\hat f$, that $d_0\hat f=0$ says $d|\hat f|^2=0$, if $\hat f=0$ at $S_i$, it is zero altogether.

  As $da$ is basic w.r.t $\reeb$, we compute its norm
  \be
   |da|^2=\int\limits_M da \wedge *_5 da=-\int\limits_M\gk\wedge da\wedge da~,\nn
  \ee
  which can be computed using equivariant localisation. The formula in sec.\ref{sec_Lffbf} says
  \bea -\int\limits_M\gk\wedge da\wedge da=\sum_i\ell_i\frac{(2\pi)^2f|_i}{e_i}\nn\eea
  with sum over $S_i$.

  Assume now the deformation $f$ is such that $f|_i=0$ at all $S_i$, then the norm of $da$ is zero. The 5-manifolds in question have trivial $\pi_1$, showing that $a$ is exact and so is a gauge transformation.

  Also $da=0$ implies $df=0$, but $f|_i=0$ and $f=0$ identically.
\end{proof}

\subsection{Determine the flux and the value of $\varphi$}\label{sec_Dtfatov}
We keep to the geometrical setting of appendix \ref{sec_Mgs}.  The forms on $X$ can be regarded as forms on $M$ basic w.r.t the vector field $\xeeb$, while integrals over $X$ can be also be done in 5D as
\be
 \int\limits_X\go=\frac{1}{2\pi}\int\limits_M\gb\wedge  \go~.\nn
 \ee
The Stokes theorem etc are still valid  thanks to the fact that $\iota_{\xeeb}d\gb=0$.

The same remark applies to integration over toric invariant cycles. The 2-cycles $C_i$ of $X$ can be chosen to descend from the toric invariant 3-cycles of $M$, which in turn correspond to the faces of the cone $C_{\mu}(M)$. So for a basic 2-form $\go$
\be
 \int\limits_{C_i} \go=\frac{1}{2\pi}\int\limits_M \gb\wedge\go\wedge u_i~,\nn
\ee
with $u_i$ being the Poincar\'e duals of the 3-cycle of $M$. One can of course directly integrate over the 3-cycle and ignore $u_i$, but since we shall do the integral using localisation, either method is equally simple.

We now apply these formulae to the curvature defined in \eqref{sigma_version}, so it makes sense to drop the superscript sigma altogether.
We take the combination
\be
 F-(d\gb)\varphi+\gb \wedge d\varphi=F_4~,~~~{\rm c.f.}~ \eqref{special_type}\nn
\ee
which is basic w.r.t $\xeeb$  and has equivariant extension
\be
 F\to F+\bra \gb,\reeb\ket\varphi~.\nn
\ee
The evaluation of $\bra \gb,\reeb\ket$ at $S_i$ is
\be
 \bra \gb,\reeb\ket\big|_i=\frac{[\vec v_{i-1},\vec v_i,\vec\reeb]}{[\vec v_{i-1},\vec v_i,\vec\xeeb]}=[\vec v_{i-1},\vec v_i,\vec\reeb][\vec v_{i-1},\vec v_i,\vec\xeeb]~.\nn
\ee
Here the vectors $\vec v_i \in \mathbb{Z}^3$ are the inward normals of the faces of $C_\mu(M)$, and $\vec \reeb$ is the 3d vector of the torus weights of $\reeb$, and the notation $[\vec a,\vec b,\vec c]$ denotes the determinant of the $3\times 3$ matrix formed by the vectors.

Since $\varphi$ and $\bra\gb,\reeb\ket$ always appear together, we call $\hat\varphi=\bra \gb,\reeb\ket\varphi$.
We want then to determine the values of $\hat\varphi$ at the torus fixed points, using equivariance.
\be
 \hat\varphi_i=\bra\gb,\reeb\ket\varphi\big|_i=\varphi_i[\vec v_{i-1},\vec v_i,\vec\reeb][\vec v_{i-1},\vec v_i,\vec\xeeb]~.\label{varphi_hat}
 \ee
Before computing $\varphi_i$ let us quickly recall some facts.

In the usual setting, the topological class of a line bundle is determined by its $c_1$. One can expand $c_1$ in terms of a basis of $H^2(X,\BB{Z})$. For the toric manifold, a (redundant) basis of $H^2(X,\BB{Z})$ is the set of faces $[i]$, each of which gives a toric invariant divisor. Note we use the same notation $[i]$ to mean the 2-cycle represented by the $i^{th}$-face and its Poincare dual.

The 2-cycles $[i]$ obey two relations
\bea{\rm Rel}_1=\sum_i[\vec v_2,\xeeb,\vec v_i][i]=0~,~~~{\rm Rel}_2=\sum_i[\xeeb,\vec v_1,\vec v_i][i]=0~.\label{relations_toric_div}\eea
Here the choice of $\vec v_1,\vec v_2$ is unimportant.
But if one works in the domain of equivariant line bundles, then it is $H^2_{eq}(X,\BB{Z})$ that one should be concerned with. As an abelian group $H^2_{eq}(X,\BB{Z})$ is \emph{freely} generated by all of $[i]_{eq}$. Considered as a ring, they satisfy only one condition: if two faces $i,j$ do not intersect, then $[i]_{eq}[j]_{eq}=0$ \cite{Fulton:1436535}.
\begin{example}
  We illustrate this with line bundles over $\BB{C}P^1$. First we know $H^2(\BB{C}P^1,\BB{Z})$ is generated by the K\"ahler form, so the two classes $[z=0]$, $[z=\infty]$ represent the same class in $H^2$.

  Now equivariantly $[0]_{eq}$ and $[\infty]_{eq}$ are independent while $[0]_{eq}[\infty]_{eq}=0$ is the sole relation. In fact, with the Cartan model, we have representatives
  \bea &&2\pi[0]_{eq}=\go+\mu~,~~~\go=\frac{idz\wedge d\bar z}{(1+|z|^2)^2}~,~~\mu=\frac{1}{1+|z|^2}~,\nn\\
  &&2\pi\left[\infty\right]_{eq}=\go+(\mu-1)~.\nn\eea
  Due to the difference in the 0-form component, $[0]_{eq}$, $[\infty]_{eq}$ are independent. But the product is exact
  \bea [0]_{eq}\cdotp[\infty]_{eq}=(2\mu-1)\go+\mu(\mu-1)=d_{eq}\frac{i(zd\bar z-\bar zdz)}{2(1+|z|^2)^2}~.\nn\eea

  Topologically line bundles over $\BB{C}P^1$ are ${\cal O}(-n)$ with $n$ the first Chern number.
  We can draw the toric diagram of the line bundles fig.\ref{fig_S2_eq_LB}, with the segment representing $\BB{C}P^1$.
  Both pictures represent ${\cal O}(-1)$, but the difference is how the $U(1)$ of $\BB{C}P^1$ lifts to the fibre. The action of $U(1)$ acts trivially at the fibre over $z=0$ but with weight 1 over $z=\infty$, the second picture is opposite.
\begin{figure}[h]
\begin{center}
\begin{tikzpicture}

\draw [-,blue] (0,1) -- (0,0) node[left] {\scriptsize{1}} -- (1,0) node[right] {\scriptsize{2}} -- (2,1);
\draw [->,blue] (0,0.5) node[left] {\scriptsize{$\vec v_1$}} -- (.4,0.5);
\draw [->,blue] (0.5,0) node[below] {\scriptsize{$\vec v_2$}} -- (0.5,0.4);
\draw [->,blue] (1.5,0.5) node[right] {\scriptsize{$\vec v_3=[-1;1]$}} -- (1.2,0.8);
\end{tikzpicture}
\begin{tikzpicture}
\draw [-,blue] (0,1) -- (1,0) node[left] {\scriptsize{1}} -- (2,0) node[right] {\scriptsize{2}} -- (2,1);
\draw [->,blue] (0.5,0.5) node[left] {\scriptsize{$\vec v_1$}} -- (.8,0.8);

\node at (0,-.35) {${~}$};
\end{tikzpicture}
\caption{Equivariant line bundles over $\BB{P}^1$. The corners $1,2$ correspond to $z=0,\infty$}\label{fig_S2_eq_LB}
\end{center}
\end{figure}
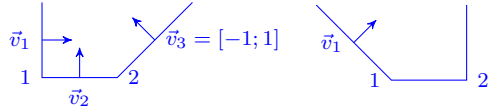
The two are therefore inequivalent as equivariant line bundles. In fact, one can read off from their equivariant curvature how does the $U(1)$ act on the fibres.
\bea \frac{1}{2\pi}F_{eq,1}=-[\infty]_{eq}~,~~~\frac{1}{2\pi}F_{eq,2}=-[0]_{eq}~,\nn\eea
where the integral of $F/(2\pi)$ is $-1$ for both cases.
The 0-form component of $F_{eq}$ gives minus the weight of $U(1)$ action on the fibre.
\end{example}

Here we are working precisely within equivariant bundles as motivated by the bound \eqref{4D_bound_eq}, so when we sum over equivariant line bundles, we must sum over all linear combinations of toric divisors, \emph{without} taking into account the relations \eqref{relations_toric_div}.
Just as in the above example, given the expansion of $F_{eq}/(2\pi)$ as $\sum_ik_i[i]_{eq}$, the equivaraint curvature can be written concretely as
\bea F_{eq}/(2\pi)=\sum_i k_iu_{i,eq}\label{sum_toric_div}\eea
with $u_{i,eq}$ being the equivariant Poincare dual to face $[i]$. The above mentioned $U(1)$ weights acting on the fibre are read off from
the 0-form $\hat\varphi$ component of $F_{eq}$. Concretely the bottom component of $u_i$ is given in \eqref{fp_val_X}, thus
\bea \hat\varphi_i=k_is_i[\vec\xeeb,\vec v_{i-1},\vec \ep]+k_{i-1}s_i[\vec v_i,\vec\xeeb,\vec \ep]~,~~~~~s_i=[\vec v_{i-1},\vec v_i,\vec \xeeb]\label{varphi_hat_i}\eea
and this gives \emph{minus} the weights.

\begin{example}
  For the case of $\BB{C}P^2$ the shifts read
  \bea \hat\varphi_i=k_i\ep_i'+k_{i-1}\ep_i,\nn\eea
  where the weights are computed from \eqref{equiv_para_i}. We list them in the case of different distributions of $+/-$ at the fixed points.
  The weights at the three fixed points are given by
  \bea \begin{array}{c|cc}
                   i & \ep_i & \ep_i' \\
                   \hline
                   2 & \ep & \ep'  \\
                   3 & \ep'-\ep & -\ep \\
                   1 & -\ep' & \ep-\ep'
                   \end{array},\nn\eea
where $\ep,\ep'$ are the weights of $U(1)^2$ acting on $z_1,z_2$ in the patch $[z_1,z_2,1]$.

\end{example}

The sum \eqref{sum_toric_div} is over all integers without taking into account the relations \eqref{relations_toric_div}. However
the relations will make an important appearance.

We can apply \eqref{F_2_cycle} and compute the integral
\bea \bra \frac{F_4}{2\pi},[i]\ket=\frac{\hat\varphi_i-\hat\varphi_{i+1}}{[\vec v_i,\vec\xeeb,\vec \reeb]}\in\BB{Z}\label{shift_I}~.\eea
This means that the values of $\hat\varphi$ at $S_i$ are tied together by the flux of $F_4$ along the 2-cycles.
But because these relative shifts are computed using integration in \eqref{shift_I}, and so is utterly oblivious of the equivariance (notwithstanding that the computation was facilitated by the equivariant localisation). This means that the shifts must obey the relations between the divisors.
In particular the two relations \eqref{relations_toric_div} imply
\bea \sum_i \frac{-s_i[\vec v_a,\vec\xeeb,\vec \reeb]\hat\varphi_i}{[\vec v_{i-1},\vec\xeeb,\vec \reeb][\vec v_i,\vec\xeeb,\vec \reeb]}=0~,~~a=1,2~.\nn\eea
So in fact the two relations give only one constraint
\bea \sum_i \frac{s_i\hat\varphi_i}{[\vec v_{i-1},\vec\xeeb,\vec \reeb][\vec v_i,\vec\xeeb,\vec \reeb]}=0~.\nn\eea
This is a good check for the equation \eqref{varphi_hat_i} we obtained earlier: indeed
\bea 0\stackrel{?}{=}\sum_i \frac{s_i\hat\varphi_i}{[\vec v_{i-1},\vec\xeeb,\vec \reeb][\vec v_i,\vec\xeeb,\vec \reeb]}
=\sum_i\big(-\frac{k_i}{[\vec v_i,\vec\xeeb,\vec \reeb]}+\frac{k_{i-1}}{[\vec v_{i-1},\vec\xeeb,\vec \reeb]}\big)~.\nn\eea

\begin{remark}
Also there is a class which is special
\bea \frac{d\gb}{2\pi}=[\vec v_1,\vec v_2,\vec\xeeb]^{-1}\sum_i[\vec v_1,\vec v_2,\vec v_i][i]~,\nn\eea
$[i]$ being a 2-cycle is considered as a 2-cocycle using Poincar\'e duality. This is the class of the $U(1)$-bundle $M$ over $X$. \end{remark}

To summarise, given any linear combination of toric divisors $\sum_ik_i[i]_{eq}$, all of $\hat\varphi_i$ are fixed using \eqref{varphi_hat_i} and these will be used to shift the Coulomb branch parameter at each of the fixed points.
Let us also compute $c_2$ using equivariance
\bea \int_X\big(\frac{F_{eq}}{2\pi}\big)^2=(\sum_ik_i[i])\cdotp(\sum_jk_j[j])=\sum_ik_i\big(k_{i-1}s_i+k_{i+1}s_{i+1}-k_is_is_{i+1}[\vec v_{i-1},\vec v_{i+1},\vec\xeeb]\big)~.\nn\eea

\section{Partition function}
So far we have not mentioned the matter sector since it requires the spin structure and yet we may be forced to deal with spin${}^c$ structure on either $M$ or $X$. In \cite{Festuccia:2016gul} we considered the restricted setting where $M$ is simply connected Sasaki-Einstein, which has a spin structure. Furthermore we restricted $M$ so that its spin structure can be reduced down the free $U(1)$ to a spin structure on $X$. In the current work we have removed both restrictions and one needs to discuss the spin issue case by case. Therefore we decide to focus on the vector multiplet only.

\subsection{Determinant in the Instanton Sector}\label{sec_DitIS}
In the instanton sector, the computation is still about the superdeterminant of the complex \eqref{key_cplx}, calculated at a given instanton background.
In principle, localisation computation for the transversally elliptic complex can be applied, \emph{provided} we know the instanton background. We do not know enough about the 5D instantons to go down this path. However, from prop.\ref{prop_local}, we learned that the cohomology of \eqref{key_cplx},
equivalently the deformation problem of instantons are controlled entirely by the data at the loci $S_i$. Recall also that the locus $S_i$ descends down the free $U(1)$ to a point $p_i$ in $X$, which is the fixed point of the vector field $\veeb$.
So we conjecture that the superdeterminant of \eqref{key_cplx} can be computed in the neighbourhood of $p_i$, and then simply multiplied together (at any rate, this is what one would have done, had one known the actual background, so this is not really a conjecture). The only 'communication' between different loci $p_i$ is through the flux present in the 2-cycle connecting $p_i$ and $p_{i+1}$, which we have determined in sec.\ref{sec_Dtfatov}.

Close to $S_i$, the local geometry is that of a twisted product $S^1\times_{\veeb}\BB{C}^2$, where the circle is the orbit of $\reeb$ (and also that of $\xeeb$ since they coincide here).
Since we will reduce along $\xeeb$, we simply forget $S^1$ and are left with the equivariant DW theory (see sec.\ref{sec_RtEDWt}) on $\BB{C}^2$, where the equivariance is provided by the vector field $\veeb$, the descendent of $\reeb$. It acts with weights
\bea \ep_i=\frac{[\vec \reeb,\vec v_i,\vec\xeeb]}{[\vec v_{i-1},\vec v_i,\vec\xeeb]},~~~\ep_i'=\frac{[\vec \xeeb,\vec v_{i-1},\vec\reeb]}{[\vec v_{i-1},\vec v_i,\vec\xeeb]}.\label{equiv_para_i}\eea

\begin{figure}[h]
\begin{center}
\begin{tikzpicture}[scale=1]
\draw [-,thick, blue] (0,0) node[left] {\small{asd}} -- (1,0) node[right] {\small{sd}} -- (1,.5) node[right] {\small{sd}} --  (0.6,1) node[right] {\small{asd}} --  (0,1) node[left] {\small{asd}} -- (0,0);
\end{tikzpicture}\caption{An almost toric 4-manifold $(S^2\times S^2)\#\overline{\BB{C}P^2}$. }\label{fig_pentagon_non_Kahler}
\end{center}
\end{figure}
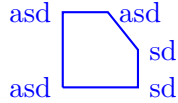

The calculation on $\BB{C}^2$ is now just the equivariant DW theory, with one subtlety. We denote by $s_i=[\vec v_{i-1},\vec v_i,\vec\xeeb]=\pm1$ that measures whether $\xeeb$ is parallel or anti-parallel with $\reeb$ at $S_i$. These are also the signs appearing in front of each of the examples \ref{example_Ypq} through \ref{ex_pentagon}. Since the 4-manifold $X$ inherits the volume form $\iota_{\xeeb}{\rm Vol}_M$, this is the volume form that determines the (anti)-self-duality in 4D. But in 5D, it is the orientation of the tangent space transverse to $\reeb$ that determines the horizontal (anti)-self-duality of the 5D instantons. This transverse orientation is given by $\iota_{\reeb}{\rm Vol}_M$ and is opposite to $\iota_{\xeeb}{\rm Vol}_M$ at the point $p_i$ should $s_i=-1$.
So taking ex.\ref{ex_pentagon} as an illustration. The $s_i$'s read $+--++$, thus if we draw the base of the moment map cone $C_{\mu}(M)$, which is a pentagon, we have fig.\ref{fig_pentagon_non_Kahler}, where the corners are numbered 1,...,5, starting from the lower left corner.
The 4D geometry is that of $(S^2\times S^2)\#\overline{\BB{C}P^2}$.
At a corner with $s_i=-1$, the corresponding DW theory actually computes anti-instantons, that is, the self-dual gauge curvature. But of course on $\BB{C}^2$, the computation of (anti)-instantons are identical, up to a conjugation of the instanton counting parameter.
\begin{remark}
  We just stated that the partition function is assembled from the instanton partition functions on $\BB{C}^2$, one copy for each corner, and that at each corner there is no substantial difference between instanton and anti-instantons. This then raises an outcry, suppose we compute DW theory on $\BB{C}P^2$ and $\overline{\BB{C}P}^2$, i.e. identical space differing only in orientation so that instanton on one is the anti-instanton on the other. Then following our assembling principle, the partition functions are simple conjugates of each other. But this cannot be correct, since $\BB{C}P^2$ is not unbiased toward (anti)-instantons, e.g. the ADHM construction only applies to anti-instantons.

 So a possible way out is that the flux between the 2-cycles that run between the points $p_i$ would make the difference. For example the Poincar\'e dual to the 2-cycles are the K\"ahler class, which is a self-dual 2-form, would certainly treat (anti)-instantons differently.
  Also in an earlier paper \cite{Festuccia:2018rew}, we have noticed that the distributions of signs will affect the possible valid fluxes. But a full treatment perhaps requires a separate paper to clarify satisfactorily.
\end{remark}

Back to the local computation on $\BB{C}^2$. The superdeterminant we need is that of the 4D complex
\bea 0\to \Go^0(ad)\stackrel{d_0}{\longrightarrow}\Go^1(ad)\xrightarrow{(d_0{\textrm-})^+}\Go^{2+}(ad)\to 0\label{key_cplx_C2}\eea
at a \emph{reducible toric equivariant} instanton background. 
We will rewrite the complex in a way that is valid not just for vector bundles but also sheaves.
It is well-known that the complex above is isomorphic to
\bea \Big(0\to \Go^{0,0}\stackrel{\bar\partial}{\longrightarrow}\Go^{0,1}\stackrel{\bar\partial}{\longrightarrow}\Go^{0,2}\to 0\Big)\oplus \Big(0\to \Go^{0,0}\stackrel{\partial}{\longrightarrow}\Go^{1,0}\stackrel{\partial}{\longrightarrow}\Go^{2,0}\to 0\Big).\nn\eea
The key to this isomorphism is the observation that a self-dual 2-form on $\BB{C}^2$ is of type $\Go^{(0,2)}\oplus \Go^{(2,0)}\oplus \BB{C}\go$ where $\go$ is the K\"ahler form. This last one can be treated as a 0-form and one gets the complex above.

It is enough to compute the first complex. The cohomology at each step equals
$H^i_{\bar\partial}(\BB{C},\End(E))$, which can be rewritten as the ext groups
\bea \ext^{0,1,2}(E,E).\nn\eea
The ext groups are defined even when $E$ are not vector bundles but sheafs.

\begin{figure}[h]
\begin{center}
\begin{tikzpicture}[scale=1]
\draw [step=0.5,thin,gray!40] (0,0) grid (2.5,1.5);

\draw [->] (0,0) -- (2.5,0) node [below] {\scriptsize$x$};
\draw [->] (0,0) -- (0,1.5) node [left] {\scriptsize$y$};

\draw [-] (0,1) -- (0.5,1) -- (0.5,0.5) -- (1.5,0.5) -- (1.5,0);
\end{tikzpicture}\caption{A Young diagram $\mu$, the monomial ideal ${\cal I}_{\mu}$ is generated by $\{x^3,xy,y^2\}$.}\label{fig_plan_part}
\end{center}
\end{figure}
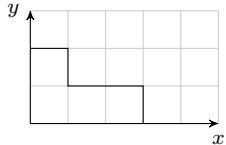

On $\BB{C}^2$, with framing fixed at infinity, the toric invariant instantons on $\BB{C}^2$ are in 1-1 correspondence with direct sums of monomial ideals \cite{Braden:2001fi} ${\cal I}_{\mu}$, here $\mu$ is a Young-diagram. For example, the ideal ${\cal I}_{\mu}=\bra x^3,xy,y^2\ket\subset\BB{C}[x,y]$ consists of monomials lying to the north-east of the contour of $\mu$, see fig.\ref{fig_plan_part}.
The ext groups above becomes $\ext^{0,1,2}_R({\cal I}_{\mu},{\cal I}_{\mu})$, where we have named $R=\BB{C}[x,y]$. The ext group can be computed by picking, say, the Taylor resolution of monomial ideals (see e.g. ex.17.11 of \cite{Eisenbud95}). The computation is standard, we shall give here only the Euler character
\bea \frac{\chi(\ext^{\sbullet}_R({\cal I}_{\mu},{\cal I}_{\gl}))}{\chi(\ext^{\sbullet}_R({\cal I}_{\emptyset},{\cal I}_{\emptyset}))}
=\prod_{\square\in\mu}\Big(2\pi\ep a_{\gl}(\square)-2\pi\ep'(l_{\mu}(\square)+1)\Big)^{-1}
\prod_{\square\in\gl}\Big(2\pi\ep'l_{\gl}(\square)-2\pi \ep(a_{\mu}(\square)+1)\Big)^{-1}\nn,\eea
where $a_{\mu},l_{\mu}$ are the arm and leg length of the YD $\mu$. Further $\ep,\ep'$ are the equivariant parameters keeping track of the weights of the two $U(1)$'s acting on the $x,y$ coordinates of $\BB{C}^2$.

Remembering that we should really be computing the superdeterminant of $-iL_{\veeb}+\phi_0$ over $\ext^{\sbullet}_R({\cal I}_{\mu},{\cal I}_{\gl})$.
The weights of $-iL_{\veeb}$ are already captured in $\chi(\ext^{\sbullet}_R({\cal I}_{\mu},{\cal I}_{\gl}))$ by $\ep,\ep'$ according to \eqref{equiv_para_i}. The determinant of $\phi_0$ over the adjoint can be done by taking a product over the roots of the Lie-algebra, i.e. write $x=\bra\phi_0,\ga\ket$ with $\ga$ being a root. So the superdeterminant we need is
\bea &&\frac{\sdet_{\mu,\gl}(-iL_{\veeb}+\phi_0)}{\sdet_{\emptyset,\emptyset}(-iL_{\veeb}+\phi_0)}
=\prod_{\square\in\mu}\Big(x+2\pi\ep a_{\gl}(\square)-2\pi\ep'(l_{\mu}(\square)+1)\Big)^{-1}\nn\\
&&\hspace{7cm}\times\prod_{\square\in\gl}\Big(x+2\pi\ep'l_{\gl}(\square)-2\pi \ep(a_{\mu}(\square)+1)\Big)^{-1}\nn,\eea
where the two super-determinants are taken over $E^{\sbullet}=\ext^{\sbullet}({\cal I}_{\mu},{\cal I}_{\gl})$, and over $E_0^{\sbullet}=\ext^{\sbullet}({\cal O},{\cal O})$. The latter is independent of ${\cal I}$ and serves merely to remove a constant factor for now. Finally one takes the product over roots and over Young-diagrams. This is the Nekrasov partition function
\bea Z_{\BB{C}^2}^{Nek}(x,\ep,\ep')\nn\eea
originally obtained from performing localisation computation on the ADHM moduli space for instantons on $\BB{C}^2$.

In the above expression, we have divided the Euler-character of $\ext^{\sbullet}_R({\cal I}_{\emptyset},{\cal I}_{\emptyset})$ as a constant background so that in the end the ratio only involves \emph{finite} products, i.e. independent of how one regulates the infinite products. What do we do with $\det\ext^{\sbullet}_R({\cal I}_{\emptyset},{\cal I}_{\emptyset})$? This is the superdeterminant at the zero instanton background, the perturbative sector. Each fixed point of $X$ contribute such an infinite product, the way one regulates the infinite product must be compatible across all the fixed points of $X$, that is, one must have a global regularisation scheme. This is the subject of sec.\ref{sec_Tps}, and the regularisation scheme is detailed in sec.\ref{sec_Teiotec}.

\begin{figure}[h]
\begin{center}
\begin{tikzpicture}[scale=1]
\draw [step=0.2,thin,gray!40] (0,0) grid (2.5,1.5);

\draw [->] (0,0) -- (2.5,0) node [below] {\scriptsize$x$};
\draw [->] (0,0) -- (0,1.5) node [left] {\scriptsize$y$};

\draw [-] (2.5,.2) -- (1,0.2) -- (1,0.6) -- (.6,0.6) -- (.6,0.8) -- (.4,.8) -- (.4,1.5);
\draw [blue,dashed] (1,0.2) -- (.4,0.2) -- (.4,1.5);

\draw [->] (0.1,0.1) -- (0.5,.3);

\node at (.2,1.7) {\scriptsize{$k_{i-1}$}};
\node at (2.7,0.1) {\scriptsize{$k_{i}$}};

\draw (-.7,0.5) node[left] {\scriptsize{shift}} edge[->, bend left =15] (.2,.25);
\end{tikzpicture}\caption{At the toric fixed point $i$, a Young diagram with infinite legs of size $k_{i-1}$ and $k_i$.}\label{fig_plan_part_inf}
\end{center}
\end{figure}
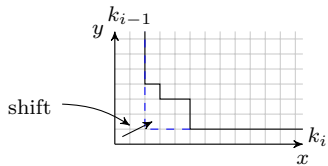

As another remark, one is probably concerned with the absence of Young-diagrams with infinite legs e.g. fig.\ref{fig_plan_part_inf}.
The infinite legs correspond to non-zero $c_1$, or non-zero flux along one of the toric invariant 2-cycles that connect this fixed point to another one.
But to compute the ext groups involving ideal sheafs of such Young-diagrams, it suffices to shift the $U(1)$ weights of each monomial so that effectively one is back at the case already dealt with. This shift of $U(1)$ weights can be absorbed as shifts of the value of $\phi_0$. In fact we have already figured out how does non-zero $c_1$ shift the values of $\phi_0$: this is exactly \eqref{varphi_hat_i} with $k_{i-1}$ and $k_i$ marked in the Young-diagram as the size of the infinite legs.

\subsection{The Perturbative Sector}\label{sec_Tps}
It is pointed out in the last section that to compute the equivariant character of the ext groups, one needs to remove an infinite product coming from $\ext^{\sbullet}_R({\cal I}_{\emptyset},{\cal I}_{\emptyset})$. These are the cohomology groups that control deformations of connections away from the zero connection i.e. the tangent space of the instanton moduli space around the trivial background. This then can be computed using the index calculation that we could not do earlier for non-trivial backgrounds, due to the lack of knowledge of how the tori acts on the bundle. But for trivial bundles, this is obvious, in fact, we can simply borrow the computation done in 5D and impose the $U(1)_{\xeeb}$ invariance in the end. The 5D computation was done using transversally elliptic index theorem in \cite{Qiu:2013pta} and subsequently a method due to Schmude \cite{Schmude:2014lfa} was used in later papers. Here to make the paper self-contained, we shall include a recount of the index theorem method in the appendix, in such a manner that requires no understanding of the equivariant K-theory, which was the foundation of the formulae derived in \cite{Ellip_Ope_Cpct_Grp}.

From our discussion in sec.\ref{sec_Sla}, we need only compute a super-determinant \eqref{conclusion_earlier}.
The complex in question is that of \eqref{instanton_cplx_5}, and we have already discussed the transversal ellipticity of the complex
of \eqref{key_cplx}. The only thing we have not mentioned is the zero modes of the ghosts, but the changes are minimal. First the cohomology of \eqref{instanton_cplx_5} is infinite dimension, but for each given $U(1)^3$ (which is the isometry of $M$) weight, the multiplicity is 1. If one labels the $U(1)^3$ weights by $\vec m\in\BB{Z}^3$, then the generators of the cohomology of \eqref{key_cplx} are in 1-1 correspondence of the lattice points
\bea \BB{Z}^3\cap \Big(C\cup(-C^{\circ})\Big),\label{lattice_cone}\eea
where $C$ denotes the moment map cone $C_{\mu}(M)$ of $M$ (see sec.\ref{sec_Mgs}), $C^{\circ}$ is the interior of $C$ and $-C$ means the negative cone.
For a point $\vec m$ inside of \eqref{lattice_cone}, its eigenvalue under $-iL_{\reeb}$ is $\vec \reeb\cdotp\vec m$, from which we can compute its eigenvalue under $V=-iL_{\reeb}$ (we included an $i$ in $V$ so as to get real eigenvalues).

Furthermore, from the discussion of sec.\ref{sec_NfAaGZM}, the effect of the ghost zero modes is that $\gd$ no longer squares to $-iL_{\reeb}$, but also to an adjoint action of the zero mode of $\phi=\gs+A_{\reeb}$ (the partner of the ghost $c$). At the background $A=0$ this zero mode equals the Coulomb branch parameter $\gs$. To summarise we need to compute the super-determinant of $V=-iL_{\reeb}+Ad_{\gs}$ over the cohomology of \eqref{key_cplx}.

The superdeterminant \eqref{conclusion_earlier} is then just a product of eigenvalues of $V$ over all points in \eqref{lattice_cone},
\bea \sdet_H(V)=\prod_{\vec m\in \BB{Z}^3\cap C}{\det}_{{\rm adj}}(\gs+\vec\reeb\cdotp\vec m)\prod_{\vec m\in \BB{Z}^3\cap -C^{\circ}}{\det}_{{\rm adj}}(\gs+\vec\reeb\cdotp\vec m).\nn\eea
In view of this one defines a generalised triple sine function associated with the cone $C=C_{\mu}(M)$ as
\bea
S_3^C ( x | \vec \go ) =\prod_{\vec m\in C\cap\BB{Z}^3}(\vec \go\cdotp\vec m+x)\prod_{\vec m\in C^{\circ}\cap\BB{Z}^3}(\vec \go\cdotp\vec m-x).\label{gen_multiple_sine}\eea
The zero-instanton partition function thus reads
\bea
Z_{5D}^{pert}
=\int_{\FR{h}}d\gs~e^{-\frac{8\pi^3 r^3}{g^2}\varrho\,\Tr[\gs^2]}\cdotp{\det}_{adj}' ~  S_3^C(i\gs| \vec\reeb),\label{Z_pert_fin}
\eea
where $\varrho = \mathrm{Vol}(M) / \mathrm{Vol}(S^5)$, and $S_3^C$ is the generalized triple sine associated to the cone $C$.

To get to the 4D partition function, instead of summing over all of \eqref{lattice_cone}, one imposes the condition that
\bea \vec\xeeb\cdotp\vec m=0~~~\stackrel{\vec\xeeb=[0;0;1]}{\To}~~~m^3=0.\nn\eea
The partition function now reads
\bea
&&Z_{4D}^{pert}
=\int_{\FR{h}}d\gs~e^{-\frac{8\pi^3 r^3}{g^2}\varrho\,\Tr[\gs^2]}\cdotp{\det}_{adj}' ~  {\Upsilon}^C(i\gs| \vec\reeb),\label{Z_pert_fin_4D}\\
&&\Gu^C( x | \vec \go ) = \prod_{\vec m \in \tilde C\cap\mathbb{Z}^2 } ( \vec \go \cdot \vec m + x )  \prod_{\vec m \in \tilde C^\circ \cap\mathbb{Z}^2 } ( \vec \go \cdot \vec m - x ),\nn\eea
where $\tilde C$ is a 2D cone obtained by restricting the 3D cone $C$ to $m^3=0$.

\subsection{Final assembling}\label{sec_Fa}
Our scheme of final assembling is that for each torus fixed point $p_i$, with $U(1)$ weights denoted $\ep_i,\ep_i'$ (read off from \eqref{equiv_para_i}), we have a factor
\bea Z_i^{pert}(x|\ep_i,\ep_i')Z^{Nek}(x|\ep_i,\ep_i',q^{s_i}),\nn\eea
where if $s_i=-1$ then it is $q^{-1}$ that is the instanton parameter signalling that it is anti-instanton at $p_i$.
The product of Young-diagrams in $Z^{Nek}$ does not permit infinite legs because these, as we discussed, capture fluxes or non-trivial $c_1$ and will be taken into account when we include shifts to $x$ during the assembling.

Now let $a=1,\cdots,\rk G$ label the basis of the Cartan subalgebra of $\FR{g}$, and the set of integers $k_i^a$ fixes the $c_1$ class. At each $p_i$ the variable $x$ above will be replaced with $\gs+\hat\varphi$, which is fixed in \eqref{varphi_hat_i}
\be
 x\mapsto \gs^a+k_i^a\ep_i'+k^a_{i-1}\ep_i ~,\nn
\ee
Here the constant $\gs^a$ is the Coulomb branch parameter.
The total partition function is a product
\be
 Z=\prod_{\{\vec k_i\}}q^{-N(\{\vec k_i\})}Z^{pert}(\vec\gs+\vec k_i\ep_i'+\vec k_{i-1}\ep_i|\ep_i,\ep_i')Z^{Nek}(\vec\gs+\vec k_i\ep_i'+\vec k_{i-1}\ep_i|\ep_i,\ep_i',q^{s_i})\nn
\ee
where $\vec k_i$ means $[k_i^1,\cdots,k_i^{\rk G}]$ and $q^{-N(\{\vec k_i\})}$ comes from the contribution of fluxes to the instanton number (in addition to the contribution of the ideal sheaves at each fixed point). The number $N(\{\vec k_i\})$ has the expression
\be
 N(\{\vec k_i\})=\sum_{a=1}^{\rk G}\sum_{i=1}^{\tt n}k_i\big(k_{i-1}s_i+k_{i+1}s_{i+1}+k_i[i]\cdotp[i]\big).\nn
 \ee
For example in the case of $\BB{C}P^2$ with all $s_i=1$ at each fixed point $p_i$, we have
\be
 N(\{\vec k_i\})=\sum_{a=1}^{\rk G}(k^a_1+k^a_2+k^a_3)^2.\nn
\ee
But if $s_2=-1$, i.e. sd at fixed point $p_2$
\be
 N(\{\vec k_i\})=-\sum_{a=1}^{\rk G}(k^a_1+k^a_2-k^a_3)^2,\nn
\ee
while if $s_2=1,\,s_1=s_3=-1$, i.e. sd at fixed point $p_{1,3}$
\be
 N(\{\vec k_i\})=\sum_{a=1}^{\rk G}(-k^a_1-k^a_2+k^a_3)^2.\nn
\ee
We remind the reader that the labeling is that the face $i$ has fixed point $p_i$ with sign $s_i$ at its starting point while $p_{i+1}$ with sign $s_{i+1}$ at its end point. For $\BB{C}P^2$ see figure
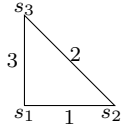
\bea\begin{tikzpicture}[scale=.8]
\draw [-] (0,0) -- (1.5,0) -- (0,1.5) -- (0,0);
\node at (0,-.15) {\scriptsize{$s_1$}};
\node at (1.45,-.15) {\scriptsize{$s_2$}};
\node at (0,1.6) {\scriptsize{$s_3$}};
\node at (0.75,-.2) {\scriptsize{$1$}};
\node at (0.85,0.85) {\scriptsize{$2$}};
\node at (-0.2,0.75) {\scriptsize{$3$}};
\end{tikzpicture}\nn\eea

\appendix
\section{Main geometric setting}\label{sec_Mgs}
In the introduction, we motivated the new instantons on $X$ by considering 5D SYM on the total space $M$ of a $U(1)$ principal bundle over $X$.
To look for good examples, we find it more fruitful to reverse the construction and look for some 5D manifolds with a free $U(1)$ action and we reduce $M$ to get $X$. For the sake of the index calculation and control of zero modes, we choose to search within the 5D toric contact manifolds with a free $U(1)$. We denote the fundamental vector field of this $U(1)$ as $\xeeb$ and $X$ as the quotient $X=M/U(1)$.

Letting $g_5$ be the metric of $M$, we define the metric $g_4$ of $X$ as follows.
Two vector fields $U,V$ of $M$ descend to vector fields $u,v$ of $X$. The inner-product on $X$ is defined as
\bea \bra u,v\ket_4=\bra U^{\perp},V^{\perp}\ket_5,\nn\eea
with $U^{\perp}=U-\bra U,\xeeb\ket_5 \gb$.
The induced 4D metric leads to a volume form that is related to $\opn{Vol}_5$ as
the volume form of $M$ is taken as
\bea {\rm Vol}_4=h^{-1}\iota_{\xeeb}{\rm Vol}_5,~~~h:=|\xeeb|_5.\label{miss_match_vol}\eea
The factor of $h$ will be important when one does a fibre-wise integration.

\subsection{Toric Sasaki 5-manifolds}\label{sec_Tc5m}
The geometry of $M$ is entirely encoded by a moment map cone $C_{\mu}(M)\subset\BB{R}^3$.
Let $\vec v_i\in\BB{Z}^3,~i=1,\cdots\tt n$ be the (primitive) inward pointing normals of the $\tt n$ faces of $C_{\mu}$. Pick $\{e_a\}$ as the fundamental vector field of $U(1)^3$ and we may express the Reeb as a linear combination $\reeb=\sum_{a=1}^3\reeb^ae_a$. We will not distinguish the Reeb and the vector in $\BB{R}^3$ that represents it. Also we need to assume that $\vec\reeb$ is within the dual cone $C_{\mu}^{\vee}$, i.e.
\bea
	\vec\reeb=\sum_{i=1}^{\tt m}\gl_i\vec v_i~,~~~\gl_i>0~.\label{dual_cone}
\eea
With this assumption, the plane (where $y^a$ are the coordinate of $\BB{R}^3$)
\bea
	\big\{\vec y\in\BB{R}^3|\vec \reeb\cdotp \vec y=\frac12\big\} \label{boundary}
\eea
intersects $C_{\mu}$ at a convex polygon $\Gd_{\mu}$ if $C_{\mu}$ is convex.
Then the geometry of $M$ is that of a $U(1)^3$ fibration over $\Gd_{\mu}$, with a certain $U(1)$ becoming degenerate at each faces of $\Gd_{\mu}$.
Indeed if the normal associated with face $i$ is $\vec v_i$, then the $U(1)$ given by $\sum_{a=1}^3v^a_ie_a$ degenerates.

It follows that at the intersection of two faces, only one $U(1)$ remains non-degenerate and its orbit is a closed Reeb orbit.
These are the only loci for closed Reeb orbits for generic $\vec\reeb$.

The cone $C_{\mu}$ is convex, and following \cite{2001math......7201L} it is also \emph{good} in the sense that
\be
 \exists \vec n_i\in\BB{Z}^3,~~{\rm such~ that}~[\vec n_i,\vec v_i,\vec v_{i+1}]=1,~~\forall i\label{goodness}
 \ee
This condition was phrased in \cite{2001math......7201L} as: $\BB{Z}^3\cap\textrm{span}_{\BB{R}}\bra\vec v_i,\vec v_{i+1}\ket=\textrm{span}_{\BB{Z}}\bra\vec v_i,\vec v_{i+1}\ket$ for all $i$.
The goodness condition ensures that $M$ is smooth.

Furthermore by the Delzant construction \cite{Delzant_1988}, such a good cone gives rise to a 6D toric K\"ahler variety $Y$. Equation \eqref{boundary} says that $M$ is the 'base' of $Y$ while $Y$ is the cone over $M$, in the sense that the metrics of $Y$ and $M$ are related $g_Y=dr^2+r^2g_M$.
In fact the most economic way of defining a Sasaki metric is to say that its cone metric is K\"ahler. We mention that the scaling vector field $r\partial_r$ is related to $\reeb$ by the complex structure of $Y$: $Jr\partial_r=\reeb$, so are their corresponding dual 1-forms $r^{-1}dr$ and $\gk=g_M\reeb$. Here $\gk$ is called the contact 1-form, it satisfies
\bea \iota_{\reeb}\gk=1,~~~\iota_{\reeb}d\gk=0,~~~\opn{Vol}=\frac18\gk\wedge d\gk\wedge d\gk.\nn\eea
As $\reeb$ is Killing, we can normalise $|\reeb|=1$, then $\gk=g\reeb$ satisfies the first two conditions automatically.

\subsection{Examples of toric contact 5-manifold with free $U(1)$}\label{sec_Eotc5mwfU}
We are also interested in free $U(1)$ actions on $M$. Let $\sum_{a=1}^3\xeeb^ae_a$, then $\xeeb$ is free on $M$ iff
\be
s_i:=\det[\vec \xeeb,\vec v_{i-1},\vec v_i]=\vec\xeeb\cdotp(\vec v_{i-1}\times\vec v_i)=\pm1~~~~\forall i,\label{freeness}
\ee
so that not only the vector field $\xeeb$ is nowhere zero, but its stability group is trivial for all points.
This also ensures the smoothness of $M$ since \eqref{freeness} implies goodness.
In \cite{Festuccia:2016gul} the convex good cones admitting such an $X$ was classified in the special case that the cone is \emph{Gorenstein}.
This condition was imposed so that the spin bundle on $M$ can be pushed down to a spin bundle on $X$ easily. But in the current work we drop this assumption. So apart from all the cases listed in prop.3.1 in \cite{Festuccia:2016gul}, we have a few new examples, the complete classification will appear elsewhere.

It suffices to list the normals of the cone and we fix $\vec\xeeb=[0;0;1]$ for all examples. We also define the \emph{sign allotment} as the collection of signs $s_{i+1}:=[\vec v_i,\vec v_{i+1},\vec\xeeb]$, for all $i$. These signs will determine eventually the insertion of (anti)instantons at each of the corner.

\begin{example}\label{example_S5}($-++$)
  First there is a sort of non-example $S^5$. The moment map cone is the first octant $y^{1,2,3}\in\BB{R}_{\geq0}$. The normals are standard,
  \bea V=[\vec v_1,\vec v_2,\vec v_3]=\left[
       \begin{array}{cccc}
         1 & 0 & 0 \\
         0 & 1 & 0 \\
         0 & 0 & 1 \\
       \end{array}\right].\nn\eea
   One has the standard Reeb $\vec\reeb=[1;1;1]$ and $\vec\reeb\cdotp\vec y=1/2$ leads to the round sphere $\sum|z_i|^2=1$. Deforming $\reeb$ gives the squashed sphere. We can choose $\xeeb=[1,1,-1]$, which clearly acts freely: we have simply flipped $z_3\to \bar z_3$ and so the quotient w.r.t. $X$ gives still $\BB{C}P^2$. The horizontal ASD instantons w.r.t. $\vec \reeb$ lead to exotic instantons on $\BB{C}P^2$ simply because we have chosen to look at it 'sideways'. This is why we called it a non-example.

   We record the observable an equivariant closed form that appears in the observable \eqref{observable}.
   \bea \go_{eq}=(4(y^1+y^2)-1)(4\go^2+2\go+1)+2(gv)\wedge d(y^1+y^2)\wedge(4\go+1),\nn\eea
   where $y^{1,2}$ are expressed with the inhomogeneous coordinates as
   \bea y^i=\frac{|z_i|^2}{2(1+|z_1|^2+|z_2|^2)},\nn\eea
   $v=i(z_1\partial_{z_1}-cc)+i(z_2\partial_{z_2}-cc)$ and $\go$ is the K\"ahler form. In particular the bottom component
   $\go_{eq0}=4(y^1+y^2)-1$ takes value $-++$ at the three corners, see the figure
 
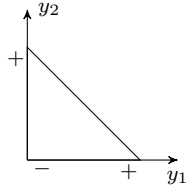
\begin{center}
   \begin{tikzpicture}[scale=1]
\draw [-] (0,0) -- (1.5,0) -- (0,1.5) -- (0,0);

\draw [->] (0,0) -- (2,0) node [below] {\scriptsize{$y_1$}};
\draw [->] (0,0) -- (0,2) node [right] {\scriptsize{$y_2$}};

\node at (.2,-.11) {\scriptsize{$-$}};

\node at (1.35,-.15) {\scriptsize{$+$}};
\node at (-.15,1.35) {\scriptsize{+}};

\end{tikzpicture}
 \end{center}
\end{example}

\begin{example}\label{example_Ypq}($++--$)
  Next up is the well-known $Y^{p,q}$-spaces, where $p>q> 0$ and $\gcd(p,q)=1$ and the normals are
  \bea V=[\vec v_1,\cdots,\vec v_4]=\left[
       \begin{array}{cccc}
         1 & 1 & 1 & 1 \\
         0 & 1 & 2 & 1 \\
         0 & 0 & p-q & p \\
       \end{array}\right]\,.\label{ex_Ypq}\eea

The base $X$ is homeomorphic to $S^2\times S^2$. We can have examples of connected sums of any number of $S^2\times S^2$.
\end{example}
\begin{example}\label{example_F1}($+--+$)
  Staying with four faces
  \bea V=\left[
       \begin{array}{cccc}
         1 & 0 & 1 & 1 \\
         0 & 1 & 0 & -1 \\
         0 & 0 & 2 & 1 \\
       \end{array}\right].\nn\eea
  The base $X$ is the Hirzebruch surface $\BB{F}_1$. This example is special in the sense that it admits a regular Reeb $\vec\reeb=[1;0;1]$.
  If we take the reduction w.r.t. the regular Reeb, the base is $S^2\times S^2$. So we have the double fibration picture fig.\ref{fig_T-dual}.
\end{example}
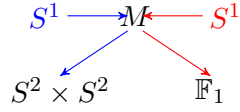
\begin{figure}[h]
\begin{center}
\begin{tikzpicture}[scale=1]
\draw [->,blue] (-0.1,-.2) -- (-1,-.8);
\draw [->,red] (0.1,-.2) -- (1,-.8);
\node at (0,0) {\small{$M$}};
\node at (-1,-1) {\small{$S^2\times S^2$}};
\node at (1,-1) {\small{$\BB{F}_1$}};
\node[red] at (1.2,0) {\small{$S^1$}};
\node[blue] at (-1.2,0) {\small{$S^1$}};
\draw [<-,red] (0.1,0) -- (.9,0);
\draw [<-,blue] (-0.1,0) -- (-.9,0);
\end{tikzpicture}\caption{The same Sasaki manifold $M$ is fibred over two 4-manifolds: $S^2\times S^2$ and the first Hirzebruch surface. }\label{fig_T-dual}
\end{center}
\end{figure}
\begin{example}($+++-$) Also with four faces
 \bea V=\left[
    \begin{array}{cccc}
      1 & 0 & -1 & -1 \\
      0 & 1 & 2 & 1 \\
      0 & 0 & p & q \\
    \end{array}\right],\nn\eea
where the cone is convex iff $q>p>0$.
The base is homeomorphic to $\overline{\BB{C}P^2}\#\overline{\BB{C}P^2}$ and this example is \emph{unrelated} to $Y^{p,q}$ as the cone is not Gorenstein (which would require $q=1$, $p=0$ which is not allowed). Also this example does not possess any regular Reeb for any choice of $p,q$.

Just a word about a quick (crude) way of recognising the homoeomorphism type of the base. The faces correspond to a toric invariant 2-cycle on $X$, and assuming the current choice of $\vec\xeeb$, the first two rows of the $V$ matrix gives relations of the 2-cycles, e.g.
\bea [1]-[3]-[4]=[2]+2[3]+[4]=0\nn\eea
for the current example. The intersection of the 2-cycles are given by
\bea & [i]\cdotp[i+1]=\sgn[\vec v_i,\vec v_{i+1},\vec\xeeb], \nn\\
& [i]\cdotp[i]=-\sgn[\vec\xeeb,\vec v_{i-1},\vec v_i]\,\sgn[\vec\xeeb,\vec v_i,\vec v_{i+1}]\,[\vec\xeeb,\vec v_{i-1},\vec v_{i+1}].\label{intersection_form}\eea
So explicitly
\bea
[1]\cdotp[1]=-1,~~[2]\cdotp[2]=2,~~[3]\cdotp[3]=-1,~~[4]\cdotp[4]=-2.\nn\eea
We choose $[1],[4]$ as the independent set leading to the intersection matrix
\bea Q=\left[
       \begin{array}{cc}
         -1 & -1 \\
         -1 & -2 \\
       \end{array}\right]\sim \left[
       \begin{array}{cc}
         -1 & 0 \\
         0 & -1 \\
       \end{array}\right],\nn\eea
showing that the geometry is that of $\overline{\BB{C}P^2}\#\overline{\BB{C}P^2}$. Note that in the previous example $\BB{F}_1\sim \BB{C}P^2\#\overline{\BB{C}P^2}$.
\end{example}
\begin{example}($++++-$) Upgrading to pentagon
  \bea V=\left[
         \begin{array}{ccccc}
           1 & 0 & -1 & -3 & -2 \\
           0 & 1 & 1 & 2 & 1 \\
           0 & 0 & p & q & r \\
         \end{array}\right].\nn\eea
This is convex iff $p,q,r>0$, $q>3p,~r>2p,~p+r>q,~3r>2q$. So e.g. $p=1,q=4,r=5$ is a valid choice. Choosing $[1],[2],[3]$ as the independent set of 2-cycles of $X$ and working out the intersection, one can show that $X\sim 3\overline{\BB{C}P^2}$.
\end{example}
\begin{example}\label{ex_pentagon}($+--++$) Finally
  \bea V=\left[
         \begin{array}{ccccc}
           1 & 0 & 1 & 1 & 2 \\
           0 & 1 & 0 & -1 & -1 \\
           0 & 0 & p & q & r \\
         \end{array}\right].\nn\eea
This is convex iff $p,q,r>0$, $p>q,~2p>r,~p+q>r,~q>r$. So e.g. $p=3,q=2,r=1$ is a valid choice.
Choosing $[1],[2],[4]$ as the independent set, the intersection form shows that $X\sim (S^2\times S^2)\#\overline{\BB{C}P^2}$.
\end{example}

\subsection{A bottom up construction}\label{sec_Abuc}
For readers who would like to start the construction from a 4 rather than 5-manifold, and arrive at a theory with flipping instantons, there is the following bottom up approach.

We take $X$ to be an almost toric 4-manifold, and we pick an \emph{equivariant $U(1)$ principal bundle} with invariant connection $\gb$, regarded as a global 1-form on the total space of the $U(1)$-bundle, which we call $M$.
We can pick a basis of the $U(1)^2$ isometry for $X$ as $e_{1,2}$ and we let $\xeeb$ be the vector field that rotates the circle fibre of $M$ over $X$.
We let
\bea \reeb=\ep_1\hat e_1+\ep_2\hat e_2+\xeeb\label{reeb_4_5}\eea
be a Killing vector field on $M$, where $\hat e_{1,2}$ are vector fields of the $U(1)$ actions on $M$ lifting $e_{1,2}$ (lifting exists since we have assumed $M$ to be equivariant). We assume of course that the metric on $M$ is invariant under all of the toric symmetries. As the norm of $\reeb$ is non-zero and toric invariant, we can rescale the metric to make $\reeb$ of unit length and keep the toric invariance of the metric intact. The last step is only for convenience.

With this done then the 1-form $\gk=g_M\reeb$ has $\iota_{\reeb}d\gk=0$, this way we have effectively gone back to the main geometric setting, except of course $\gk$ may not be contact.
Consider a vector bundle $E$ on $X$ and pulled back to $M$, then the connection of $E$ on $M$ would have the form $A_5=A_4+\gb\varphi$ for some adjoint scalar $\varphi$. As usual we impose the following equations
\be
\iota_{\reeb}F_5=0,~~~\iota_{\reeb}*_5F_5=-F_5.\nn
\ee
The second equation says that $F_5$ is anti-self-dual transverse to $\reeb$. From the point of view of $X$, the curvature $F_4$ is then an extrapolation of ASD (where $\reeb$ and $\xeeb$ are parallel) and SD where $\reeb$ and $\xeeb$ are anti-parallel.

But one may wonder whether $\xeeb$ and $\reeb$ will ever flip from parallel to anti-parallel, seeing that in \eqref{reeb_4_5} we allowed only constant $\ep_{1,2}$.
This is where the choice of the $U(1)$ bundle, or the choice of $\gb$, comes in. As $M$ is non-trivially fibred over $X$, it is the transition function of $M$ not the non-constancy of $\ep_{1,2}$ that provides the flipping.
\begin{example}
  Consider $\BB{C}P^2$ with $u_{1,2}$ rotating the phases of $\xi_{1,2}$ with $[\xi_1,\xi_2,1]$ being the inhomogeneous coordinates. Take the $U(1)$ bundle to be ${\cal O}(-1)$ and the total space is $S^5$. We let $\reeb=\ep_1u_1+\ep_2u_2+X$, that is, in this patch we lift $u_{1,2}$ to the total space trivially. But their lift in other patches will follow from the transition functions.
  Indeed in the patch $[\eta_1,1,\eta_3]$, this transforms to
  $(\ep_1-\ep_2)v_1-\ep_2v_3+(1+\ep_2)X$, where $v_{1,3}$ rotate the phases of $\eta_{1,3}$. Finally in the patch $[1,\zeta_2,\zeta_3]$, we have
  $(\ep_2-\ep_1)w_2-\ep_1w_3+(1+\ep_1)X$ with $w_{2,3}$ rotating the phases of $\zeta_{2,3}$. So depending on the sizes of $\ep_{1,2}$ one has the following picture
  
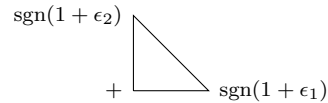
\begin{center}
\begin{tikzpicture}[scale=1]
\draw [-] (0,0) node[left] {\scriptsize{$+$}} -- (1,0) node[right] {\scriptsize{$\sgn(1+\ep_1)$}} -- (0,1) node[left] {\scriptsize{$\sgn(1+\ep_2)$}} -- (0,0);

\end{tikzpicture}
\end{center}
namely one can flip the sign assignments by changing the size of $\ep_{1,2}$. In particular, letting $\ep_1=-2,\,\ep_2=0$ corresponds to the non-example \ref{example_S5} where one merely used $[\bar z_1,z_2,z_3]$ as the homogeneous coordinates for $\BB{C}P^2$.
\end{example}
\begin{example}
One can have a similar example for $S^2\times S^2$ where one chooses $M$ to be the $U(1)$ bundle of degree $-1$ over both $S^2$. We denote by $y^{1,2}\in[0,1/2]$, $\phi_{1,2}$ the radial and angle coordinates of the two $S^2$. Let also $\gt$ be the fibre angle coordinate close to the north pole of the two $S^2$'s., we pick
\bea \gb=d\gt+2y^1d\phi_1+2y^2d\phi_2,\nn\eea
corresponding to the bundle specified above. The vector field $\xeeb$ is still $\partial_{\gt}$ and we choose
\bea \reeb=\xeeb-\ep_1\partial_{\phi_1}-\ep_2\partial_{\phi_2},\nn\eea
The transition function would possibly flip the sign of $\xeeb$, i.e. the following distribution of signs
\begin{center}
\begin{tikzpicture}[scale=1]
\draw [-] (0,0) node[left] {\scriptsize{$+$}} -- (1,0) node[right] {\scriptsize{$\sgn(1-\ep_1)$}} -- (1,1) node[right] {\scriptsize{$\sgn(1-\ep_1-\ep_2)$}} -- (0,1) node[left] {\scriptsize{$\sgn(1-\ep_2)$}} -- (0,0);
\end{tikzpicture}
\end{center}
\end{example}

\subsection{A quasi-topological bound}
Writing $A_5=A_4+\varphi\gb$, the YM term of the 5D curvature $F_5=F_4+\varphi d\gb-\gb D_4\varphi$ has the decomposition
\be
 |F_5|^2=-\gk \wedge F_5\wedge F_5+2|(F_5)_H^+|^2+|\iota_{\reeb}F_5|^2~,\label{5D_bound_I}
\ee
so the configuration $\iota_{\reeb}F_5=0=(F_5)_H^+$ minimises the YM term to be $-\gk\wedge  F_5\wedge F_5$, a quasi-topological term.
Is is also a part of the supersymmetric observable
\be
{\cal O}_5=\Tr\int\limits_M\gk\wedge ((d\gk)^2+d\gk+1)\wedge (F_5+\Psi+\gs)^2~,\nn
\ee
where the combination ${\cal F}=F_5+\Psi+\gs$ is the universal curvature mentioned in the introduction.
The rest of \eqref{5D_bound_I} can be arranged into $\gd$-exact terms, that is,
the 5D action can be written in the form $-{\cal O}_5+\gd\textrm{-exact}$,
\bea
&&\hspace{3cm} S_5=\Tr\int\limits_M-{\cal O}_5+\gd W~\label{5D_action}\\
&&W=\Psi\wedge *(-\iota_{\reeb}F_5-D\gs)-\frac12\chi\wedge * H+2\chi \wedge *F_5+ \gs \gk \wedge d\gk \wedge \chi,\nn\eea
the details of $W$ will not be important here.

Now we rewrite this equation in pure 4D terms, which would give us the structure of 4D action and a bound of the YM term on some equivariant cohomology class. First the 5D observable ${\cal O}_5$ is composed of two pieces $\go_5$ and ${\cal F}$
\bea
&\go_5=\gk\wedge ((d\gk)^2+d\gk+1)~,~~~(d-\iota_{\reeb})\go_5=0~,\nn\\
&{\cal F}=F_5+\Psi+\gs~,~~~\gd{\cal F}=(D_5-\iota_{\reeb}){\cal F}~.\label{used_VIII}\eea
Here we have combined $\gd$, $\gdh$ and simply called it $\gd$. One can also ignore the ghost part since our manipulation is explicitly gauge covariant.
We decompose $\Psi=\psi+\gb\eta$ i.e. into horizontal and vertical piece w.r.t. $\xeeb$ and likewise for ${\cal F}$
\bea &{\cal F}=(F_4+\varphi d\gb+\psi+\gs)+\gb(-D_4\varphi+\eta)~.\nn\eea
Reducing similarly \eqref{used_VIII}
\bea  &\gd(F_4+\varphi d\gb+\psi+\gs)=(D_4-\iota_{\veeb})(F_4+\varphi d\gb+\psi+\gs)+\nu_{eq}(\eta-D_4\varphi)~,~~~\nu_{eq}=d\gb-\rho/h^2~,\nn\\
&\gd(\eta-D_4\varphi)=-(D_4-\iota_{\veeb})(\eta-D_4\varphi)+[F_4+\varphi d\gb+\psi+\gs,\varphi]~.\nn\eea
At the same time $\go_5$ decomposes into
\bea &\go_5=(\go_5)^{\perp}+\gb\iota_{\xeeb}\go_5:=\tilde\go+\gb\go_{eq}~,\nn\\
&(d-\iota_{\veeb})\go_{eq}=0,~~~(d-\iota_{\veeb})\tilde\go=-\nu_{eq}\go_{eq}~,\nn\eea
both relations following from $(d-\iota_{\reeb})\go_5=0$.

Note that $\go_{eq}$ is a 4D equivariantly closed form, and using these relations one can write
\be
 \frac{1}{2\pi}{\cal O}_5=\Tr\int\limits_X\go_{eq}\wedge (F_4+\varphi\rho/h^2+\psi+\gs)^2={\cal O}_4\nn
\ee
up to the following $D_{eq}$ and $\gd$-exact terms
\be
 -2\Tr\big[(\gd+D_{eq})\tilde\go\varphi(F+d\gb \varphi+\psi+\gs)\big]+\Tr[D_{eq}(\tilde\go\go_{eq} \varphi^2)\big] ~. \nn
\ee
Rewriting \eqref{5D_action} in 4D terms is also straightforward and gives us the structure of 4D action as $-{\cal O}_4+\gd$-exact.
The actual expression of the $\gd$-exact terms, i.e. the 4D rewriting of $W$ is hardly instructive and we omit it.

Next we focus on writing a bound similar to \eqref{5D_bound_I},
in fact what we write next is applicable to any equivariant curvature, coming from supersymmetry or not makes no difference.
First, the 5D relation $\iota_{\reeb}F-D\gs=0$ implies $L^A_{\reeb}\gs=0$ and so by shifting $\tilde A_5=A_5+\gs\gk$ one has $\iota_{\reeb}\tilde F_5=0$. Thus we can proceed assuming $\gs=0$ and $\iota_{\reeb}F_5=0$. Writing as before $A_5=A_4+\varphi\gb$, $F_5=F_4+\varphi d\gb-\gb D_4\varphi$, with $A_4,\varphi$ etc invariant under $\xeeb$ for reduction to 4D. That $\iota_{\reeb}F_5=0$ implies
$\iota_{\veeb}(F_4+\varphi d\gb)-(\rho/h^2)D\varphi=0$ or more neatly
\bea
D_{eq}F_{eq}=0~,~~F_{eq}=F_4+\varphi\rho/h^2,~~D_{eq}=D_4-\iota_{\veeb}~.\nn\eea

The above discussion suggests that the correct setting for writing a bound is that of equivariant bundles with equivariant curvature.
We record next two bounds, one for the 4D exotic DW
\bea
 &&\int\limits_Xh(|F_4+\varphi d\gb|^2+h^{-2}|D_4\varphi|^2)=\int\limits_X -\go_{eq}\wedge F_{eq}\wedge F_{eq}+h(1+f^2)|B|^2~,\nn\\
&&\hspace{2cm}B=P\big(F+\varphi d\gb+h^{-1}*_4\big((g_4\veeb)\wedge D_4\varphi\big)\big)~,\nn
\eea
where the extra $h$ factor comes from fibrewise integration and \eqref{miss_match_vol}.
The last two terms give the exotic instanton equation
\be
P\big(F_{eq}+\go_{eq}\varphi+h^{-1}*_4\big((g_4\veeb)\wedge D_4\varphi\big)\big)=0~.\nn
\ee
This construction does not work on $S^4$, except of course for the case of usual ASD instantons. In particular, there is no non-trivial equivariant $U(1)$ bundle on $S^4$ and what we called $\go_{eq}$ will fail to be equivariantly closed.

In fact one could have started from a 4 manifold $X$ and use the reverse engineering in sec.\ref{sec_Abuc} and state a bound purely in terms of 4D geometry. We only record such a bound for the equivariant DW theory.
For this one can simply take $M=X\times S^1$, we get
\begin{proposition}
Let $X$ be a 4-manifold with a Killing vector field $\veeb$, and $E\to X$ be a vector bundle such that its curvature has equivariant extension $F_{eq}=F+\varphi$, $(D-\iota_{\veeb})F_{eq}=0$, then we have the bound
\bea \int_Xh(|F|^2+|D\varphi|^2)=\int_X-\go_{eq}F_{eq}\wedge F_{eq}+2h|C|^2\nn,\eea
where $h^2=(1+|v|^2)^{-1}$ and
\bea C=\frac{1}{2}(1+*)\big(F+\frac{h}{1+h}*(gv\wedge D\varphi)\big).\nn\eea
The equivariant form $\go_{eq}$ has expression
\bea \go_{eq}=h^2((d\ga)^2+d\ga+1)+\ga(2d\ga+1)dh^2,~~~\ga=h^2(g\veeb).\nn\eea
\end{proposition}

\section{Localisation formula for basic forms}\label{sec_Lffbf}

First we recall the more familiar localisation formula on an even dimensional manifold $X^{2n}$.
Let $\veeb$ be a Killing vector field on $X$ with isolated zeros. Let $\ga_{2n}$ be a top form admitting an extension
\bea &\ga_{2n}\to \ga_{eq}=\ga_{2n}+\ga_{2n-2}+\cdots+\ga_2+\ga_0~,\nn\\
&d_{\veeb}\ga_{eq}=(d-\iota_{\veeb})\ga_{eq}=0~.\nn\eea
Then the we have
\bea \int\limits_X\ga_{2n}=\sum_i\frac{(2\pi)^n\ga_0|_i}{e_i}~,\nn\eea
where the sum is over the fixed points of $\veeb$ and $e_i$ is the equivariant Euler class evaluated at the $i^{th}$ point. More concretely, at a fixed point $x_i$, the first derivative $d\veeb$ acts as an orthogonal transformation on $T_{x_i}X$, the pfaffian gives $e$. If one further assumes that $X$ is complex and $\veeb$ preserves the complex structure, then one can find local complex coordinates such that $d\veeb$ acts as product of $U(1)$'s. The product of $U(1)$ weights gives $e$.
For a proof of this formula, see e.g. sec. 7 of \cite{Libine_equiv}.

Progressing to odd dimension $M^{2n+1}$ with a non-vanishing Killing vector field $\reeb$ (not assumed to be related to contact structure). One defines basic forms as
 \be
  \Go^{\sbullet}_B(M):~~\{\ga\in\Go^{\sbullet}(M)\,|\,\iota_{\reeb}\ga=0=L_{\reeb}\ga\}.\nn
 \ee
Let $\gk=(g\reeb)/|\reeb|^2$ be the 1-form dual to $\reeb$ i.e. $\iota_{\reeb}\gk=1$ and also one notices $\iota_{\reeb}d\gk=0=L_{\reeb}d\gk$, i.e. $d\gk$ is basic.

One then repeats the previous construction on the basis forms: let $\ga_{2n}$ be basic admitting an equivariant extension $\ga_{eq}$ s.t. $d_{\xeeb}\ga_{eq}=0$. Here we assume that $\xeeb$ is Killing with $[\xeeb,\reeb]=0$ and that at the isolated closed $\reeb$ orbits $S_i$, one has $\xeeb\|\reeb$.
One has the formula
\be
 \int\limits_{M}\gk\ga_{2n}=\sum_i\ell_i\frac{(2\pi)^n\ga_0|_i}{e_i}\label{equiv_loc_ctc}.
 \ee
where the sum is over the loci $S_i$ while $\ell_i$ is the integral of $\gk$ over $S_i$, finally $e_i$ is the Euler class transverse to $\reeb$. To obtain $e_i$, one can again choose local coordinates $\{\gt,x^1,\cdots,x^n\}$ close to $S_i$, such that $\reeb=\partial_{\gt}$ and $\vec x=0$ is the locus $S_i$. Then $d\xeeb$ acts as orthogonal transformation on $\vec x$ and one obtains $e$ as the pfaffian.

\begin{figure}[h]
\begin{center}
\begin{tikzpicture}[scale=1]
\draw[-] (0,0) node[right] {\scriptsize{$S_i$}} -- (.2,1.5) node[right] {\scriptsize{$S_{i+1}$}} -- (-1,1.2) -- (-1.4,.2) -- (0,0);
\draw [red,->] (-0.7,0.1) node [below] {\scriptsize{$\vec v_{i-1}$}} -- (-0.6,0.5);
\draw [red,->] (0.1,0.75) node [right] {\scriptsize{$\vec v_i$}} -- (-0.3,0.8);
\end{tikzpicture}\caption{Only the base of the cone is draw.}\label{fig_area_polygon}
\end{center}
\end{figure}
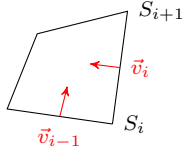

\begin{example}
  We use the geometrical setting of toric Sasaki 5-manifolds, see appendix \ref{sec_Mgs}, and we want to integrate
  \be
   \frac{1}{8}\int\limits_M\gk\wedge (d\gk)^2~ , \nn
  \ee
  which is in fact the volume. Let $\xeeb$ be a Killing that is the linear combination of the three $U(1)$ isometries,
  the equivariant extension reads
  \be
  \frac18(d\gk)^2\To \frac18(d\gk)^2-\frac14\bra\xeeb,\gk\ket d\gk+\frac18\bra\xeeb,\gk\ket^2~.\nn
  \ee
  In this case, transverse to $\reeb$, there is an almost complex (in fact K\"ahler) structure, the equivariant Euler class is the product of the $U(1)$ weights acting on this complex coordinates. The weights are obtained combinatorially.
  Indeed, the loci $S_i$ correspond to the edges of the moment cone $C_{\mu}(M)$ (see fig.\ref{fig_area_polygon}), where since there is only one effective $U(1)$, $\reeb$ must be (anti)-parallel to $\xeeb$. Close to $S_i$, $d\xeeb$ acts with weights
  \bea \frac{[\vec\reeb,\vec v_{i-1},\vec \xeeb]}{[\vec\reeb,\vec v_{i-1},\vec v_i]}~,~~~\frac{[\vec v_i,\vec\reeb,\vec \xeeb]}{[\vec\reeb,\vec v_{i-1},\vec v_i]}~,\nn  \eea
  where $[\vec\reeb,\vec v_{i-1},\vec v_i]=(\vec\reeb\times\vec v_{i-1})\cdotp\vec v_i$ for 3-vectors.

  We also need to evaluate $\bra\xeeb,\gk\ket$, $\ell_i$ at $S_i$
  \bea \bra\xeeb,\gk\ket|_i=\frac{[\xeeb,\vec v_{i-1},\vec v_i]}{[\reeb,\vec v_{i-1},\vec v_i]}~,~~~\ell_i=\frac{2\pi}{[\reeb,\vec v_{i-1},\vec v_i]}~.\nn\eea
  Putting everything together
  \bea
   \textrm{Vol}=\pi^3\sum_i\frac{-[\vec \xeeb,\vec v_i,\vec v_{i+1}]^2}{[\vec\reeb,\vec v_i,\vec v_{i+1}][\vec\reeb,\vec v_i,\vec\xeeb][\reeb,\vec v_{i+1},\vec \xeeb]}~.\nn\eea
  The volume in fact equals
  \bea 48\pi^3{\rm Vol}_{\Gd}\nn\eea
  with $\Gd$ obtained from chopping off the cone perpendicular to $\vec\reeb$ at height $1/(2|\vec\reeb|)$, in particular it is independent of $\xeeb$.
  So for $S^5$ whose cone is the first octant of $\BB{R}^3$, and $\vec\reeb=[1;1;1]$, then ${\rm Vol}_{\Gd}=1/48$ and ${\rm Vol}_{S^5}=\pi^3$ as expected.
\end{example}

In the main text, we will use formula \eqref{equiv_loc_ctc} for forms basic w.r.t. $\reeb$ or $\xeeb$, where $\xeeb$ is from the free $U(1)$.
So we record some formulae that might come in handy.
We shall need the Poincar\'e dual $u_i\in\Go^2(M)$ of the 3-cycle given by the $i^{th}$-faces of the cone $C_{\mu}(M)$. One can choose $u_i$ basic w.r.t $\reeb$ or $\xeeb$ (though not simultaneously), we shall use the same symbol for both choices. Let $\vec\ep$ represent another Killing, then one has equivariant extensions $u_i\to u_i+f_{i,\reeb}$ for $u_i$ basic w.r.t $\reeb$ and $u_i\to u_i+f_{i,\xeeb}$ basic w.r.t $\xeeb$. We only need to know the evaluation of $f_{i,\reeb/\xeeb}$ at the edges to perform integrals
\bea &&2\pi f_{i,\reeb}\big|_i=\frac{[\vec \ep,\vec\reeb,\vec v_{i-1}]}{[\vec \reeb,\vec v_{i-1},\vec v_i]}~,
~~~2\pi f_{i,\reeb}\big|_{i+1}=\frac{[\vec \ep,\vec v_{i+1},\vec \reeb]}{[\vec \reeb,\vec v_i,\vec v_{i+1}]}~,
~~~2\pi f_{i,\reeb}\big|_{j\neq i,i+1}=0~,\label{fp_val_R}\\
&&2\pi f_{i,\xeeb}\big|_{i}=[\vec v_{i-1},\vec v_i,\vec\xeeb][\vec\xeeb,\vec v_{i-1},\vec \ep]~,~~~
2\pi f_{i,\xeeb}\big|_{i+1}=[\vec v_i,\vec v_{i+1},\vec\xeeb][\vec v_{i+1},\vec\xeeb,\vec \ep]~.\label{fp_val_X}\eea
\begin{example}
  We apply these formulae to compute the class $d\gb$, note we always take the orientation of $M$ as $\gk(d\gk)^2/8$.

  We want to compute
  \bea \bra \frac{d\gb}{2\pi},[i]\ket=\frac{1}{2\pi}\int\limits_M \frac{\gb}{2\pi} d\gb u_i\nn\eea
  which can be thought of as the integral of $d\gb$ over the 2-cycle on $X$ represented by the $i^{th}$ face.
  We use another Killing vector $\ep$ with $[\ep,\xeeb]=0$ to localise, typically $\ep$ is chosen to be $\reeb$.
  Note $d\gb$ has equivariant extension $d\gb-\bra\ep,\gb\ket$
  The localisation formula receives contribution at corner $i,i+1$ only, at $i$ one has
  \bea &&\ell_i=2\pi[\vec v_{i-1},\vec v_i,\vec\xeeb]~,~~e_i=[\vec\xeeb,\vec v_{i-1},\vec \ep][\vec v_i,\vec\xeeb,\vec \ep]~,~~
  \bra \ep,\gb\ket|_i=\frac{[\vec v_{i-1},\vec v_i,\vec\ep]}{[\vec v_{i-1},\vec v_i,\vec \xeeb]}~,\nn\\
  &&f_{i,\xeeb}|_i=\frac{1}{2\pi}[\vec v_{i-1},\vec v_i,\vec\xeeb][\vec\xeeb,\vec v_{i-1},\vec \ep]~,~~~
  f_{i,\xeeb}|_{i+1}=\frac{1}{2\pi}[\vec v_i,\vec v_{i+1},\vec\xeeb][\vec v_{i+1},\vec\xeeb,\vec \ep]~,\nn\eea
  and putting the contributions together
  \bea \bra \frac{d\gb}{2\pi},[i]\ket=-[\vec v_{i-1},\vec v_i,\vec\xeeb]\frac{[\vec v_{i-1},\vec v_i,\vec\ep]}{[\vec v_i,\vec \xeeb,\vec\ep]}-[\vec v_i,\vec v_{i+1},\vec\xeeb]\frac{[\vec v_i,\vec v_{i+1},\vec\ep]}{[\vec \xeeb,\vec v_i ,\vec\ep]}
  =[\vec v_{i-1},\vec v_i,\vec\xeeb][\vec v_i,\vec v_{i+1},\vec\xeeb][\vec v_{i-1},\vec v_i,\vec v_{i+1}].\nn\eea

       More generally if we have a class $F+\varphi$ with $d_{\ep}(F+\varphi)=0$ and basic w.r.t $\xeeb$, then
   \bea \bra \frac{F}{2\pi},[i]\ket=\frac{\varphi_i}{[\vec v_i,\vec\xeeb,\vec \ep]}
   +\frac{\varphi_{i+1}}{[\vec\xeeb,\vec v_i,\vec \ep]}=\frac{\varphi_i-\varphi_{i+1}}{[\vec v_i,\vec\xeeb,\vec \ep]}\label{F_2_cycle}~.\eea
\end{example}
\begin{remark}
  As a remark of the formulae \eqref{equiv_loc_ctc}, one actually does not need $\reeb$ to be Killing, but only a 1-form $\xi$ defined in the complement of the loci with $\reeb\|\xeeb$, such that $\iota_{\xeeb}\xi=1$, $\iota_{\reeb}\xi=0$, $\iota_{\xeeb}d\xi=0$ and $\iota_{\reeb}d\xi=0$. For example, when $\reeb$ \emph{is} Killing, we can use
  \bea \xi=\frac{g\xeeb-\bra \xeeb,\reeb\ket\gk}{|\xeeb|^2-\bra \xeeb,\reeb\ket^2}~.\nn\eea
  From $\xi$, we can construct for $\ga_{eq}=\ga_4+\ga_2+\ga_0$ its primitive
  \bea \ga_{eq}=-(d-\iota_{\xeeb})\big(\xi\ga_0+\xi\ga_2+\xi d\xi\ga_0\big)\nn\eea
  \emph{away} from the loci $\xeeb\|\reeb$. Then one can use the Stokes theorem to reduce the calculation to some local computation around the loci $\xeeb\|\reeb$. What is behind all this is that the Stokes theorem is still valid for the basic forms thanks to $d\gk$ being basic. For details as well as the typical scenario where our setting arises, see \cite{contact_loc}.
\end{remark}

\section{The equivariant index of transversally elliptic complex}\label{sec_Teiotec}
\subsection{A toy model on $\BB{C}$}\label{sec_AtmoC}
As a baby model consider the operator $\bar\partial:~E_0\to E_1$ where $E_i\sim\Go^{0,i}$ are two trivial complex line bundles over $\BB{C}$.
The symbol of this operator is simply
\bea \gs(\bar\partial)=\xi^{0,1}\nn\eea
where $\xi$ is the fibre coordinate of $T^*\BB{C}$. One regards $\gs(\bar\partial)$ as a bundle map $\pi^*E_0\to \pi^*E_1$ with $\pi$ the projection $T^*\BB{C}\to\BB{C}$. Simply put $\gs(\bar\partial)$ is a map $E_0\to E_1$ depending linearly on the fibre of $T^*\BB{C}$.
This symbol is an isomorphism whenever $\xi\neq 0$, i.e. away from the zero section of $T^*\BB{C}$, the map $\gs(\bar\partial):\,\pi^*E_0\to \pi^*E_1$ is an isomorphism. One says that the complex $\pi^*E_0\stackrel{\gs(\bar\partial)}{\to} \pi^*E_1$ has support only along the zero section. One can use a group action to further reduce the support of this complex, i.e. localisation.

Let $v$ be the vector field
\bea v=i\ep(z\partial_z-\bar z\bar\partial_{\bar z})\nn\eea
with $\ep\in\BB{C}$, $\re\ep\neq0,\,\im\ep\neq0$. Applying the flat metric to $v$: $gv=i\ep(zd\bar z-\bar zdz)$, we then use $i\ep z$ (the coefficient of $d\bar z$) to deform $\bar\partial$
\bea \bar\partial\to \bar\partial-d\bar z \bra v,\bar\partial\ket=\bar\partial-i\ep zd\bar z.\label{deform_symbol}\eea
Its symbol is deformed to
\bea \xi^{0,1}=\gs(\bar\partial)\to \gs(\bar\partial_{\ep})=\xi^{0,1}-i\ep z.\nn\eea
One wonders why do we keep the non-derivative term $-i\ep z$ in the symbol, seeing that the symbols by definition care only about the highest derivative terms.
This is because we now regard $\gs(\bar\partial_{\ep})$ as a bundle map $\pi^*E_0\to \pi^*E_1$ over $T_{U(1)}^*\BB{C}$, instead of $T^*\BB{C}$. Here the notation $T_{U(1)}^*\BB{C}$ denotes the covetors of $\BB{C}$ that have zero paring with the fundamental vector field of the $U(1)$ action.
Simply put $\xi$ is restricted so that $\xi_{\gt}=0$ where $\gt$ is the angle coordinate of $\BB{C}$. On the other hand the deformation $-i\ep z$ comes solely from the $\gt$ component of $gv$. Under normal circumstance one would have discarded terms of order 0 since they can be homotoped to zero in the presence of an order 1 term $\xi_{\gt}$. But now $\xi_{\gt}=0$, so even the terms of order 0 is dominant and cannot be homotoped away. Another way of understanding this is to look at the vanishing locus of the symbol: that $\xi^{0,1}=i\ep z$ implies $\xi_{\gt}=2|z|^2\re\ep$, which does not vanish unless at $z=0$ where $v=0$, or $\re\ep=0$. This means that the support of the complex $\pi^*E_0\stackrel{\gs(\bar\partial_{\ep})}{\to} \pi^*E_1$ is reduced to the origin of $\BB{C}$ if $\re\ep\neq0$.

What then is the role of $\im\ep$? It turns out the same type of manoeuver can be applied to a compact manifold where one has a vector field $v$ preserving the differential operator in question. One can then deform the symbol as we did above and reduce the support of the complex to the zeros of $v$. Then one would like to do a local calculation of the index and insert the result back to the big manifold, this requires a regularisation (Thom isomorphism) to ensure that the insertion has support only close to the zeros of $v$. Here is where $\im\ep$ is used, as we shall see next.

Back to our example on $\BB{C}$, and we compute the kernel and cokernel of $\bar\partial_{\ep}$.
\bea (\bar\partial-i\ep z) \psi=0~\To~\psi_n=z^ne^{i\ep|z|^2}\label{kernel},\eea
while for the cokernel we need the adjoint operator $(\bar\partial+i\ep z)^{\dag}=\partial+i\bar\ep\bar z$
\bea (\partial-i\bar\ep\bar z)\chi=0~\To~\chi_n=\bar z^ne^{i\bar\ep|z|^2}d\bar z.\label{cokernel}\eea
The treating of $\BB{C}^n$ is exactly the same. Now we assume that $\bar\partial$ is defined on a compact complex manifold $X$ with a Killing vector $v$. We assume that the neighbourhood of a zero of $v$ looks like $\BB{C}^n$, and we would like to insert the modes $\psi_n$ and $\chi_n$ back into $X$.
For this we must select only the exponentially decaying modes.
Note the crucial point that within such decaying modes, only one of the $\psi_n$ or $\chi_n$ is valid, depending on the sign of $\im\ep$
\bea \im\ep>0:&&\ker=\{\psi_n\},~~\coker=\emptyset,\nn\\
\im\ep<0:&&\coker=\{\chi_n\},~~\ker=\emptyset.\nn\eea
We are ultimately interested in the decomposition of the kernel or cokernel into the $U(1)$ weights, i.e. the equivariant index. So the local contribution to the index from the above is
\bea \im\ep>0:&& ~~\ind=1+t+t^2+\cdots=\Big[\frac{1}{1-t}\Big]^+,\label{positive_reg}\\
\im\ep<0:&& ~~\ind=-t^{-1}-t^{-2}-\cdots=\Big[\frac{1}{1-t}\Big]^-,\label{minus_reg}\eea
where $t=e^{i\ep}$ and where $+/-$ is a useful short hand meaning expansion into geometric series in $t$ or $t^{-1}$. The second line, i.e. the modes $\chi_n$ contribute $-$ to the index because it belongs to $E_1$ and so is placed at degree 1.

For a compact manifold, it may happen that with a good choice of $v$, the support of the deformed symbol will be localised to the isolated zeros of $v$ and so the index can be calculated from the local contribution plus or minus regulated depending on the sign of $\ep$.

\subsection{The $\bar\partial_H$ operator on toric Sasaki manifolds}
We apply the machinery above to compute the equivariant index of the transverse $\bar\partial_H$ operator.

Recall from \cite{2010arXiv1004.2461S} that on a Sasaki 5-manifold, one has the local adapted coordinate system
\bea \vgt,\xi^1,\xi^2,\bar\xi^1,\bar\xi^2\label{adapted_coord}\eea
such that the Reeb $\reeb$ is simply $\partial_{\vgt}$ and under a change of patch
\bea \vgt'=\vgt'(\vgt,\xi,\bar\xi),~~\xi'=\xi'(\xi).\label{change_coord}\eea
From this one sees that $\xi^{1,2}$ are complex coordinates transverse to $\reeb$.
In fact there is a K\"ahler structure transverse to $\reeb$, using which one can write down a transverse Dolbeault operator more easily (see the appendix of \cite{Qiu:2016dyj}). But this is not strictly necessary.

From the change of coordinates, one checks that the operator
\bea\bar\partial_H=d\bar \xi^i\partial_{\bar \xi^i}\label{del_bar}\eea
is well-defined. But the adapted coordinates $\xi^i,\vgt$ are not easy to use seeing that they are local and have wrong periods.

It is therefore helpful to have a coordinate independent way to write \eqref{del_bar} shown in \cite{Kohn_Rossi}, which we review quickly.

Our toric 5D Sasaki-manifold $M$ is the boundary of a 6D toric K\"ahler variety $Y$ as follows. The moment map of $Y$ is the cone denoted $C_{\mu}(M)$ in sec.\ref{sec_Mgs} (note $Y$ is singular at the tip of the cone, but smooth otherwise). Recall from \eqref{boundary} $M$ is described as:
\bea r^2=1/2,~~r^2:=\vec\reeb\cdotp\vec y\nn\eea
The $\bar\partial_H$ will be derived from $\bar\partial$ of $Y$ through a series of maps.
Take any form $\ga$ on $M$, one extends it into a small neighbourhood of $M$ in $Y$, where it can be decomposed into types $\Go^{p,q}(Y)$. Continue using the same letter $\ga$, one applies to it the $\bar\partial$ operator of $Y$ (as $M$ is bounded away from the singularity of $Y$, there is no smoothness issue in this operation). Then one projects $\bar\partial\ga$ to the component perpendicular to $(dr)^{0,1}$ (those forms perpendicular to $(dr)^{0,1}$ are called complex tangential to $M$ in \cite{Kohn_Rossi}). The composition of operations
\bea \Go(M)\stackrel{\rm ext}{\to}\Go^{p,q}(Y) \stackrel{\bar\partial}{\to}\Go^{p,q+1}(Y)\stackrel{\textrm{prj}}{\to}\Go^{p,q+1}\stackrel{\rm res}{\to}\Go(M)\label{composition}\eea
defines an operator $\bar\partial_H$. It squares to zero because $\bar\partial^2=0$ and when acting on forms of $M$. Furthermore, the operator is independent of the choice of extension, nor the deformations of $r$, provided that $r$ has gradient transverse to $M$.
To see how this is related to the first definition \eqref{del_bar}, recall from sec.\ref{sec_Tc5m} that $dr$ and $\gk=g\reeb$ are related by the complex structure of $Y$, and $dr$ is transverse to $M$. Thus from the perspective of $M$, applying $\bar\partial$ and projecting away $(dr)^{(0,1)}$ is the same as applying $\bar\partial$ and removing components along $\gk$. Such procedure normally would not have produced anything nilpotent, but what is crucial for $\bar\partial_H^2=0$ is that $\gk$ is exact in $Y$.

To summarise, from whichever perspective above, one gets an operator $\bar\partial_H$ acting on the complex
\bea \Go^{0,0}_H\stackrel{\bar\partial_H}{\to}\Go^{0,1}_H\stackrel{\bar\partial_H}{\to}\Go^{0,2}_H\nn\eea
where ${}_H$ means transversely to $\reeb$.

\subsection{The $\bar\partial_H$ operator near a closed Reeb orbit}

There is a $U(1)^3$-action on $M$ preserving $\partial_H$ and so we can compute the fully $U(1)^3$-equivariant index of $\bar\partial$.
This operator is certainly elliptic transverse to the $U(1)^3$ actions, in fact, the only direction in which it fails to be elliptic is along $\gk$.
We shall pick a vector field $\ep$ which is a generic linear combination of the $U(1)^3$, and deform the symbol $\gs(\bar\partial_H)$ (as described in sec.\ref{sec_AtmoC}) along $\ep^{\perp}$, where $\perp$ means perpendicular to $\gk$. Thanks to the ellipticity of $\bar\partial_H$ perpendicular to $\gk$, such a deformation reduces the support of the symbol to the locus $\ep^{\perp}=0$, which happens when $\ep$ and $\reeb$ are parallel.
This only occurs at the closed Reeb orbits since these are the loci where two of $U(1)^3$ have collapsed and so $\ep$ $\reeb$ both are proportional to the remaining $U(1)$.

This discussion above shows that the index of $\bar\partial_H$ can be computed from the local geometry round the closed Reeb orbits. It suffices thus to find a simplified description of $\bar\partial_H$ round such loci and apply sec.\ref{sec_AtmoC}).

Recall from sec.\ref{sec_Mgs}, the isolated closed Reeb orbits are located at the edges of $C_{\mu}$.
In the following we will not distinguish between the base of $C_{\mu}$, which is a convex polygon, and $C_{\mu}$ itself.

As before we denote the primitive normals of the cone $C_{\mu}$ as $\vec v_1,\cdots,\vec v_{\tt n}\in\BB{Z}^3$.
At the neighbourhood of the corner between face 1 and 2, we pick $\vec n\in\BB{Z}^3$ such that
\bea [\vec n,\vec v_1,\vec v_2]=1\label{n_choice}\eea
thanks to the good cone condition.
We also let $\vec u^1=\vec v_2\times\vec n$, $\vec u^2=\vec n\times\vec v_1$ and $\vec m=\vec v_1\times\vec v_2$.
We can define three angle coordinates
\bea \phi^1=\vec u^1\cdotp\vec\gt,~~\phi^2=\vec u^2\cdotp\vec\gt,~~\gt=\vec m\cdotp\vec\gt\label{adapted_coord_good}\eea
where $\gt_{1,2,3}$ are the angle coordinates of the $U(1)^3$ fibre\footnote{since $M$ is a $U(1)^3$ fibration cover the base polygon,
we have three angle coordinates in a dense open subset of $X$. Along the faces of the polygon, certain angles become ill defined e.g. at face 1, $d\phi^1=\vec u^1\cdotp d\gt$ is ill defined and so must be accompanied by quantities vanishing sufficiently fast at face 1}.
These angles have $2\pi$ periods thanks to the condition $\det[\vec u^1,\vec u^2,\vec m]=1$.

The three angles above are the arguments of three complex coordinates $z^1,z^2,z^3$ of $Y$ with $z^1,z^2$ vanishing at the corner and $z^3$ non-vanishing.  However $(z^1,z^2,\gt)$ are not the adapted coordinates of \eqref{adapted_coord}, for the reason that $L_{\reeb}\phi^{1,2}\neq0$ while from \eqref{adapted_coord} one must have $L_{\reeb}\xi^{1,2}=0$. In fact the angles coordinates of \eqref{adapted_coord} are
\bea
\psi^1=\frac{[\vec v_2,\vec\reeb,\vec \gt]}{[\vec v_1,\vec v_2,\vec\reeb]},~~
\psi^2=\frac{[\vec \reeb,\vec v_1,\vec \gt]}{[\vec v_1,\vec v_2,\vec\reeb]},~~
\vgt=\frac{[\vec v_1,\vec v_2,\vec \gt]}{[\vec v_1,\vec v_2,\vec\reeb]}.\nn\eea
On can check that $\psi^{1,2}$ are 'perpendicular' to $\vec\reeb$ as an easy consequence of the cross product, and so $\psi^{1,2}$ invariant under $L_{\reeb}$.
For the record the two sets of angle coordinates are related as
\bea (\phi^1,\phi^2,\gt)=(\psi^1+(\vec\reeb\cdotp\vec u^1)\vartheta,\psi^2+(\vec\reeb\cdotp\vec u^2)\vartheta,
(\vec\reeb\cdotp\vec m)\vartheta).\label{local_wghts}\eea

Recall that one can deform $r^2$ while maintaining that its derivative be transverse to $M$. At the neighbourhood of the corner, one can use a simpler choice of $r^2=|z^3|^2$. This way the local geometry is $\BB{C}^2\times S^1$ with $z^{1,2}$ parameterising $\BB{C}^2$ and $S^1$ is the angle of $z^3$.
Following earlier discussion on the construction of $\bar\partial_H$, i.e. applying $\bar\partial$ and projecting out $d\bar z^3$ simply leaves us with
\bea \bar\partial_H=\sum_{i=1,2}d\bar z^i\partial_{\bar z^i}\nn\eea
and we get back to the toy model of sec.\ref{sec_AtmoC}, with the addition of an extra circle.

We need to deform $\bar\partial_H$ as in \eqref{deform_symbol} with a vector field $\ep$ which is a (complex) linear combination of the three $U(1)$'s. As usual we denote $\ep$ with its weights
\bea \vec\ep=[\ep_1,\ep_2,\ep_3].\nn\eea
The deformation reads
\bea \bar\partial_H\mapsto \bar\partial_{H,\ep}=\sum_{i=1}^2 \big(d\bar z^i\partial_{\bar z^i}-d\bar z^i\bra \ep^{\perp},\partial_{\bar z^i}\ket\big),\nn\eea
where we recall that $\ep^{\perp}$ means the component perpendicular to $\reeb$.

The tricky bit is to compute the inner product, as we certainly \emph{cannot} assume that the metric near the corner is the flat one when expressed in $z^{1,2}$ coordinates. The obvious reason is that $z^{1,2}$ coordinates are not even canonically defined, but rather involves a choice of $\vec n$ in \eqref{n_choice}. Different choices of $\vec n$ amounts to adding extra twists of $z^{1,2}$ as one goes along the Reeb orbit.

What is canonically defined are the adapted coordinates $\xi^{1,2}$ and in fact the metric \emph{transverse} to $\reeb$ is the flat metric in the $\xi^{1,2}$ coordinates. Showing this requires using techniques developed in \cite{Martelli:2005tp} for the explicit metric and complex coordinates, which we will not go into. The mismatch between $\partial_{\bar z^{1,2}}$ and $\partial_{\bar \xi^{1,2}}$ are purely along the $\reeb$ direction (an inkling of which can be seen from \eqref{local_wghts}), and so will not affect the inner product with $\ep^{\perp}$. To summarise, the inner products
$\bra \ep^{\perp},\partial_{\bar z^i}\ket$ are given by the $U(1)$ weights of $\xi^{1,2}$ under $\ep$
\bea \bra v,\partial_{\bar z^1}\ket=iz^1\frac{[\vec v_2,\vec\reeb,\vec \ep]}{[\vec v_1,\vec v_2,\vec\reeb]},~~\bra v,\partial_{\bar z^2}\ket=iz^2 \frac{[\vec\reeb,\vec v_1,\vec \ep]}{[\vec v_1,\vec v_2,\vec\reeb]},\nn\eea
where we have used the weights \eqref{local_wghts}. We get then that the deformation terms read
\bea -d\bar z^1 z^1 (i\frac{[\vec v_2,\vec\reeb,\vec \ep]}{[\vec v_1,\vec v_2,\vec\reeb]})-d\bar z^2 z^2 (i\frac{[\vec\reeb,\vec v_1,\vec \ep]}{[\vec v_1,\vec v_2,\vec\reeb]})\nn\eea

As we saw in sec.\ref{sec_AtmoC} it was the sign of the imaginary parts of
\bea s^1=\im([\vec\ep,\vec v_2,\vec\reeb]),~~~s^2=\im([\vec\ep,\vec \reeb,\vec v_1])\nn\eea
that determine the plus/minus regulation. Note also that $[\vec v_1,\vec v_2,\vec\reeb]>0$ at all corners.

The toy model computation shows that the cohomology of this operator is given by the modes (where $p,q\in\BB{Z}_{\geq0}$, $r\in\BB{Z}$)
\bea s^1>0,\,s^2>0:~~\ep^2e^{ir\gt}(z^1)^p(z^2)^q,~~~~s^1<0,\,s^2>0:~~e^{ir\gt}(\bar z^1)^p(z^2)^qd\bar z^1,\nn\\
s^1>0,\,s^2<0:~~e^{ir\gt}(z^1)^p(\bar z^2)^qd\bar z^2,~~~~s^1<0,\,s^2<0:~~e^{ir\gt}(\bar z^1)^p(\bar z^2)^qd\bar z^1\wedge d\bar z^2,\nn\eea
where we have omitted the exponentially decaying factor since they do not contribute anything to the $U(1)$ weights.
We also remark that since the local geometry is now $\BB{C}^2\times S^1$ instead of $\BB{C}^2$, we have included a complete tower of Kaluza-Klein modes from the $S^1$ factor. The summation over $r$ erases the ambiguity of the weights of $z^{1,2}$. Indeed $\arg z^{1,2}$ were only defined up to integer shifts of $\gt$ (as a result of the non-uniqueness of $\vec n$ in \eqref{n_choice}). But we are also summing over all shifts of $r\gt$, the ambiguity drops.

The treatment of regulation at other corners is of course the same. To summarise the vector field $\ep$ gives us a prescription of regularisation at all
corners, and we shall see that the regularisation at different corners orchestrates globally so that there is no over/under counting.

\subsection{Collection of the local contributions}
\begin{figure}[h]
\begin{center}
\begin{tikzpicture}[scale=1]
\draw [step=0.3,thin,gray!40] (-.8,-0.2) grid (2,1.4);

\draw [-] (0,0) -- (1.8,0);
\draw [-] (0,0) -- (-0.6,1.2);

\draw [->] (-.3,.6) node [left] {\scriptsize{$\vec v_1$}} -- (0.1,.8);
\draw [->] (.75,0) node [below] {\scriptsize{$\vec v_2$}} -- (0.75,.45);

\node at (-.2,.15) {\scriptsize{+}};
\node at (.2,-.11) {\scriptsize{+}};

\foreach \x in {0,1,...,6}
\node at (0.3*\x,0) [blue] {$\sbullet$};
\foreach \x in {0,1,...,6}
\node at (0.3*\x,0.3) [blue] {$\sbullet$};
\foreach \x in {0,1,...,7}
\node at (-0.3+0.3*\x,.6) [blue] {$\sbullet$};
\foreach \x in {0,1,...,7}
\node at (-0.3+0.3*\x,.9) [blue] {$\sbullet$};
\foreach \x in {0,1,...,8}
\node at (-0.6+0.3*\x,1.2) [blue] {$\sbullet$};
\end{tikzpicture}

\begin{tikzpicture}[scale=1]

\draw [step=0.3,thin,gray!40] (-2,-0.2) grid (1.3,1.4);

\draw [-] (0,0) -- (1,0);
\draw [-] (0,0) -- (-0.7,1.4);

\draw [->] (-.3,.6) -- (0.1,.8) node [above] {\scriptsize{$\vec v_1$}};
\draw [->] (.75,0)  -- (0.75,.45) node [right] {\scriptsize{$\vec v_2$}};

\node at (-.2,.15) {\scriptsize{$-$}};
\node at (.2,-.11) {\scriptsize{+}};

\foreach \x in {1,2,...,6}
\node at (-0.3*\x,0) [blue] {\tiny$\circ$};
\foreach \x in {1,2,...,6}
\node at (-0.3*\x,0.3) [blue] {\tiny$\circ$};
\foreach \x in {2,3,...,6}
\node at (-0.3*\x,0.6) [blue] {\tiny$\circ$};
\foreach \x in {2,3,...,6}
\node at (-0.3*\x,0.9) [blue] {\tiny$\circ$};
\foreach \x in {3,4,...,6}
\node at (-0.3*\x,1.2) [blue] {\tiny$\circ$};
\end{tikzpicture}
\caption{Two corners with $++$ and $+-$ regularisation respectively. We only draw the base of the positive cone.
The solid dot correspond to the $+$ sign in Eq.\ref{positive_reg}, while the hollow circles the negative sign in Eq.\ref{minus_reg}.}\label{fig_plus_minus}
\end{center}
\end{figure}
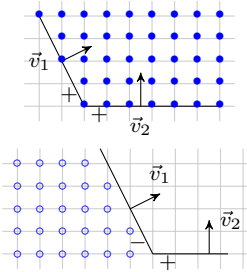

We will now collect the contributions to the index from each corner. At corner 12, and $s^1>0$, $s^2>0$ we read off the $U(1)$ weights as
$p\vec u^1+q\vec u^2+r\vec m$, with $p,q\in\BB{Z}_{\geq0}$, $r\in\BB{Z}$. Each mode contributes
\bea \sum_{\vec w\in\BB{Z}^3,\,\vec w\cdotp\vec v^1\geq0,\,\vec w\cdotp\vec v^2\geq0}\vec t\;{}^{\vec w}\nn\eea
where $\vec t=[t_1,t_2,t_3]$ is the equivariant parameter for the three $U(1)$'s and $\vec t\;{}^{\vec w}$ is a short hand
\bea \vec t\;{}^{\vec w}:=(t_1)^{w_1}(t_2)^{w_2}(t_3)^{w_3}.\nn\eea
This simply means the sign of $s^i$ determines where the $U(1)$ weights are to the positive or negative side of face $i$.
The other three combination of $s^1,s^2$ give
\bea &&s^1>0,\,s^2>0:~~\sum_{\vec w\in\BB{Z}^3,\,\vec w\cdotp\vec v^1\geq0,\,\vec w\cdotp\vec v^2\geq0}\vec t\;{}^{\vec w},~~~~s^1<0,\,s^2>0:~~(-1)\sum_{\vec w\in\BB{Z}^3,\,\vec w\cdotp\vec v^1\lneq0,\,\vec w\cdotp\vec v^2\geq0}\vec t\;{}^{\vec w},\nn\\
&&s^1>0,\,s^2<0:~~(-1)\sum_{\vec w\in\BB{Z}^3,\,\vec w\cdotp\vec v^1\geq0,\,\vec w\cdotp\vec v^2\lneq0}\vec t\;{}^{\vec w},~~~~
s^1<0,\,s^2<0:~~\sum_{\vec w\in\BB{Z}^3,\,\vec w\cdotp\vec v^1\lneq0,\,\vec w\cdotp\vec v^2\lneq0}\vec t\;{}^{\vec w}.\nn\\
\label{wghts_12}\eea
We can illustrate in picture the first two scenario, and we have only drawn the base of the \emph{positive half} cone in fig.\ref{fig_plus_minus}.
Here $\circ$ means one should \emph{remove} the corresponding the lattice points due to the negative sign in \eqref{wghts_12}.
Also it is important to remember that for the negative half of the cone, the situation is opposite: which we illustrate with the 2D section of a cone
\begin{figure}[h]
\begin{center}

\begin{tikzpicture}[scale=0.9]

\draw [step=0.3,thin,gray!40] (-.8,-1.4) grid (1.0,1.4);

\draw [-] (.6,-1.2) -- (-0.6,1.2);

\draw [->] (-.3,.6) node [left] {\scriptsize{$\vec v_1$}} -- (0.1,.8);

\node at (-.6,.15) {\scriptsize{$s^1>0$}};

\foreach \x in {0,1}
\node at (0.6+0.3*\x,-1.2) [blue] {$\sbullet$};
\foreach \x in {0,1}
\node at (0.6+0.3*\x,-0.9) [blue] {$\sbullet$};
\foreach \x in {0,1}
\node at (0.6+0.3*\x,-0.6) [blue] {$\sbullet$};
\foreach \x in {0,1,...,2}
\node at (0.3+0.3*\x,-0.6) [blue] {$\sbullet$};
\foreach \x in {0,1,...,2}
\node at (0.3+0.3*\x,-0.3) [blue] {$\sbullet$};
\foreach \x in {0,1,...,3}
\node at (0.3*\x,0) [blue] {$\sbullet$};
\foreach \x in {0,1,...,3}
\node at (0.3*\x,0.3) [blue] {$\sbullet$};
\foreach \x in {0,1,...,4}
\node at (-0.3+0.3*\x,.6) [blue] {$\sbullet$};
\foreach \x in {0,1,...,4}
\node at (-0.3+0.3*\x,.9) [blue] {$\sbullet$};
\foreach \x in {0,1,...,5}
\node at (-0.6+0.3*\x,1.2) [blue] {$\sbullet$};

\node at (1.4,0) {\scriptsize{$+$}};
\end{tikzpicture}

\begin{tikzpicture}[scale=0.9]

\draw [step=0.3,thin,gray!40] (-0.8,-1.4) grid (1,1.4);

\draw [-] (.6,1.2) -- (-0.6,-1.2);

\draw [->] (0.3,0.6) -- (-0.1,0.8) node [left] {\scriptsize{$\vec v_2$}};

\node at (-.6,.15) {\scriptsize{$s^2<0$}};

\foreach \x in {0}
\node at (0.9-0.3*\x,1.2) [blue] {\tiny$\circ$};
\foreach \x in {0,1}
\node at (0.9-0.3*\x,0.9) [blue] {\tiny$\circ$};
\foreach \x in {0,1}
\node at (0.9-0.3*\x,0.6) [blue] {\tiny$\circ$};
\foreach \x in {0,1}
\node at (0.9-0.3*\x,0.6) [blue] {\tiny$\circ$};
\foreach \x in {0,1,2}
\node at (0.9-0.3*\x,0.3) [blue] {\tiny$\circ$};
\foreach \x in {0,1,2}
\node at (0.9-0.3*\x,0) [blue] {\tiny$\circ$};
\foreach \x in {0,1,...,3}
\node at (0.9-0.3*\x,-0.3) [blue] {\tiny$\circ$};
\foreach \x in {0,1,...,3}
\node at (0.9-0.3*\x,-.6) [blue] {\tiny$\circ$};
\foreach \x in {0,1,...,4}
\node at (0.9-0.3*\x,-.9) [blue] {\tiny$\circ$};
\foreach \x in {0,1,...,4}
\node at (0.9-0.3*\x,-1.2) [blue] {\tiny$\circ$};

\node at (1.4,0) {\scriptsize{$=$}};
\end{tikzpicture}

\begin{tikzpicture}[scale=0.9]

\draw [step=0.3,thin,gray!40] (-0.8,-1.4) grid (0.8,1.4);

\draw [-] (.6,1.2) -- (-0.6,-1.2);
\draw [-] (.6,-1.2) -- (-0.6,1.2);

\foreach \x in {0,1,2}
\node at (-0.3+0.3*\x,-1.2) [blue] {\tiny$\circ$};
\foreach \x in {0,1,2}
\node at (-0.3+0.3*\x,-0.9) [blue] {\tiny$\circ$};
\foreach \x in {0}
\node at (0+0.3*\x,-0.6) [blue] {\tiny$\circ$};
\foreach \x in {0}
\node at (0+0.3*\x,-0.3) [blue] {\tiny$\circ$};
\foreach \x in {0}
\node at (0+0.3*\x,0) [blue] {$\sbullet$};
\foreach \x in {0}
\node at (0+0.3*\x,0.3) [blue] {$\sbullet$};
\foreach \x in {0,1,2}
\node at (-0.3+0.3*\x,.6) [blue] {$\sbullet$};
\foreach \x in {0,1,2}
\node at (-0.3+0.3*\x,.9) [blue] {$\sbullet$};
\foreach \x in {0,1,...,4}
\node at (-0.6+0.3*\x,1.2) [blue] {$\sbullet$};
\end{tikzpicture}
\caption{A 2D cone. After cancellation, one gets all lattice points within the positive half cone and minus the lattice points in the interior of the negative half cone. But for a 3D (or odd D) cone, one gets all the lattice points in the positive cone, and plus all the lattice points in the interior of the negative cone.}\label{fig_2D_section}
\end{center}
\end{figure}
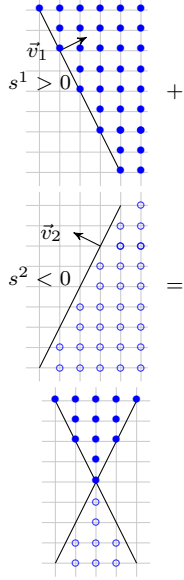

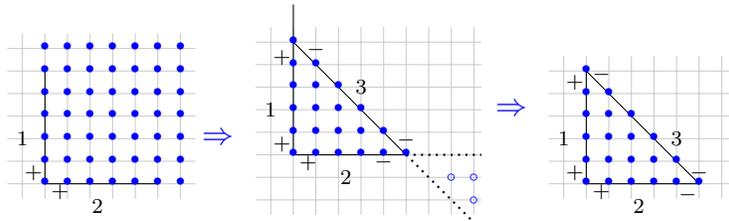
\begin{figure}[h]
\begin{center}
\begin{tikzpicture}[scale=1]
\draw [step=0.3,thin,gray!40] (-.5,-0.2) grid (2,2.0);

\draw [-] (0,1.5) -- (0,0) -- (1.5,0);

\node at (-.3,.6) {\scriptsize{$1$}};
\node at (.7,-.3) {\scriptsize{$2$}};

\node at (-.15,.15) {\scriptsize{+}};
\node at (.2,-.11) {\scriptsize{+}};

\foreach \x in {0,1,...,6}
\node at (0.3*\x,0) [blue] {$\sbullet$};
\foreach \x in {0,1,...,6}
\node at (0.3*\x,0.3) [blue] {$\sbullet$};
\foreach \x in {0,1,...,6}
\node at (0.3*\x,.6) [blue] {$\sbullet$};
\foreach \x in {0,1,...,6}
\node at (0.3*\x,.9) [blue] {$\sbullet$};
\foreach \x in {0,1,...,6}
\node at (0.3*\x,1.2) [blue] {$\sbullet$};
\foreach \x in {0,1,...,6}
\node at (0.3*\x,1.5) [blue] {$\sbullet$};
\foreach \x in {0,1,...,6}
\node at (0.3*\x,1.8) [blue] {$\sbullet$};

\node at (2.3,.6) [blue] {$\To$};
\end{tikzpicture}
\begin{tikzpicture}[scale=1]
\draw [step=0.3,thin,gray!40] (-.5,-0.8) grid (2.5,1.7);

\draw [-] (0,1.5) -- (0,0) -- (1.5,0);
\draw [-] (0,1.5) -- (1.5,0);
\draw [thick,dotted] (1.5,0) -- (2.4,-0.9);
\draw [-] (0,1.5) -- (0,2);
\draw [thick,dotted] (1.5,0) -- (2.5,0);

\node at (-.3,.6) {\scriptsize{$1$}};
\node at (.7,-.3) {\scriptsize{$2$}};
\node at (0.9,0.9) {\scriptsize{$3$}};

\node at (-.15,.15) {\scriptsize{+}};
\node at (.2,-.11) {\scriptsize{+}};
\node at (-.15,1.3) {\scriptsize{$+$}};
\node at (.3,1.4) {\scriptsize{$-$}};
\node at (1.2,-.1) {\scriptsize{$-$}};
\node at (1.5,0.2) {\scriptsize{$-$}};

\foreach \x in {0,1,...,5}
\node at (0.3*\x,0) [blue] {$\sbullet$};
\foreach \x in {0,1,...,4}
\node at (0.3*\x,0.3) [blue] {$\sbullet$};
\foreach \x in {0,1,...,3}
\node at (0.3*\x,.6) [blue] {$\sbullet$};
\foreach \x in {0,1,...,2}
\node at (0.3*\x,.9) [blue] {$\sbullet$};
\foreach \x in {0,1}
\node at (0.3*\x,1.2) [blue] {$\sbullet$};
\foreach \x in {0}
\node at (0.3*\x,1.5) [blue] {$\sbullet$};

\node at (2.1,-0.3) [blue] {\tiny$\circ$};
\node at (2.4,-0.3) [blue] {\tiny$\circ$};
\node at (2.4,-0.6) [blue] {\tiny$\circ$};

\node at (2.9,.6) [blue] {$\To$};
\end{tikzpicture}
\begin{tikzpicture}[scale=1]
\draw [step=0.3,thin,gray!40] (-.5,-0.2) grid (2,1.7);

\draw [-] (0,0) -- (1.5,0) -- (0,1.5) -- (0,0);

\node at (-.3,.6) {\scriptsize{$1$}};
\node at (.7,-.3) {\scriptsize{$2$}};
\node at (1.2,.6) {\scriptsize{$3$}};

\node at (-.15,.15) {\scriptsize{+}};
\node at (.2,-.11) {\scriptsize{+}};

\node at (1.35,-.15) {\scriptsize{$-$}};
\node at (1.5,.15) {\scriptsize{$-$}};

\node at (-.15,1.35) {\scriptsize{+}};
\node at (.2,1.45) {\scriptsize{$-$}};

\foreach \x in {0,1,...,5}
\node at (0.3*\x,0) [blue] {$\sbullet$};
\foreach \x in {0,1,...,4}
\node at (0.3*\x,0.3) [blue] {$\sbullet$};
\foreach \x in {0,1,...,3}
\node at (0.3*\x,.6) [blue] {$\sbullet$};
\foreach \x in {0,1,...,2}
\node at (0.3*\x,.9) [blue] {$\sbullet$};
\node at (0,1.2) [blue] {$\sbullet$};
\node at (0.3,1.2) [blue] {$\sbullet$};
\node at (0,1.5) [blue] {$\sbullet$};
\end{tikzpicture}

\caption{A valid assignment of $+/-$ regularisation. Note that the two signs that flank the same side must be opposite.}\label{fig_reg_trig}
\end{center}
\end{figure}

From the illustration fig.\ref{fig_2D_section} we see that we can focus on the positive half of the cone, and we only draw the base of it.
We take a simpler case, a cone whose 3 faces are just the coordinate planes in $\BB{R}^3$ and we take $\vec\reeb=[\reeb_1,\reeb_2,\reeb_3]$ such that $\reeb_i>0$. If we take $\im\vec\ep=[1,-1,-1]$, then at the corner $1,2$, we have $\im([\vec\ep,\vec\reeb,\vec v_1])=-\reeb_3+\reeb_2$, $\im([\vec\ep,\vec\reeb,\vec v_2])=-\reeb_3-\reeb_1$, $\im([\vec\ep,\vec\reeb,\vec v_3])=\reeb_2+\reeb_1$. Then if we pose $\reeb_2>\reeb_3$, then the regularisations at each corner (that is $+$ or $-$) are as in fig.\ref{fig_reg_trig}.
The 12 corner (left panel) contributes to all the lattice points at the 1st quadrant. Adding the corner from 13, one removes all the lattice points outside face 3, but it removes too many lattice points (the circles in the middle panel). But the contribution from corner 23 exactly corrects this error. In the end, we get all the lattice points within the triangle.

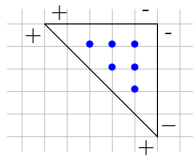
\begin{figure}[h]
\begin{center}
\begin{tikzpicture}[scale=1]

\draw [step=0.3,thin,gray!40] (-.5,-0.2) grid (2,1.7);

\draw [-] (0,1.5) -- (1.5,1.5) -- (1.5,0) -- (0,1.5);

\node at (1.65,1.35) {\scriptsize{-}};
\node at (1.35,1.65) {\scriptsize{-}};

\node at (1.65,.15) {\scriptsize{$-$}};
\node at (1.35,-.15) {\scriptsize{$+$}};

\node at (-.15,1.35) {\scriptsize{$+$}};
\node at (.2,1.65) {\scriptsize{$+$}};

\node at (.6,1.2) [blue] {$\sbullet$};
\node at (0.9,1.2) [blue] {$\sbullet$};
\node at (1.2,1.2) [blue] {$\sbullet$};
\node at (0.9,.9) [blue] {$\sbullet$};
\node at (1.2,0.9) [blue] {$\sbullet$};
\node at (1.2,0.6) [blue] {$\sbullet$};
\end{tikzpicture}\caption{For the negative cone, one gets all the lattice points in the interior of the cone. As for whether one gets solid dot or hollow dot depends on the dimension of the cone: solid for odd and hollow for even.}\label{fig_reg_trig_min}
\end{center}
\end{figure}

So far our discussion is from the perspective of the positive cone, the negative cone is similar but one gets all the lattice points in the interior of the negative cone fig.\ref{fig_reg_trig_min}.

\subsection{The General Cancellation Scheme}\label{sec_TGCS}
The behaviour of cancellation from different corners is completely general: taking any generic $\ep$, one might get different $+/-$ regularisation schemes, but one gets the same result for all polygons, as we show next \footnote{We thank Maksim Maydanskiy for providing the following proof.}

Take a plane perpendicular to $\vec\reeb$, it cuts off the cone at a base which is a convex polygon. For a lattice point within this cut off cone, we draw a ray from the origin passing this point, this ray then pierces the base at some point. So long as we have a cut off, there are finite number of such piercing points. Then one is able to push the sides of the polygon out slightly so that all piercing points are in the interior of the polygon, and yet no extra points are encompassed. Such technicality is for handling the points that were exactly on the edge of the faces, as we shall later use winding number to count points. When all is done, we shall take the cut off to infinity and finish the proof.

For convenience of drawing, we rotate the cone with an $SO(3)$ element so that $\vec\reeb=[0,0,1]$, and we shall only draw the base. Since now $\vec\reeb$ is the $z$-direction, in computing $s^{1,2}$, the $z$-component of $\vec\ep$ may well be set to zero. After a further rotation, we set $\vec\ep=[0,-1,0]$, arriving at the figure
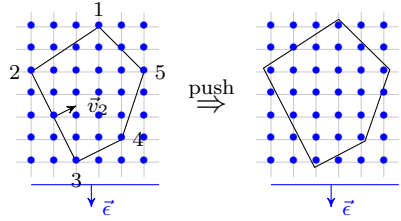
\begin{center}
\begin{tikzpicture}[scale=1]
\draw [step=0.3,thin,gray!40] (-.2,-0.2) grid (1.7,2.0);

\draw [-] (0.9,1.8) node [above] {\scriptsize{$1$}} -- (0,1.2) node [left] {\scriptsize{$2$}}
-- (0.6,0) node [below] {\scriptsize{$3$}} -- (1.2,0.3) node [right] {\scriptsize{$4$}} -- (1.5,1.2) node [right] {\scriptsize{$5$}} -- (0.9,1.8);
\draw [->] (0.3,.6) -- (0.6,.75) node[right] {\scriptsize{$\vec v_2$}};

\foreach \x in {0,1,...,5}
\node at (0.3*\x,0) [blue] {$\sbullet$};
\foreach \x in {0,1,...,5}
\node at (0.3*\x,0.3) [blue] {$\sbullet$};
\foreach \x in {0,1,...,5}
\node at (0.3*\x,.6) [blue] {$\sbullet$};
\foreach \x in {0,1,...,5}
\node at (0.3*\x,.9) [blue] {$\sbullet$};
\foreach \x in {0,1,...,5}
\node at (0.3*\x,1.2) [blue] {$\sbullet$};
\foreach \x in {0,1,...,5}
\node at (0.3*\x,1.5) [blue] {$\sbullet$};
\foreach \x in {0,1,...,5}
\node at (0.3*\x,1.8) [blue] {$\sbullet$};

\node at (2.4,0.9) [black] {$\stackrel{{\rm push}}{\To}$};

\draw [-,blue] (0,-0.3) -- (1.7,-0.3);
\draw [->,blue] (0.8,-0.3) -- (0.8,-.6) node[right] {\scriptsize{$\vec\ep$}};
\end{tikzpicture}
\begin{tikzpicture}[scale=1]

\draw [step=0.3,thin,gray!40] (-.2,-0.2) grid (1.7,2.0);

\draw [-] (0.9,1.9) -- (-0.1,1.25) -- (0.58,-0.07) -- (1.25,0.28) -- (1.58,1.23) -- (0.9,1.9);

\foreach \x in {0,1,...,5}
\node at (0.3*\x,0) [blue] {$\sbullet$};
\foreach \x in {0,1,...,5}
\node at (0.3*\x,0.3) [blue] {$\sbullet$};
\foreach \x in {0,1,...,5}
\node at (0.3*\x,.6) [blue] {$\sbullet$};
\foreach \x in {0,1,...,5}
\node at (0.3*\x,.9) [blue] {$\sbullet$};
\foreach \x in {0,1,...,5}
\node at (0.3*\x,1.2) [blue] {$\sbullet$};
\foreach \x in {0,1,...,5}
\node at (0.3*\x,1.5) [blue] {$\sbullet$};
\foreach \x in {0,1,...,5}
\node at (0.3*\x,1.8) [blue] {$\sbullet$};

\draw [-,blue] (0,-0.3) -- (1.7,-0.3);
\draw [->,blue] (0.8,-0.3) -- (0.8,-.6) node[right] {\scriptsize{$\vec\ep$}};
\end{tikzpicture}
\end{center}
We assume that $\vec\ep$ is generic, so that no edge of the polygon is horizontal.

The pushing off is designed to agree with \eqref{wghts_12}. For example, at the top corner one has $[\ep,\vec v_1,\vec\reeb]>0$, $[\ep,\vec\reeb,\vec v_5]>0$, and \eqref{wghts_12} says one counts all points on the positive side of (and including those on) the two edges. The pushing off then puts all points strictly to the positive side of the edges. At the corner 2, one has $[\ep,\vec v_2,\vec\reeb]>0$ and one counts points to the positive side of (and including those on) edge 12. But $[\ep,\vec\reeb,\vec v_1]<0$, so one counts points strictly to the negative side of edge 23, again the pushing off excludes exactly those points on edge 23.

To summarise the situation, with the pushing off, we need to show that, by our prescription \eqref{wghts_12}, all points inside the polygon are counted exactly once and points outside are counted zero times. To show this, consider the following propagator
\bea \gt(x,y)=\frac{-ydx+xdy}{x^2+y^2}\nn\eea
which, when integrated along any contour, gives the winding number of the contour round the origin. Let $(x_0,y_0)$ be any point on the plane, consider the integral
\bea \oint_C\gt(x_0-x,y_0-y)\nn\eea
with $C$ going round the edges of the polygon counterclockwise once. We know that the integral is zero for $(x_0,y_0)$ outside the polygon and one for those inside. Next we evaluate this integral in a different way.

Since $\gt$ is closed, and so is locally exact, we can construct its local primitive $d^{-1}\gt(x_0-x,y_0-y)$ with the following prescription.
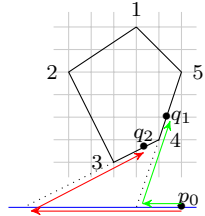
\begin{center}
\begin{tikzpicture}[scale=1]

\draw [step=0.3,thin,gray!40] (-.2,-0.2) grid (1.7,2.0);

\draw [-] (0.9,1.8) node [above] {\scriptsize{$1$}} -- (0,1.2) node [left] {\scriptsize{$2$}}
-- (0.6,0) node [left] {\scriptsize{$3$}} -- (1.2,0.3) node [right] {\scriptsize{$4$}} -- (1.5,1.2) node [right] {\scriptsize{$5$}} -- (0.9,1.8);

\draw [-,blue] (-0.8,-0.6) -- (1.7,-0.6);
\node at (1.5,-0.6) {\scriptsize{$\sbullet$}};
\node at (1.6,-0.5) {\scriptsize{$p_0$}};

\draw[dotted] (1.2,0.3) -- (.9,-0.6);
\draw [->,green] (1.5,-0.55) -- (0.98,-0.55);
\draw [->,green] (0.98,-0.55) -- (1.35,0.55);
\node at (1.3,0.6) {\scriptsize{$\sbullet$}};
\node at (1.5,0.6) {\scriptsize{$q_1$}};

\draw[dotted] (0.6,0) -- (-0.6,-0.6);
\draw [->,red] (1.5,-0.65) -- (-0.5,-0.65);
\draw [->,red] (-0.5,-0.65) -- (1.0,0.13);

\node at (1.0,0.2) {\scriptsize{$\sbullet$}};
\node at (1.0,0.35) {\scriptsize{$q_2$}};
\end{tikzpicture}
\end{center}
One fixes an arbitrary base point $p_0$ on the blue line.
For a point $(x,y)$ on an edge, say 45, one extends the edge downwards until it meets the blue line. Then the value of $d^{-1}\gt(x_0-x,y_0-y)$ is given by integrating $\gt(x_0-x,y_0-y)$ from $p_0$ along the blue line to the point of intersection with edge 45, then one goes upwards along 45 till point $(x,y)$. This local primitive is discontinuous at the corners, in fact, by Stokes theorem, the contour integral of $\gt$ along $C$, and so the winding number of $C$ round $(x_0,y_0)$, is given by the sum of the discontinuities at all corners.

To calculate the discontinuity, e.g. for $q_1,q_2$ very close to corner 4, but on two different edges, it is easy to see that the discontinuity equals the integral of $\gt$ \emph{clockwise} along the triangle bounded by the two dotted lines and the blue line. Such integral gives winding number $-1$ for all points within the said triangle and zero otherwise.
In contrast, for two points close on either sides of corner 1, the discontinuity is given by the integral of $\gt$ \emph{counterclockwise} along the triangle bounded by edges 12, 15 and the blue line. This integral gives winding number $+1$ for all points in the said triangle and zero outside.
More generally, at a corner $i$ between two edges with normals $\vec v_{i-1},\vec v_i$, the discontinuity is always the integral of $\gt$ in the triangle bounded by the (extension of) the two edges and the blue line, we just need to find the orientation. The edge with normal $\vec v_{i-1}$ (resp. $\vec v_i$) is extended in the direction $\sgn([\vec v_{i-1},\vec \reeb,\vec\ep])\vec v_{i-1}\times\vec \reeb$ (resp. $\sgn([\vec v_i,\vec \reeb,\vec\ep])\vec v_i\times\vec \reeb$) to meet the blue line, the orientation is thus the sign
\bea &\sgn\big(-\sgn([\vec v_{i-1},\vec \reeb,\vec\ep])\vec v_{i-1}\times\vec \reeb\big)\times\big( \sgn([\vec v_i,\vec \reeb,\vec\ep])\vec v_i\times\vec \reeb\big)\cdotp\vec\reeb\big)\nn\\
&=-\sgn\big([\vec v_{i-1},\vec \reeb,\vec\ep][\vec v_i,\vec \reeb,\vec\ep][\vec v_{i-1},\vec v_i,\vec \reeb]\big)=\sgn\big([\vec\reeb,\vec v_{i-1},\vec\ep][\vec v_i,\vec \reeb,\vec\ep]\big).\nn\eea
This agrees completely with \eqref{wghts_12}.

In summary the equivariant index of $\bar\partial_H$ reads
\bea \ind(\bar\partial_H)=\sum_{\vec m\in\BB{Z}^3\cap C}(\vec t\;)^{\vec m}+\sum_{\vec m\in\BB{Z}^3\cap(-C^{\circ})}(\vec t\;)^{\vec m},\label{equiv_index}\eea
where $C$ is the cone encoding the toric Sasaki geometry, $-C$ is its negative cone and $C^{\circ}$ means the interior.

\section{Weil/Kalkman Model for Equivariant Cohomology}\label{sec_WKMfEC}
We give a quick review of the Weil/Kalkman model for equivariant cohomology.

Let a Lie group $G$ with Lie algebra $\FR{g}$ act on a manifold $M$. We denote with $\xi\in\FR{g}$ also the fundamental vector field of the action by $\xi$. If the $G$-action is free then $M/G$ is smooth and the differential forms of $M/G$ can be extracted as the basic forms
\bea \Go^{\sbullet}(M/G)\simeq \Go_{\rm basic}^{\sbullet}(M).\nn\eea
Recall that a form $\ga$ is basic iff 1. it is horizontal: $\iota_{\xi}\ga=0$ for $\xi\in\FR{g}$, and 2. it is $G$-invariant.

However if the $G$-action is not free, one has a replacement of $\Go^{\sbullet}(M/G)$ as $\Go_G^{\sbullet}(M):=\Go^{\sbullet}(EG\times_GM)$, where $EG$ is the universal $G$-bundle over the classifying space $BG$. For readers unfamiliar with such jargons, it suffices to think of $EG$ as a principal $G$-bundle such that $EG$ is contractible. As we are not so much concerned with the topology as with the de Rham algebra $\Go_G^{\sbullet}$, we just say that the differential forms of $EG$ are generated by $\FR{g}$-valued odd/even generators $c,\phi$ with degree 1 and 2. These two generators are meant to model abstractly $\vgt$ and $\vrho$, indeed one has two differentials
\bea
\begin{array}{l|l}
  \gd c=\phi, & \gdh c=c^2,\\
  \gd\phi=0, & \gdh \phi=[c,\phi], \\
  \end{array},\nn\eea
with the sum $\gd+\gdh$ satisfies
\bea (\gd+\gdh)c=\phi+c^2,~~(\gd+\gdh)\phi=[c,\phi].\nn\eea
The first mimics universal relation between connection and curvature $\vrho=d\vgt+\vgt\vgt$, while the second mimics the Biancchi identity of the curvature. Therefore $c/\phi$ are often called the universal connection/curvature.

The de Rham algebra of $EG$ interacts with that of $M$ via the $G$-action resulting in the Weil/Kalkman differential. To write it, we pick a basis $\{\xi_a\}$ for $\FR{g}$
\bea d_w\ga=d\ga-L_c\ga+\iota_{\phi}\ga:=d\ga-c^aL_{\xi_a}\ga+\phi^a\iota_{\xi_a}\ga,~~~\ga\in\Go^{\sbullet}.\label{Kalkman_diff}\eea
This differential is isomorphic to the usual de Rham via a conjugation by $e^{c^a\iota_{\xi_a}}$ see \cite{kalkman1993}, and the above algebra is called the Kalkman algebra. It is a variation of (and isomorphic to) the Weil algebra, but more adapted for physicists due to its affinity with BRST.

With these preparations one extracts $\Go_G^{\sbullet}(M)$ from the $G$-basic part of $\Go^{\sbullet}(EG\times M)$.
In contrast to Weil algebra, in the Kalkman algebra context, horizontal means simply the absence of $c$. This is familiar to physicist since $c$ is the ghost and should not appear in physical quantities.
The $G$-invariant part of $\Go^{\sbullet}(EG\times M)$ can usually be enforced by averaging over $G$.

When the $G$-action is free one can explicitly realise the isomorphism $\Go_G^{\sbullet}(M)\simeq\Go^{\sbullet}(M/G)$ as follows. Since $G$ acts freely, one regards $M$ as a principal bundle over $M/G$ and therefore one can get the connection and curvature of this bundle. One replaces $c,\phi$ in $\Go^{\sbullet}_G(M)$ with the actual connection and curvature, in particular $\phi$ is no longer a free generator but becomes torsion. This way one gets a closed basic form on $M$ i.e. a closed form on $M/G$. This is nothing but the Chern-Weil construction of characteristic classes of bundles.
While if the action is not free, there will be some $\phi$ that cannot not be replaced with the curvature and remain a free generator in the de Rham algebra, and such free generator always signals non-free action.

\subsection{Equivariance formulated on the principal bundle}\label{sec_Efotpb}
Here we discuss the technical issue that arises when the associated bundle $P\times_GV$ is $\Gc$-equivariant, but $P$ is not.

We have introduced the complex \eqref{dR_Weil_equiv} in the main text involving
$C_{\sbullet,\sbullet}:\,\textrm{Lie}\,\Gc\times \textrm{Lie}\,G\to \textrm{Lie}\,G$ that measures the non-commutativity between $\Gc$-'action' and the right $G$-action on $P$. It satisfies
\bea [R_u,Y_{\xi}]=R_{C_{\xi,u}},~~~u\in{\rm Lie}\,G,~~\xi\in{\rm Lie}\,\Gc\label{infts_RC}.\eea
Furthermore $T_{\xi}$ is the 'action' of $\xi\in \textrm{Lie}\,\Gc$ on $V$, satisfying
\bea (R_u\circ T_{\xi})+[u,T_{\xi}]=-C_{\xi,u}.\label{infts_RTC}\eea
As $Y_{\xi}$ and $T_{\xi}$ are not from bona fide actions, the commutativity fails according to
\bea &[Y_{\xi_2},Y_{\xi_1}]=Y_{[\xi_1,\xi_2]}+R_{Z_{\xi_1,\xi_2}},\nn\\
    &Y_{\xi_2}\circ T_{\xi_1}-[1\leftrightarrow2]+[T_{\xi_1},T_{\xi_2}]=-R_{Z_{\xi_1,\xi_2}}+T_{[\xi_1,\xi_2]},\nn\eea
where $Z_{\xi_1,\xi_2}=-Z_{\xi_2,\xi_1}$ and it in turn satisfies the coherence relation
\bea C_{\xi,[u,v]}=[C_{\xi,u},v]+R_u\circ C_{\xi,v}-[u\leftrightarrow v].\label{infts_C}\eea
\begin{remark}
  The coherence relations, e.g. \eqref{infts_RC} are derived by noting that the noncommutativity of $\Gc$ and $G$ actions can be fixed by another $G$-action (since $P\times_GV$ does have a $\Gc$ action), then by differentiating this coherence one can get the infinitesimal version \eqref{infts_RC}.

  Secondly, the list above is far from complete, since what we have trying to write down here is essentially the coherence relations of a 2-category, i.e. think of the points of $P\times V$ as objects, and $\Gc$ action as morphisms. The morphisms only compose correctly only up to $G$ action (2-morphisms), in the same way associativity of composition holds up to $G$ actions etc.
\end{remark}

\providecommand{\href}[2]{#2}\begingroup\raggedright

\bibliographystyle{utphys}
\bibliography{Bib_common}{}

\end{document}